\documentclass{article}
\PassOptionsToPackage{numbers, compress}{natbib}

\usepackage[final]{neurips_2022}
\usepackage{amsmath}
\usepackage{amssymb}
\usepackage{enumitem}
\usepackage{mathtools}
\usepackage{amsthm}
\usepackage{mathrsfs}
\usepackage{multirow}
\usepackage{subfigure}
\usepackage{titletoc}
\usepackage{bm}
\usepackage{colortbl}
\usepackage{wrapfig}
\usepackage{tabularx}
\usepackage{caption}
\usepackage{multirow}
\usepackage{booktabs}
\usepackage{url} 
\usepackage{tcolorbox}     
\usepackage[framemethod=tikz]{mdframed}

\usepackage[colorlinks,
linkcolor=blue,
anchorcolor=pink,
citecolor=orange
]{hyperref}

\usepackage[page,header]{appendix}

\theoremstyle{plain}
\newtheorem{theorem}{Theorem}

\newtheorem{lem}{Lemma}
\newtheorem{coro}{Corollary}
\theoremstyle{definition}
\newtheorem{defi}{Definition}
\newtheorem{assu}{Assumption}
\newtheorem{rem}{Remark}

\newtheorem*{rthm1}{\textbf{Restate of Theorem} \ref{them1}}
\newtheorem*{rlem1}{\textbf{Restate of Lemma} \ref{covering_lem}}

\newtheorem*{rdef2}{\textbf{Restate of Definition} \ref{def2}}
\newtheorem*{rdef3}{\textbf{Restate of Definition} \ref{def3}}
\newtheorem*{rassu1}{\textbf{Restate of Assumption} \ref{assu1}}
\newtheorem*{rcor1}{\textbf{Restate of Corollary} \ref{cor1}}

\DeclareMathOperator*{\expe}{\mathbb{E}} 
\DeclareMathOperator*{\mcu}{\mathcal{U}} 
\DeclareMathOperator*{\mci}{\mathcal{I}}
\DeclareMathOperator*{\mcd}{\mathcal{D}}
\DeclareMathOperator*{\mcl}{\mathcal{L}}
\DeclareMathOperator*{\mcr}{\mathcal{R}}
\def \bmg{\boldsymbol{g}}

\usepackage{xcolor}
\definecolor{ballblue}{HTML}{338EA7}
\definecolor{lightseagreen}{HTML}{759D39}
\definecolor{lightred}{HTML}{DD7769}
\definecolor{org}{HTML}{F8A145}
\definecolor{blu}{HTML}{63ACE5}
\definecolor{lightseagreen}{HTML}{759D39}

\usepackage[textsize=tiny]{todonotes}

\tcbuselibrary{skins}
\newenvironment{qbox}
{\begin{tcolorbox}[enhanced jigsaw, drop shadow=black!50!white,colback=white, width=0.95\linewidth, center, left=2pt,right=2pt,top=1pt,bottom=1pt]}
{\end{tcolorbox}}

\title{The Minority Matters: A Diversity-Promoting Collaborative Metric Learning Algorithm}

\author{\parbox{14cm}
  {\centering
    {\large  Shilong Bao$^{1,2}$ \ \ \ \ \ \ \ \ \  Qianqian Xu$^{3}$\thanks{Corresponding authors.} \ \ \ \ \ \ \ \ \ Zhiyong Yang$^{4}$ \ \ \ \ \ \ \ \ \  Yuan He$^{5}$  \\ \quad\quad Xiaochun Cao$^{6}$ \ \ \ \ \ \ \ \ \ Qingming Huang$^{3,4,7,8*}$ }\\
    {\normalsize
    $^1$ State Key Laboratory of Information Security, Institute of Information Engineering, CAS \\
    $^2$ School of Cyber Security, University of Chinese Academy of Sciences\\
    $^3$ Key Lab. of Intelligent Information Processing, Institute of Computing Technology, CAS\\
    $^4$ School of Computer Science and Tech., University of Chinese Academy of Sciences\\
    $^5$ Alibaba Group\\
    $^6$ School of Cyber Science and Technology, Shenzhen
Campus, Sun Yat-sen University\\
    $^7$ Key Laboratory of Big Data Mining and Knowledge Management, CAS \\
    $^8$ Peng Cheng Laboratory\\
    }
    {\tt\small baoshilong@iie.ac.cn, xuqianqian@ict.ac.cn, heyuan.hy@alibaba-inc.com, \quad\quad \\ caoxiaochun@mail.sysu.edu.cn, \{yangzhiyong21,qmhuang\}@ucas.ac.cn ~~ 
    }
  }
}

\begin{document}
	\maketitle

	\begin{abstract} 
		Collaborative Metric Learning (CML) has recently emerged as a popular method in recommendation systems (RS), closing the gap between metric learning and Collaborative Filtering. Following the convention of RS, existing methods exploit unique user representation in their model design. This paper focuses on a challenging scenario where a user has multiple categories of interests. Under this setting, we argue that the unique user representation might induce preference bias, especially when the item category distribution is imbalanced. To address this issue, we propose a novel method called \textit{Diversity-Promoting Collaborative Metric Learning} (DPCML), with the hope of considering the commonly ignored minority interest of the user. The key idea behind DPCML is to include a multiple set of representations for each user in the system. Based on this embedding paradigm, user preference toward an item is aggregated from different embeddings by taking the minimum item-user distance among the user embedding set. Furthermore, we observe that the diversity of the embeddings for the same user also plays an essential role in the model. To this end, we propose a \textit{diversity control regularization} term to accommodate the multi-vector representation strategy better. Theoretically, we show that DPCML could generalize well to unseen test data by tackling the challenge of the annoying operation that comes from the minimum value. Experiments over a range of benchmark datasets speak to the efficacy of DPCML.
		
		
		
		
	\end{abstract}
	\section{Introduction}\label{intro}
	Recommender system (RS) is a well-known building block in eCommerce, which can assist buyers to find products they wish to purchase by giving them the relevant recommendations. The key recipe behind RS is to learn from user-item interaction records \cite{DBLP:conf/pakdd/WangZ0HC20, DBLP:conf/nips/MaZ0Y019, DBLP:conf/kdd/MaZYCW020, DBLP:journals/aei/LvZWWW20,DBLP:journals/tkde/JiangCCW0Y15}. In practice, since user preferences are hard to collect, such records often exist as implicit feedback \cite{DBLP:conf/nips/WangGZZ18,DBLP:conf/sigir/AskariSS21,DBLP:conf/www/TogashiKOS21} where only indirect actions are provided (say clicks, collections, reposts, and etc.). Such a property of implicit feedback raises a great challenge to RS-targeted machine learning methods and thus stimulates a wave of relevant studies along this course \cite{DBLP:conf/icml/XuRKKA21,DBLP:conf/icml/ZhengTDZ16,DBLP:conf/aaai/WangWSSL20}. 
	
	
	Over the past two decades, most literature follows a typical paradigm known as the One-Class Collaborative Filtering (OCCF) \cite{DBLP:conf/icdm/PanZCLLSY08}, where the items not being observed are usually assumed to be of less interest for the user and labeled as negative instances. In the early days, the vast majority of studies in the OCCF community focus on Matrix Factorization (MF) based algorithms, where the preference of a specific user to an item is conveyed by the inner product between their embeddings \cite{DBLP:journals/ijon/ZhangR21, DBLP:conf/aaai/ChenL019}. Recently, a milestone study known as \textit{Collaborative Metric Learning} (CML) \cite{hsieh2017collaborative} pointed out that the inner product involved in MF violates the triangle inequality, resulting in a sub-optimal topological embedding space. To fix this, CML proposes a novel framework to overcome such a problem by borrowing the strength from metric learning. Practically, CML has achieved promising performance over a series of RS benchmark datasets. Hereafter, many efforts have been made along the research direction to improve CML
	\cite{tran2019improving,DBLP:conf/www/TayTH18, DBLP:conf/icdm/ParkKXY18, DBLP:conf/mm/BaoXMYCH19,DBLP:journals/nn/WuZNC20,wang2019group, DBLP:conf/ijcai/ZhouLL019,DBLP:conf/dasfaa/ZhangZLXF0SC19, DBLP:conf/recsys/TranSHM21}. More discussions of the related work are presented in Appendix.\ref{rel_work}.
	
	However, through the lens of a critical example in the practical scenarios (shown in Sec.\ref{Sec3.2}), we notice that users usually have multiple categories of preferences in real-world RS. Moreover, such interest groups are often not equally distributed, where the amount of some groups dominates the others.  Unfortunately, as shown in Fig.\ref{motivation}, in this case, the existing studies might induce preference bias since they tend to meet the majority interest while missing the other potential preference.  Therefore, in this paper, we ask:

\begin{qbox}
    \begin{center}
      \textit{How to develop an effective CML-based algorithm to accommodate the diversity of user preferences?}
    \end{center}
\end{qbox}
	
\textbf{Contributions.}	In search of an answer, we propose a novel algorithm called \textit{Diversity-Promoting Collaborative Metric Learning (DPCML)}. The key idea is to explore the diversity of user interest which spans multiple groups of items. To this end, we propose a multi-vector user representation strategy, where each user has a set of $C$ embeddings. To find out the score of a given item embedding $\boldsymbol{g}_v$, we aggregate the results from the user embeddings $\boldsymbol{g}^1_u,\boldsymbol{g}^2_u,\cdots, \boldsymbol{g}^C_u$ by taking the minimum distance  $s(u,v) =\min\limits_{c} \|\boldsymbol{g}^c_u - \boldsymbol{g}_v\|^2$. Then we will recommend the item with the smallest $s$ value. In this way, we can focus on all potential items that fit one of the users' interests well, both for the majority and the minority interests. Meanwhile, we observe that the diversity of the embeddings among the same user representation set also plays an important role in better achieving our goal. Therefore, we further present a novel diversity control regularization scheme.
	
	Taking a step further, we continue to ask the following question:
	
\begin{qbox}
    \begin{center}
      \textit{Could CML generalize well under the multi-vector representation strategy?}
    \end{center}
\end{qbox}
	To the best of our knowledge, such a problem remains barely explored in the existing literature.  To solve the problem, we then proceed to explore the generalization bound for DPCML algorithm. Here the major challenges fall into two aspects: 1) The risk of DPCML could not be expressed as a sum of independently identically distributed (i.i.d.) loss terms, making the standard Rademacher Complexity-based \cite{DBLP:conf/colt/BartlettM01, DBLP:books/daglib/0034861} theoretical arguments unavailable; 2) The annoying minimum operation are not continuous, which cannot be analyzed easily in the Rademacher complexity framework. Facing these challenges, we employ the covering number and $\epsilon$-net arguments to derive the generalization bound. On top of this, we show that DPCML could induce a small generalization error with high probability. This supports the effectiveness of DPCML from a theoretical perspective.
	
	Finally, we conduct empirical studies over a range of RS benchmark datasets that demonstrate the superiority of DPCML.  
	
	
	

	\section{Methodology}
	\subsection{Preliminary}
	In this paper, we focus on how to develop an effective CML-based recommendation system on top of the implicit feedback signals (say clicks, browses, and bookmarks).
	Assume there are a pool of users and items in the system, denoted by $\mathcal{U}=\{u_1, u_2, \dots, u_{|\mathcal{U}|}\}$ and $\mathcal{I} = \{v_1, v_2\, \dots, v_{|\mathcal{I}|}\}$, respectively. For each user $u_i \in \mathcal{U}, i=1, 2, \dots, |\mathcal{U}|$, let $\mathcal{D}^+_{u_i} = \{v_1^+, v_2^+, \dots, v_{n_{i}^+}^+\}$ denote the set of items that user $u_i$ has interacted with  (i.e., observed user-item interactions) and the rest of the items (i.e., unobserved interactions) are denoted by $\mathcal{D}^{-}_{u_i} = \{v_1^-, v_2^-, \dots, v_{n_{i}^-}^-\}$, where $n_i^+,n_i^-$ are the number of observed/unobserved interactions of user $u_i$. We have $\mathcal{I} = \mathcal{D}_{u_i} = \mathcal{D}^+_{u_i} \cup \mathcal{D}^-_{u_i}$ and $|\mathcal{I}| = n_i^+ + n_i^-$. In the standard settings of OCCF, one usually assumes that users tend to have a higher preference for the items contained in $\mathcal{D}^+_{u_i}$ than the items in $\mathcal{D}^-_{u_i}$. Therefore, given a target user $u_i \in \mathcal{U}$ and his/her historical interaction records, the goal of RS is to discover the most interested $N$ items by recommending the items with the top-$N$ (bottom-$N$) score. The top-$N$ item list is denoted as $\mathcal{I}_N^{u_i}$.

		\begin{figure}[!t]
	\centering
	\begin{minipage}[b]{0.45\columnwidth}
		\subfigure[MovieLens-1M]{
			\includegraphics[width=0.45\columnwidth]{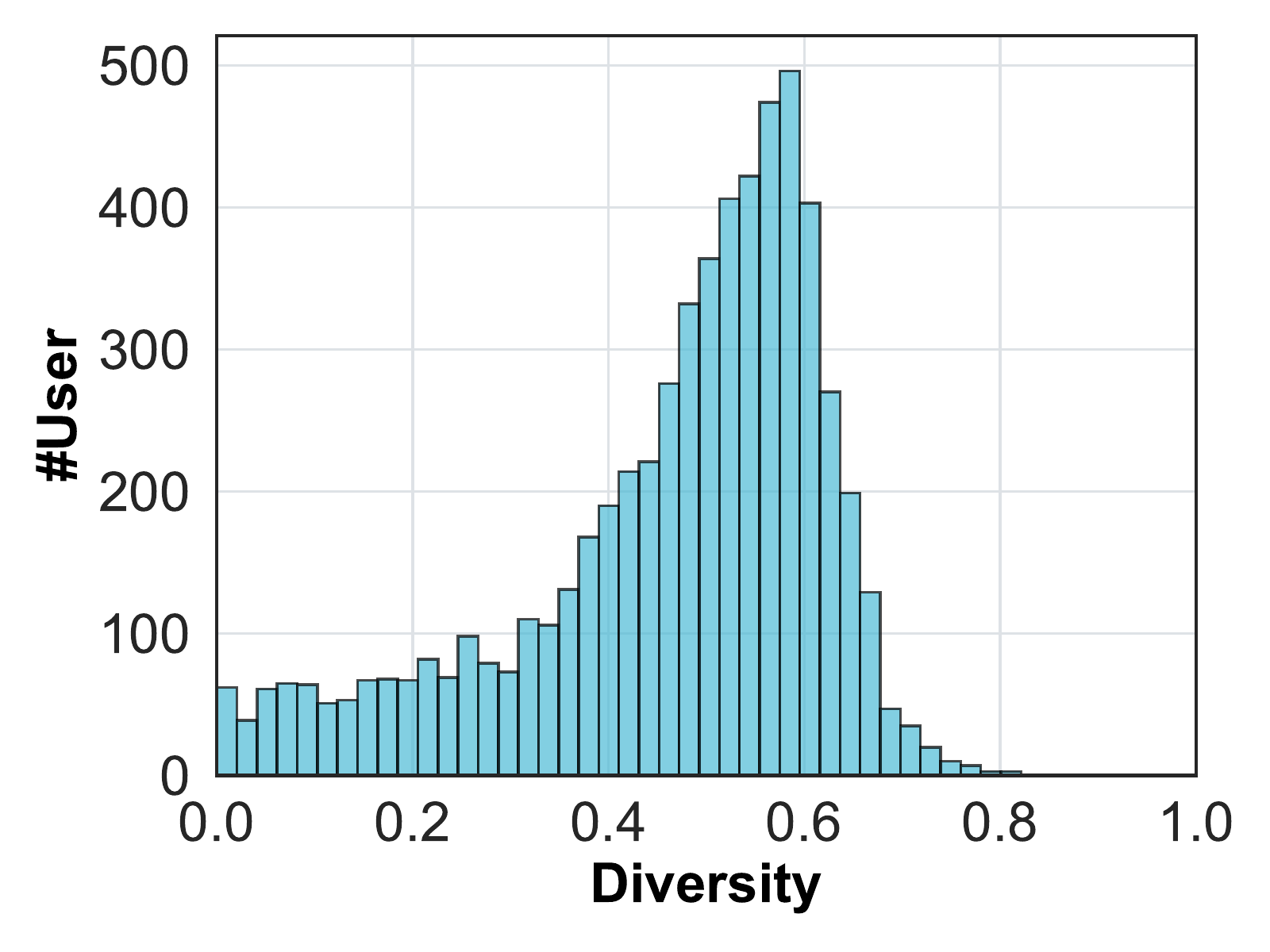}
			\label{ml-1m_div}
		}
		\subfigure[MovieLens-10M]{
			\includegraphics[width=0.45\columnwidth]{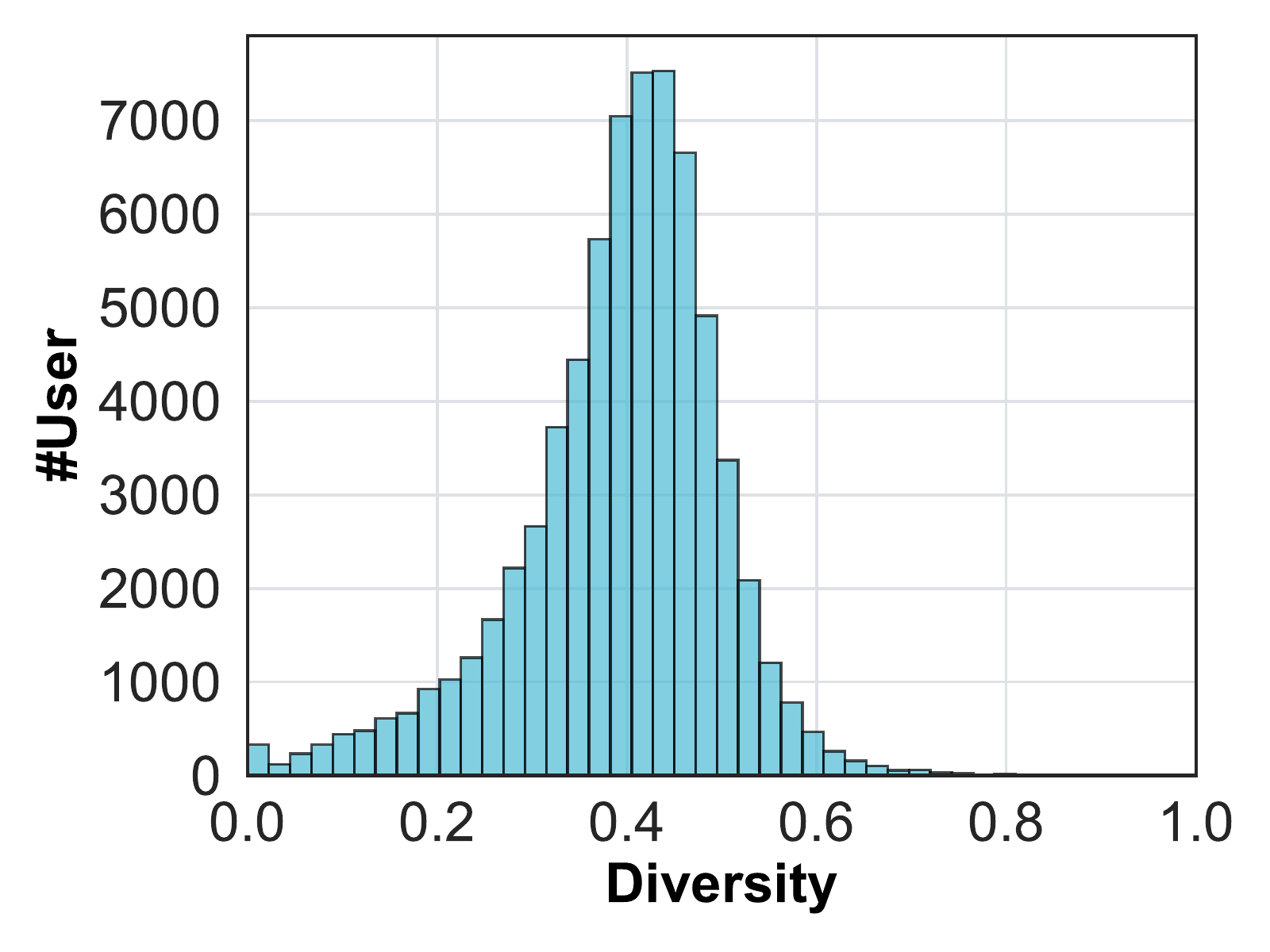}
			\label{ml-10m_div}
		}
		\caption{Statistics of preference diversity.}
		\label{diversity_10m}
	\end{minipage}
	\hspace{0.1cm}
	\begin{minipage}[b]{0.45\columnwidth}
		\centering
		\subfigure[MovieLens-1M]{
			\includegraphics[width=0.45\columnwidth]{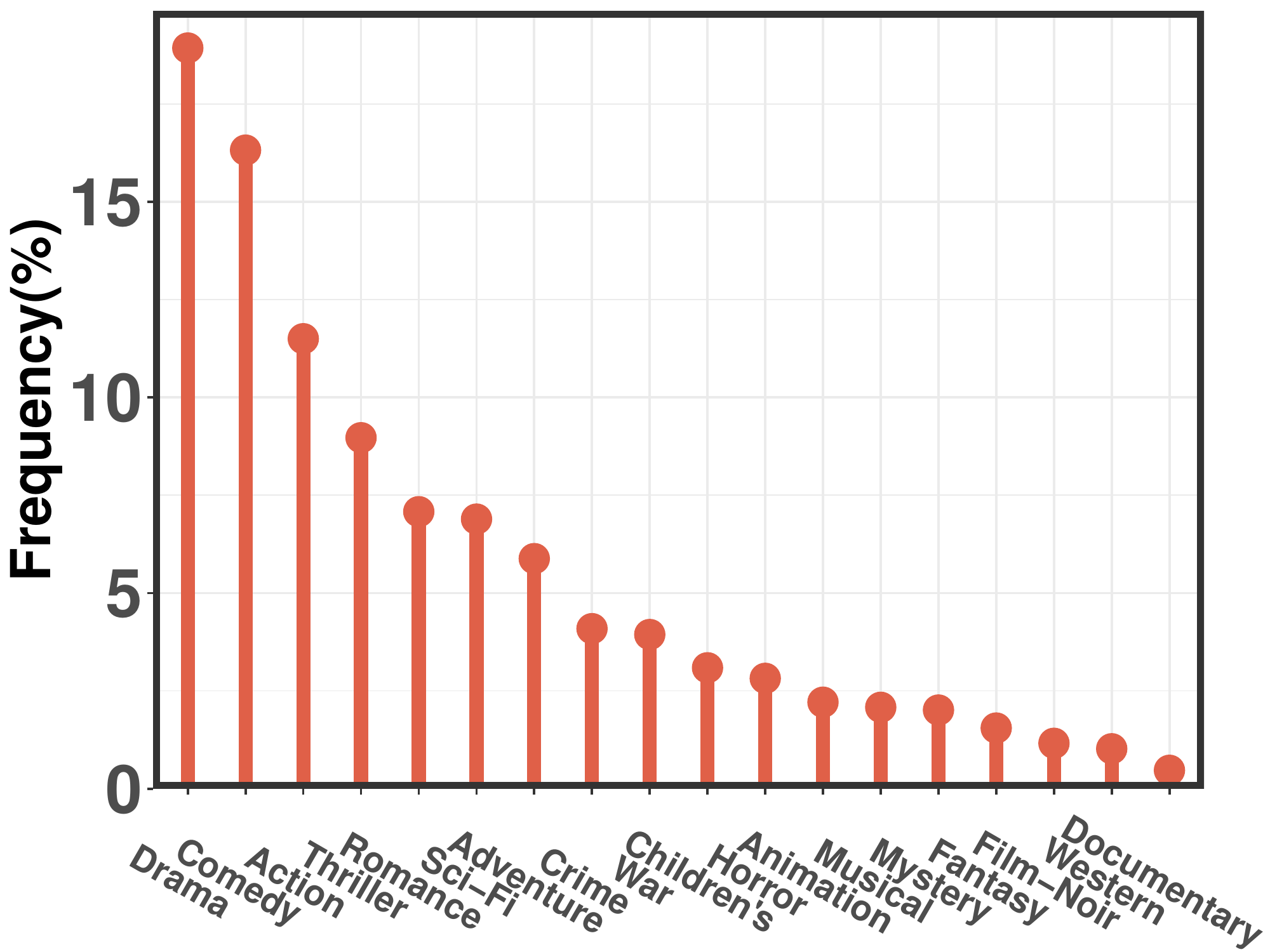}
			\label{ml-1m_freq}
		}
		\subfigure[MovieLens-10M]{
			\includegraphics[width=0.45\columnwidth]{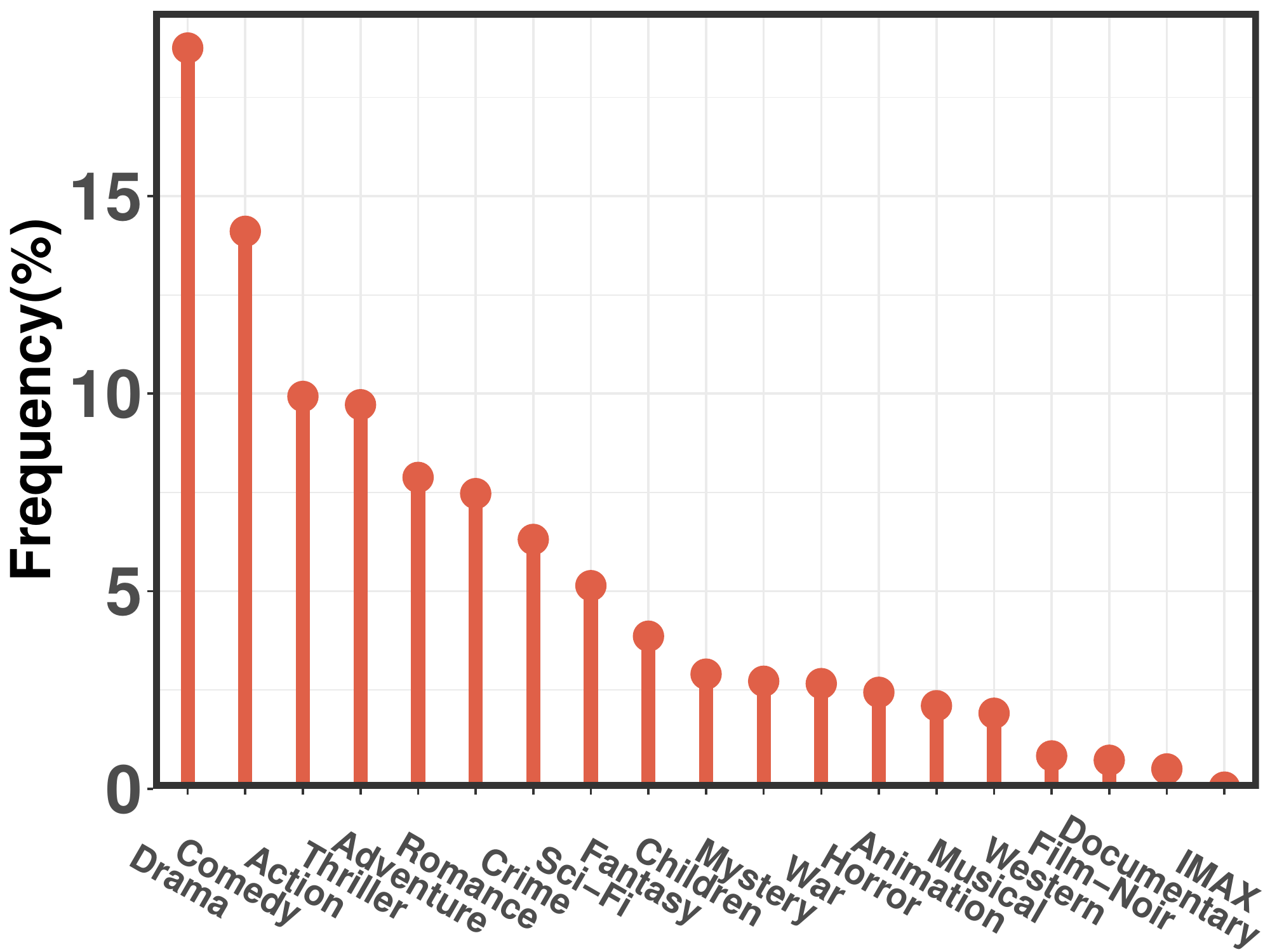}
			\label{ml-10m_freq}
		}
		\caption{The item category distribution.}
		\label{freq}
	\end{minipage}
	\label{fig:all}
\end{figure}
		  \begin{wrapfigure}{r}{0.45\columnwidth}
      \begin{center}
        \includegraphics[width=0.45\columnwidth]{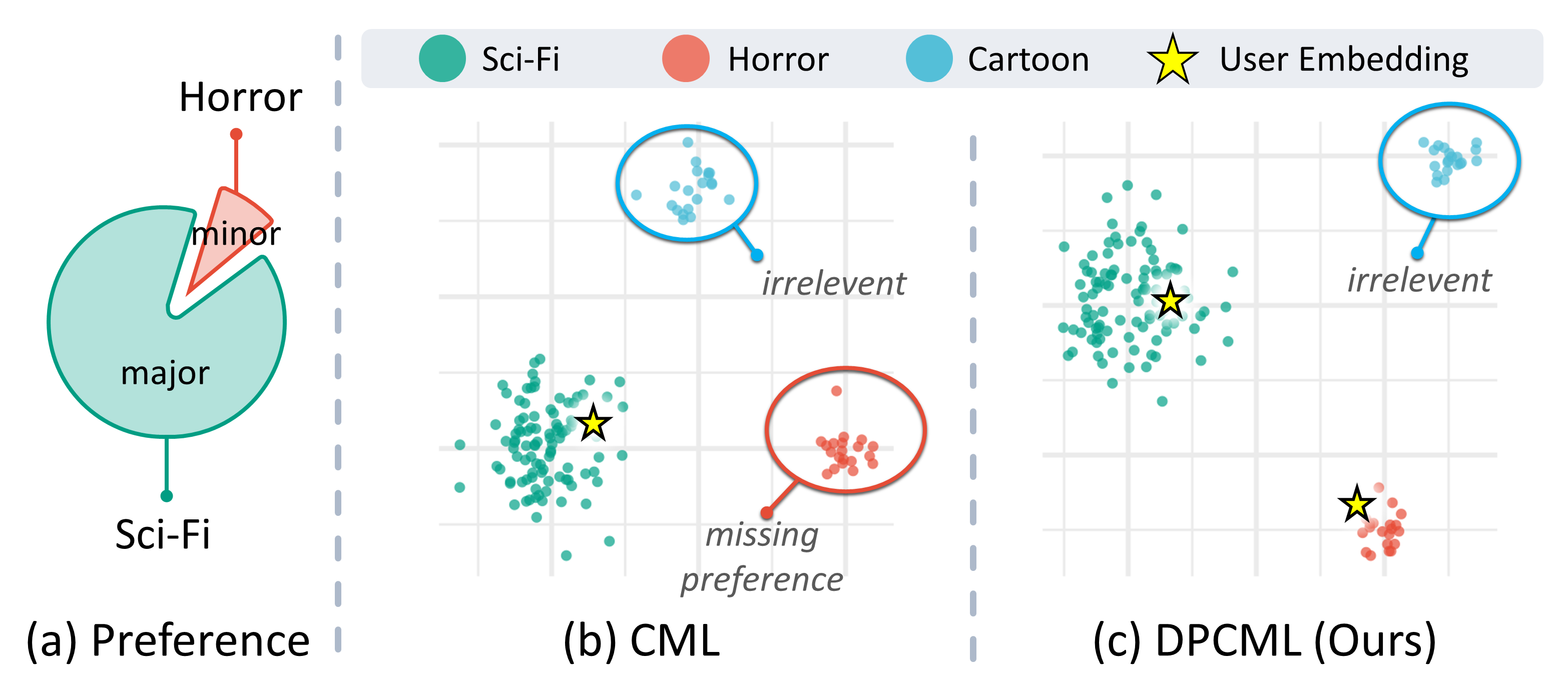}
      \end{center}
     \caption{An illustration shows the benefit of our proposed algorithm when a user has multiple categories of preferences. Taking movies as an example, we assume that Sci-Fi/Horror is the majority/minority interest of the user while Cartoon is an irrelevant movie type. It is easy to see that if the item embeddings are distributed as shown in the figure, we can hardly find a single user embedding that simultaneously captures both interests.}
		\label{motivation}
    \end{wrapfigure}
	\subsection{Motivating Example} \label{Sec3.2}
	We start by a definition of the preference diversity of users.
	    
		\begin{defi} [Preference Diversity] \label{div_defi} Assume that there exists an attribute set $\mathcal{T} = \{\mathcal{T}(v_1), \mathcal{T}(v_2), \dots, \mathcal{T}(v_{|\mci|})\}$ in a typical RS, where $\mathcal{T}(v_j) = \{t_1, t_2, \dots, t_{T_j}\}$ contains the attribute information of item $v_j$ (e.g., the genres of a movie) and $T_j$ is the number of attributes. Given a user $u_i$ and interaction records $\mathcal{D}_{u_i}^+$, the preference diversity is defined as follows:
			\begin{equation} \nonumber
				\begin{aligned}
					\mathsf{Div}(u_i) = \frac{\sum\limits_{v_j, v_k \in\mathcal{D}_{u_i}^+, v_j \neq v_k} \mathbb{I}\left[\mathcal{T}(v_j) \cap \mathcal{T}(v_k) = \varnothing \right]}{|\mathcal{D}_{u_i}^+| (|\mathcal{D}_{u_i}^+| - 1)},
				\end{aligned}
			\end{equation}
			where $\mathbb{I}(x)$ is an indicator function, i.e., returns $1$ if the condition $x$ holds, otherwise $0$ is returned.
		\end{defi}
	  
		\begin{rem}
			Intuitively, the range of $\mathsf{Div}(u_i)$ is among $[0, 1]$, and its value measures the diversity of $u_i$'s preference to a certain extent. That is to say, if items among the historical interaction records of users are irrelevant, there should induce a large value (e.g., $\mathsf{Div}(u_i) = 1$), implying the diversity of their preferences. If the opposite is the case, the value is small. This means users may have narrow interests where only some unique attributes appeal to them.
		\end{rem}
    
	Based on Def.\ref{div_defi}, we visualize the user preferences on two real-world benchmark datasets, including \textbf{MovieLens-1m} and \textbf{MovieLens-10m}. The detailed information of datasets is listed in Tab.\ref{table1} in Appendix.\ref{exp_supp}. Here we adopt the movie genres as the attribute set $\mathcal{T}$ because such information is easy to obtain. The results are shown in Fig.\ref{diversity_10m}. From the results, we can make the following observations. First, only a few users have limited interest. Moreover, most of the users have a diversity value spaning $(0, 0.8]$, suggesting that they have multiple categories of interests. Finally, one can notice that there are very few users with high preference diversity (at the lower-right corner) in both figures. This is a convincing case in the real-world recommendation since most users usually have interests in a couple of movie genres but not all.
	
	\textbf{Motivation and Discussion}. Through the above example, the key information is that users usually have multiple categories of preference in real-world recommendations. This poses a critical challenge to the current CML framework. Specifically, following the convention of RS, the existing CML-based methods leverage unique representations of users to model their preferences. Facing the multiplicity of user intentions, such a paradigm may induce preference bias due to the limited expressiveness, especially when the item category distribution is imbalanced. Fig.\ref{freq} visualizes the item distribution on MovieLens-1m and MovieLens-10m datasets. We see that both of them are imbalanced. In this case, as shown in Fig.\ref{motivation}-(b), CML would pay more attention to the \textbf{majority} interest of users making the unique user embedding close to the items with the science fiction (Sci-Fi) category. In this way, the \textbf{minority}  interest of the user (i.e., Horror movies) would be ignored by the method, inducing performance degradation. This motivates us to explore diversity-promoting strategies on top of CML.

	\subsection{Diversity-Promoting Collaborative Metric Learning}
	Recall that the critical recipe behind CML-based algorithms is to seek a metric space (usually adopting the Euclidean space) such that user preferences could be naturally specified by their distance toward different items. To do this, the traditional CML-based methods usually represent each user and each item as a vector, respectively. Different from them, taking the preference diversity of users into account, we propose to adopt $C$ ($C > 1$) different embeddings for each user and represent each item as one single vector in a joint Euclidean space. 
	
	Concretely, each user $u_i$ is projected into the metric space via the following lookup transformations \cite{DBLP:conf/wsdm/WangF0NC21, DBLP:conf/sigir/Wang0WFC19, DBLP:conf/icml/WuZGYZ21}:
	
	\begin{equation}
		\boldsymbol{g}_{u_i}^c = \boldsymbol{P}_{c}^\top\boldsymbol{u}_i, \ \ \forall c, u_i, \ c \in [C], \ \  u_i \in \mcu,
	\end{equation}
	where $\boldsymbol{g}_{u_i}^c \in \mathbb{R}^d$ is a representation vector of user $u_i$; $[C]$ is the set $\{1, 2, \dots, C\}$; $\boldsymbol{P}_{c} \in \mathbb{R}^{|\mathcal{U}| \times d}$ is a learned transformation weight; $d$ is the dimension of space and $\boldsymbol{u}_i \in \mathbb{R}^{|\mcu|}$ is a one-hot encoding that the nonzero elements correspond to its index of a particular user ${u}_i$.
	
	Similarly, we apply the following transformation to each item $v_j$:
	\begin{equation}
		\boldsymbol{g}_{v_j} = \boldsymbol{Q}^\top \boldsymbol{v}_j, \ \ \forall v_j \in \mci,
	\end{equation}
	where $\boldsymbol{g}_{v_j} \in \mathbb{R}^d$ is the embedding of item $v_j$; $\boldsymbol{Q} \in \mathbb{R}^{|\mathcal{I}| \times d}$ is the learned transformation weight and $\boldsymbol{v}_j \in \mathbb{R}^{|\mathcal{I}|}$ is a one-hot embedding of item $v_j$.
	
	In what follows, given a target user $u_i$, we need to find out a score function to express the user preference toward an item in the context of multiple representations of users. Here we define the score function by taking the minimum item-user Euclidean distance among the user embedding set:
	\begin{equation}\label{eq3}
		s(u_i, v_j) = \min\limits_{c \in [C]} \|\boldsymbol{g}_{u_i}^c - \boldsymbol{g}_{v_j}\|^2, \forall v_j \in \mci.
	\end{equation}
	
	Equipped with this formulation, we focus on the potential items that fit one of the user preferences. If user $u_i$ has interacted with item $v_j$, there should be a small value with respect to $s(u_i, v_j)$. If the opposite is the case, we then expect to see a large $s(u_i, v_j)$. Mathematically, the following inequality should be satisfied to reflect the relative preference of $u_i$ in the learned Euclidean space:
	\begin{equation}
		\begin{aligned}
			\ \  s(u_i, v_j^+) < s(u_i, v_k^-), & \ \forall v_j^+ \in \mathcal{D}_{u_i}^+, \ \forall v_k^- \in \mathcal{D}_{u_i}^-. \\
		\end{aligned}\label{Eq3ss}
	\end{equation}  
	
	Therefore, given the whole sample set $\mathcal{D} = \mathop{\cup}\limits_{u_i \in \mathcal{U}} \ \mathcal{D}_{u_i}$, we adopt the following pairwise learning problems \cite{hsieh2017collaborative, tran2019improving,DBLP:conf/nips/LeiLK20,DBLP:journals/corr/abs-2203-15046} to achieve such goal:
	\begin{equation} \label{eqn4}
		\min\limits_{\boldsymbol{g}} \ \ \hat{\mathcal{R}}_{\mcd, \bmg},
	\end{equation}
	where, $\forall v_j^+ \in \mathcal{D}_{u_i}^+, \ \forall v_k^- \in \mathcal{D}_{u_i}^-$, we have
	\begin{equation}\nonumber
		\hat{\mathcal{R}}_{\mcd, \bmg} = \frac{1}{|\mcu|} \sum_{u_i \in \mcu} \frac{1}{n_i^+n_i^-} \sum_{j=1}^{n_i^+} \sum_{k=1}^{n_i^-} \ell^{(i)}_g(v_j^+, v_k^-),
	\end{equation}
	\begin{equation}\label{margin_eq}
		\ell^{(i)}_g(v_j^+, v_k^-) = \max (0, \lambda + s(u_i, v_j^+) - s(u_i, v_k^-)).
	\end{equation}
	and $\lambda > 0$ is a safe margin.
	
	According to (\ref{eqn4}), we have the following explanations. At first, optimizing the above problem could pull the observed items close to the users and push the unobserved items apart from the observed items. This achieves our goal of preserving user preferences in the Euclidean space. Then, as shown in Fig.\ref{motivation}-(c), equipped with a multiple set of representations for each user, DPCML would exploit different user vectors to focus on different interest groups. In this sense, the minority interest groups can also be modeled well. Last but not least, one appealing property is that, DPCML also preserves the triangle inequality for the items falling into the same interest group.
	
	
	
	
	\subsection{Diversity Control Regularization Scheme} \label{Sec.3.4}
	In practice, we note that a proper regularization scheme is crucial to accommodate the multi-vector representation strategy. Here we focus on the diversity within the embedding sets of a given user. Such diversity is defined as the average pairwise distance among the $C$ user embeddings for user $u_i$, i.e.,
	\begin{equation*}
		\delta_{\bmg, u_i} = \frac{1}{2C(C-1)} \sum_{c_{1}, c_2 \in [C]} \|\boldsymbol{g}_{u_i}^{c_1} - \boldsymbol{g}_{u_i}^{c_2}\|^2. \\
	\end{equation*}
	Based on the definition, we argue that one should attain a proper $\delta_{\bmg, u_i}$ to get a good performance since extremely large/small values of $\delta_{\boldsymbol{g}, u_i}$ might be harmful for the generalization error. It is easy to see that if $\delta_{\boldsymbol{g}, u_i}$ is extremely small, the embeddings for a given user are very close to each other such that the multi-vector representation strategy degenerates to the original single-vector representation. This increases the model complexity with few performance gains and obviously will induce overfitting. On the other hand, a too large diversity might also induce overfitting. It might be a bit confusing at first glance. But, imagine that when some noise observations or extremely rare interests far away from the normal patterns exist in the data, having a large diversity will make it easier to overfit such data. Moreover, it is also a natural assumption that a user's interests should not be too different, as validated in Fig.\ref{diversity_10m}. In this sense, the distance across different user embeddings should remain at a moderate magnitude. 
	
	Therefore, controlling a proper diversity is essential for the multi-vector representation. To do this, we put forward the following diversity control regularization scheme (DCRS):
	\begin{equation} \label{eqn5}
		\hat{\Omega}_{\mcd, \bmg} = \frac{1}{|\mcu|} \sum_{u_i \in \mcu} \psi_{\bmg}(u_{i}),
	\end{equation}
	where, we have
	\begin{equation}\nonumber
		\begin{aligned}
			\psi_{\bmg}(u_{i}) &= \max\left(0, \delta_1 - \delta_{\bmg, u_i}\right) + \max\left(0, \delta_{\bmg, u_i} - \delta_2\right), \\
		\end{aligned}
	\end{equation}
	and $\delta_1$, $\delta_2$ are two threshold parameters with $\delta_1 \le \delta_2$. Intuitively, optimizing (\ref{eqn5}) ensures that the diversity of user's vectors lies between $\delta_1$ and $\delta_2$. 
	
	\subsection{Optimization}
	Finally, we arrive at the following optimization problem for our proposed DPCML:
	\begin{equation}
		\begin{aligned}
			\min\limits_{\bmg} \hat{\mcl}_{\mcd}(\bmg),
		\end{aligned}
	\end{equation}
	where
	
	\begin{equation}
		\hat{\mcl}_{\mcd}(\bmg) = \hat{\mathcal{R}}_{\mcd, \bmg} + \eta \cdot 	\hat{\Omega}_{\mcd, \bmg},
	\end{equation}
	and $\eta$ is a trade-off hyper-parameter.
	
	When the training is completed,  one can easily carry out recommendations by choosing the items with the smallest $s(u_i, v_j), \forall v_j, v_j \in \mci$. 
	
	\subsection{General Framework of Joint Accessibility} \label{sec.3.6}
	Now, we expect to provide another intriguing perspective of our proposed method. As we discussed in Sec.\ref{Sec.2.2}, equipped with a multiple set of representations for each user, our proposed algorithm could be treated as a generalized framework against the joint accessibility issue. To see this, if we restrict the user and item embeddings within a unit sphere, then the score function (\ref{eq3}) degenerates to :
	\begin{equation}\label{eq3123}
		\begin{aligned}
			s(u_i, v_j) &= \min\limits_{c \in [C]} \left(1 - \hat{\bmg}^c_{u_i} \boldsymbol{g}_{v_j}\right), \\
			s.t. \ \ \|\boldsymbol{g}^c_{u_i}\| &= 1,  \forall u_i \in \mcu,\\
			\ \ \|\bmg_{v_j}\| &= 1,  \forall v_j \in \mci,
		\end{aligned}
	\end{equation}
	where $\hat{\bmg}^c_{u_i} \in \mathbb{R}^{1 \times d}$ represents the transpose vector of $\boldsymbol{g}^c_{u_i} \in \mathbb{R}^d$. Therefore, to minimize (\ref{eq3123}), one only needs to maximize the following equivalent problem:
	
	\begin{equation}\label{mf_3123}
		\begin{aligned}
			\hat{s}(u_i, v_j) &= \max\limits_{c \in [C]} \  \hat{\bmg}^c_{u_i} \boldsymbol{g}_{v_j}, \\
			s.t. \ \ \|\hat{\bmg}^c_{u_i}\| &= 1,  \forall u_i \in \mcu,\\
			\ \ \|\bmg_{v_j}\| &= 1,  \forall v_j \in \mci,
		\end{aligned}
	\end{equation}
	which is exactly the original form of the joint accessibility model.

	\section{Generalization Analysis}
	In this section, we present a systematic theoretical analysis of the generalization ability of our proposed algorithm. Following the standard learning theory, deriving a uniform upper bound of the generalization error relies on the proper measure of its complexity over the given hypothesis space $\mathcal{H}$. The most common complexity to achieve this is the Rademacher complexity \cite{DBLP:conf/colt/BartlettM01, DBLP:books/daglib/0034861, DBLP:journals/ijon/LeiDB16}, which is derived from the symmetrization technique as an upper bound for the largest deviation over a given hypothesis space $\mathcal{H}$:
	
	\[
	\mathbb{E}_{\mathcal{D}}\left[\sup_{f \in \mathcal{H}} \mathbb{E}_{\mathcal{D}}(\hat{\mathcal{R}}_{\mathcal{D}}) - \hat{\mathcal{R}}_{{\mathcal{D}}} \right].
	\]
	
	However, the standard symmetrization technique requires the empirical risk $\hat{\mathcal{R}}_{\mathcal{D}}$ to be a sum of independent terms, which is not applicable for the CML-based methods since they usually involve a sum of pairwise terms in (\ref{eqn4}). For instance, with respect to (\ref{eqn4}), the terms $\ell^{(i)}_g(v_j^+, v_k^-)$ and $\ell^{(i)}_g(\tilde{v}_j^+, \tilde{v}_k^-)$ are interdependent as long as one of them is the same (i.e., $v_j^+=\tilde{v}_j^+$ or $v_k^- = \tilde{v}_k^-$).
	
	Therefore, we turn to leverage another complexity measure, i.e., covering number, to overcome this difficulty. The necessary notations are summarized as follows. 
	
		\begin{defi} [$\epsilon$-Covering] \label{def2} \cite{ledoux1991probability} Let $(\mathcal{F}, \rho)$ be a (pseudo) metric space, and $\mathcal{G} \subseteq \mathcal{F}$. $\{f_1, \dots, f_K\}$ is said to be an $\epsilon$-covering of $\mathcal{G}$ if $\mathcal{G} \subseteq \mathop{\cup}\limits_{i=1}^K \mathcal{B}(f_i, \epsilon)$, i.e., $\forall g \in \mathcal{G}$, $\exists i$ such that $\rho(g, f_i) \le \epsilon$.
		\end{defi}

		\begin{defi} [Covering Number] \label{def3} \cite{ledoux1991probability} According to the notations in Def.\ref{def2}, the covering number of $\mathcal{G}$ with radius $\epsilon$ is defined as:
			\begin{equation} \nonumber
				\mathcal{N}(\epsilon;\mathcal{G}, \rho) = \min\{n: \exists \epsilon-covering \ over \ \mathcal{G} \ with \ size \ n\}
			\end{equation}
		\end{defi}
	
	With the above definitions, we further have the following assumption and lemma to help us derive the generalization bound.
		\begin{assu} [Basic Assumptions] \label{assu1} We assume that all the embeddings of users and items are chosen from the following embedding hypothesis space:
			\begin{equation}
				\mathcal{H}_R = \left\{\bmg: \bmg \in \mathbb{R}^d, \|\bmg\| \le r\right\},
			\end{equation}
			where $\boldsymbol{g}^c_{u_i} \in \mathcal{H}_R, u_i \in \mcu, c \in [C]$ and $\boldsymbol{g}_{v_j} \in \mathcal{H}_R, v_j \in \mci$.
		\end{assu}

		\begin{lem} \label{covering_lem} \cite{DBLP:conf/iclr/LongS20,DBLP:conf/icml/LiL21, DBLP:journals/jc/Zhou02}
			The covering number of the hypothesis class $\mathcal{H}_R$ has the following upper bound: 
			\begin{equation}
				\log \mathcal{N}(\epsilon;\mathcal{H}_R, \rho) \le d \log \left(\frac{3r}{\epsilon}\right),
			\end{equation}
			where $d$ is the dimension of embedding space. 
		\end{lem}

	Based on the above introductions, we have the following results. \textbf{\textit{Due to space limitations, please refer to Appendix.\ref{supp.sec.b} for all proofs in detail.}}
	
		\begin{theorem} [Generalization Upper Bound of DPCML] \label{them1} Let $\expe [\hat{\mathcal{L}}_{\mcd}(\bmg)]$ be the population risk of $\hat{\mathcal{L}}_{\mcd}(\bmg)$. Then, $\forall \ \bmg \in \mathcal{H}_R$, with high probability, the following inequation holds:
			\begin{equation}
				\begin{aligned} \label{eq:them1}
					\left| \hat{\mathcal{L}}_{\mcd}(\bmg) - \expe [\hat{\mathcal{L}}_{\mcd}(\bmg)]\right| \le \sqrt{\frac{2d\log \left(3r \tilde{N}\right)}{\tilde{N}}},
				\end{aligned}
			\end{equation}
			where we have
			\begin{small}
				\begin{equation} \nonumber
					\tilde{N} = \left(4r^2\sqrt{\left(\frac{(4 + \eta)^2}{|\mcu|} + \frac{2}{|\mcu|^2} \sum_{u_i \in \mcu} \left(\frac{1}{n_i^+} + \frac{1}{n_i^-}\right)\right)}\right)^{-2}
				\end{equation}
			\end{small}
		\end{theorem}
		Intriguingly, we see that our derived bound does not depend on $C$. This is consistent with the over-parameterization phenomenon \cite{DBLP:journals/corr/abs-2109-02355, DBLP:conf/iclr/NakkiranKBYBS20}. On top of Thm.\ref{them1}, we have the following corollary.
		\begin{coro} \label{cor1} DPCML could enjoy a smaller generalization error than CML.
		\end{coro}
		Therefore, we can conclude that DPCML generalizes to unseen data better than single-vector CML and thus improves the recommendation performance. This supports the superiority of our proposed DPCML from a theoretical perspective. In addition, we also empirically demonstrate this in the experiment Sec.\ref{qqaa} and Appendix.\ref{exp:cor1}.

	\begin{table*}[!h]
		\centering
		\setlength{\abovecaptionskip}{6pt}    
		\setlength{\belowcaptionskip}{15pt}    
		\setlength{\tabcolsep}{9pt}
		\caption{Basic Information of the Datasets. \%Density is defined as $\frac{\#Ratings}{\#Users \times \#Items} \times 100\% $.}	
		\label{table1}
		\begin{tabular}{c|ccccc}
			\toprule
			Datasets & MovieLens-1M & Steam-200k & CiteULike-T & MovieLens-10M\\
			\midrule
			Domain   & Movie & Game & Paper & Movie \\
			\#Users   & 6,034  & 3,757 & 5,219 & 69,167\\
			\#Items   & 3,953 & 5,113 & 25,975 & 10,019\\
			\#Ratings   & 575,271  & 115,139 & 125,580 & 5,003,437\\
			\%Density  & 2.4118\% & 0.5994\% & 0.0926\%  & 0.7220\%\\
			\bottomrule
		\end{tabular}
	\end{table*}
	
	\section{Experiments} \label{exp}
	
	In this section, our proposed method is applied to a wide range of real-world recommendation datasets to show its superiority. \textbf{\textit{Please refer to Appendix.\ref{exp_supp} for more results about experiments}}. 
	
	\begin{table*}[!t]
		\centering
		\caption{Performance comparisons on MovieLens-1m and Steam-200k datasets. }
		\label{tab:addlabel}%
		\scalebox{0.75}{
			\begin{tabular}{c|c|c|cccccccc}
				\toprule
				& Type & Method & P@3 & R@3 & NDCG@3 & P@5 & R@5 & NDCG@5 & MAP & MRR \\
				\midrule
				\multirow{14}[8]{*}{MovieLens-1m} & Item-based & itemKNN & \cellcolor[rgb]{ .988,  .957,  .957}12.24  & \cellcolor[rgb]{ .984,  .949,  .949}2.90  & \cellcolor[rgb]{ .988,  .961,  .965}12.41  & \cellcolor[rgb]{ .984,  .945,  .945}12.43  & \cellcolor[rgb]{ .969,  .929,  .918}4.29  & \cellcolor[rgb]{ .973,  .945,  .933}12.79  & \cellcolor[rgb]{ .965,  .91,  .902}8.34  & \cellcolor[rgb]{ .984,  .953,  .953}26.16  \\
				\cmidrule{2-11}      & \multirow{5}[2]{*}{MF-based} & GMF & \cellcolor[rgb]{ .98,  .933,  .933}14.10  & \cellcolor[rgb]{ .984,  .953,  .953}2.81  & \cellcolor[rgb]{ .98,  .937,  .937}14.33  & \cellcolor[rgb]{ .976,  .922,  .925}14.28  & \cellcolor[rgb]{ .969,  .937,  .925}4.08  & \cellcolor[rgb]{ .969,  .922,  .914}14.73  & \cellcolor[rgb]{ .965,  .91,  .902}8.29  & \cellcolor[rgb]{ .976,  .925,  .925}29.51  \\
				&   & MLP & \cellcolor[rgb]{ .98,  .933,  .933}13.95  & \cellcolor[rgb]{ .984,  .953,  .957}2.78  & \cellcolor[rgb]{ .98,  .937,  .941}14.22  & \cellcolor[rgb]{ .976,  .925,  .925}14.06  & \cellcolor[rgb]{ .973,  .937,  .929}3.98  & \cellcolor[rgb]{ .969,  .925,  .918}14.56  & \cellcolor[rgb]{ .965,  .91,  .902}8.30  & \cellcolor[rgb]{ .976,  .925,  .925}29.39  \\
				&   & NeuMF & \cellcolor[rgb]{ .969,  .902,  .902}16.43  & \cellcolor[rgb]{ .98,  .933,  .933}3.20  & \cellcolor[rgb]{ .969,  .906,  .906}16.87  & \cellcolor[rgb]{ .965,  .894,  .894}16.73  & \cellcolor[rgb]{ .965,  .918,  .906}4.68  & \cellcolor[rgb]{ .961,  .894,  .886}17.40  & \cellcolor[rgb]{ .957,  .89,  .882}9.69  & \cellcolor[rgb]{ .965,  .894,  .894}33.23  \\
				&   & M2F & 8.61  & 1.84  & 9.36  & 7.60  & \cellcolor[rgb]{ .984,  .992,  .976}2.30  & \cellcolor[rgb]{ .984,  .992,  .976}8.67  & \cellcolor[rgb]{ .984,  .992,  .976}2.95  & 20.40  \\
				&   & MGMF & \cellcolor[rgb]{ .965,  .89,  .89}17.38  & \cellcolor[rgb]{ .973,  .918,  .922}3.51  & \cellcolor[rgb]{ .965,  .89,  .89}18.08  & \cellcolor[rgb]{ .961,  .882,  .886}17.63  & \cellcolor[rgb]{ .961,  .902,  .898}5.05  & \cellcolor[rgb]{ .957,  .878,  .875}18.52  & \cellcolor[rgb]{ .957,  .882,  .878}10.12  & \cellcolor[rgb]{ .961,  .878,  .878}35.15  \\
				\cmidrule{2-11}      & \multirow{8}[2]{*}{CML-based} & UniS & \cellcolor[rgb]{ .965,  .886,  .89}17.56  & \cellcolor[rgb]{ .973,  .91,  .91}3.71  & \cellcolor[rgb]{ .965,  .89,  .894}17.89  & \cellcolor[rgb]{ .961,  .875,  .875}18.34  & \cellcolor[rgb]{ .957,  .886,  .882}5.60  & \cellcolor[rgb]{ .953,  .875,  .871}18.79  & \cellcolor[rgb]{ .945,  .847,  .847}12.40  & \cellcolor[rgb]{ .957,  .871,  .875}35.77  \\
				&   & PopS & \cellcolor[rgb]{ .984,  .945,  .949}12.96  & \cellcolor[rgb]{ .98,  .937,  .941}3.11  & \cellcolor[rgb]{ .984,  .949,  .953}13.30  & \cellcolor[rgb]{ .98,  .941,  .941}12.82  & \cellcolor[rgb]{ .969,  .925,  .918}4.41  & \cellcolor[rgb]{ .973,  .937,  .929}13.40  & \cellcolor[rgb]{ .969,  .922,  .914}7.59  & \cellcolor[rgb]{ .98,  .933,  .933}28.61  \\
				&   & 2stS & \cellcolor[rgb]{ .949,  .843,  .847}21.07  & \cellcolor[rgb]{ .953,  .851,  .855}4.84  & \cellcolor[rgb]{ .949,  .847,  .847}21.35  & \cellcolor[rgb]{ .945,  .831,  .835}21.81  & \cellcolor[rgb]{ .945,  .839,  .835}7.07  & \cellcolor[rgb]{ .945,  .835,  .835}22.29  & \cellcolor[rgb]{ .937,  .816,  .816}14.42  & \cellcolor[rgb]{ .945,  .831,  .835}40.36  \\
				&   & HarS & \cellcolor[rgb]{ .933,  .792,  .796}\underline{24.88}  & \cellcolor[rgb]{ .933,  .8,  .804}\underline{5.86}  & \cellcolor[rgb]{ .933,  .792,  .796}\underline{25.38}  & \cellcolor[rgb]{ .933,  .796,  .8}\underline{24.89}  & \cellcolor[rgb]{ .933,  .8,  .8}\underline{8.25}  & \cellcolor[rgb]{ .933,  .796,  .8}\underline{25.77}  & \cellcolor[rgb]{ .933,  .796,  .8}\underline{15.74}  & \cellcolor[rgb]{ .933,  .792,  .796}\underline{45.15}  \\
				&   & TransCF & \cellcolor[rgb]{ .996,  .984,  .984}10.03  & 1.84  & \cellcolor[rgb]{ .996,  .988,  .988}10.31  & \cellcolor[rgb]{ .988,  .961,  .965}10.90  & \cellcolor[rgb]{ .98,  .969,  .957}3.09  & \cellcolor[rgb]{ .976,  .965,  .953}11.20  & \cellcolor[rgb]{ .969,  .929,  .922}7.07  & \cellcolor[rgb]{ .992,  .973,  .976}23.66  \\
				&   & LRML & \cellcolor[rgb]{ .965,  .894,  .894}17.15  & \cellcolor[rgb]{ .973,  .918,  .918}3.52  & \cellcolor[rgb]{ .965,  .894,  .898}17.56  & \cellcolor[rgb]{ .965,  .886,  .886}17.45  & \cellcolor[rgb]{ .961,  .902,  .894}5.12  & \cellcolor[rgb]{ .957,  .886,  .878}18.08  & \cellcolor[rgb]{ .957,  .878,  .875}10.42  & \cellcolor[rgb]{ .961,  .882,  .886}34.36  \\
				& & AdaCML & \cellcolor[rgb]{ .957,  .867,  .871}19.06  & \cellcolor[rgb]{ .965,  .886,  .89}4.12  & \cellcolor[rgb]{ .957,  .871,  .875}19.31  & \cellcolor[rgb]{ .953,  .859,  .859}19.74  & \cellcolor[rgb]{ .953,  .867,  .863}6.23  & \cellcolor[rgb]{ .949,  .859,  .859}20.20  & \cellcolor[rgb]{ .941,  .835,  .831}13.30  & \cellcolor[rgb]{ .953,  .859,  .859}37.36  \\
& &    HLR   & \cellcolor[rgb]{ .949,  .843,  .847}21.10  & \cellcolor[rgb]{ .953,  .855,  .855}4.80  & \cellcolor[rgb]{ .949,  .843,  .847}21.53  & \cellcolor[rgb]{ .945,  .835,  .839}21.61  & \cellcolor[rgb]{ .945,  .839,  .835}7.06  & \cellcolor[rgb]{ .945,  .835,  .835}22.28  & \cellcolor[rgb]{ .941,  .824,  .824}13.95  & \cellcolor[rgb]{ .929,  .788,  .792}40.71  \\
				\cmidrule{2-11}      & \multirow{2}[2]{*}{Ours} & DPCML1 & \cellcolor[rgb]{ .957,  .867,  .871}19.12  & \cellcolor[rgb]{ .965,  .886,  .89}4.14  & \cellcolor[rgb]{ .957,  .871,  .875}19.34  & \cellcolor[rgb]{ .953,  .855,  .859}19.90  & \cellcolor[rgb]{ .953,  .863,  .859}6.27  & \cellcolor[rgb]{ .949,  .859,  .855}20.29  & \cellcolor[rgb]{ .945,  .835,  .835}13.24  & \cellcolor[rgb]{ .953,  .855,  .859}37.55  \\
				&   & DPCML2 & \cellcolor[rgb]{ .929,  .788,  .792}\textbf{25.18} & \cellcolor[rgb]{ .929,  .788,  .792}\textbf{6.06} & \cellcolor[rgb]{ .929,  .788,  .792}\textbf{25.64} & \cellcolor[rgb]{ .929,  .788,  .792}\textbf{25.35} & \cellcolor[rgb]{ .929,  .788,  .792}\textbf{8.51} & \cellcolor[rgb]{ .929,  .788,  .792}\textbf{26.16} & \cellcolor[rgb]{ .929,  .788,  .792}\textbf{16.09} & \cellcolor[rgb]{ .929,  .788,  .792}\textbf{45.32} \\
				\midrule
				\multirow{14}[8]{*}{Steam-200k} & Item-based & itemKNN & \cellcolor[rgb]{ .984,  .98,  .957}12.58  & \cellcolor[rgb]{ .996,  .937,  .898}9.47  & \cellcolor[rgb]{ 1,  .988,  .98}13.23  & 6.47  & \cellcolor[rgb]{ .984,  .992,  .976}3.90  & \cellcolor[rgb]{ .984,  .992,  .976}7.23  & \cellcolor[rgb]{ .996,  .945,  .91}11.74  & 23.33  \\
				\cmidrule{2-11}      & \multirow{5}[2]{*}{MF-based} & GMF & \cellcolor[rgb]{ .984,  .98,  .957}12.57  & \cellcolor[rgb]{ 1,  .992,  .988}6.17  & \cellcolor[rgb]{ 1,  .988,  .98}13.29  & \cellcolor[rgb]{ .992,  .929,  .886}14.22  & \cellcolor[rgb]{ .984,  .937,  .894}6.86  & \cellcolor[rgb]{ .984,  .925,  .871}15.39  & \cellcolor[rgb]{ .996,  .965,  .945}9.72  & \cellcolor[rgb]{ .996,  .965,  .945}28.38  \\
				&   & MLP & \cellcolor[rgb]{ .984,  .933,  .886}17.07  & \cellcolor[rgb]{ .992,  .933,  .894}9.63  & \cellcolor[rgb]{ .996,  .941,  .906}17.49  & \cellcolor[rgb]{ .992,  .906,  .843}16.89  & \cellcolor[rgb]{ .984,  .91,  .847}8.49  & \cellcolor[rgb]{ .984,  .902,  .839}17.67  & \cellcolor[rgb]{ .992,  .91,  .851}15.15  & \cellcolor[rgb]{ .992,  .922,  .875}34.54  \\
				&   & NeuMF & \cellcolor[rgb]{ .984,  .929,  .878}17.36  & \cellcolor[rgb]{ .992,  .933,  .89}9.65  & \cellcolor[rgb]{ .996,  .937,  .898}17.95  & \cellcolor[rgb]{ .988,  .898,  .835}17.41  & \cellcolor[rgb]{ .984,  .902,  .835}8.79  & \cellcolor[rgb]{ .98,  .898,  .827}18.45  & \cellcolor[rgb]{ .992,  .91,  .851}15.11  & \cellcolor[rgb]{ .992,  .914,  .863}35.55  \\
				&   & M2F & \cellcolor[rgb]{ .984,  .992,  .976}11.33  & 5.69  & 11.95  & \cellcolor[rgb]{ .996,  .957,  .925}11.44  & \cellcolor[rgb]{ .984,  .961,  .925}5.73  & \cellcolor[rgb]{ .984,  .945,  .902}12.98  & 6.43  & \cellcolor[rgb]{ 1,  .988,  .98}25.05  \\
				&   & MGMF & \cellcolor[rgb]{ .984,  .98,  .961}12.51  & \cellcolor[rgb]{ 1,  .996,  .988}6.14  & \cellcolor[rgb]{ 1,  .988,  .98}13.25  & \cellcolor[rgb]{ .992,  .925,  .882}14.45  & \cellcolor[rgb]{ .984,  .937,  .89}6.88  & \cellcolor[rgb]{ .984,  .922,  .867}15.55  & \cellcolor[rgb]{ .996,  .969,  .945}9.63  & \cellcolor[rgb]{ .996,  .965,  .945}28.40  \\
				\cmidrule{2-11}      & \multirow{8}[2]{*}{CML-based} & UniS & \cellcolor[rgb]{ .98,  .894,  .824}20.71  & \cellcolor[rgb]{ .988,  .894,  .827}11.97  & \cellcolor[rgb]{ .988,  .902,  .839}21.42  & \cellcolor[rgb]{ .984,  .867,  .784}20.92  & \cellcolor[rgb]{ .98,  .875,  .792}10.36  & \cellcolor[rgb]{ .98,  .871,  .784}21.61  & \cellcolor[rgb]{ .984,  .867,  .788}18.88  & \cellcolor[rgb]{ .988,  .882,  .808}40.10  \\
				&   & PopS & \cellcolor[rgb]{ .984,  .922,  .867}18.05  & \cellcolor[rgb]{ .988,  .902,  .839}11.58  & \cellcolor[rgb]{ .992,  .929,  .886}18.76  & \cellcolor[rgb]{ .992,  .922,  .875}14.94  & \cellcolor[rgb]{ .984,  .918,  .859}7.98  & \cellcolor[rgb]{ .984,  .922,  .863}15.78  & \cellcolor[rgb]{ .992,  .91,  .851}15.13  & \cellcolor[rgb]{ .992,  .925,  .878}34.04  \\
				&   & 2stS & \cellcolor[rgb]{ .98,  .847,  .753}25.20  & \cellcolor[rgb]{ .984,  .847,  .753}14.62  & \cellcolor[rgb]{ .984,  .851,  .757}26.20  & \cellcolor[rgb]{ .984,  .839,  .737}23.97  & \cellcolor[rgb]{ .98,  .843,  .745}11.91  & \cellcolor[rgb]{ .98,  .839,  .737}25.35  & \cellcolor[rgb]{ .984,  .839,  .741}21.48  & \cellcolor[rgb]{ .984,  .839,  .741}46.17  \\
				&   & HarS & \cellcolor[rgb]{ .98,  .831,  .725}\underline{26.66}  & \cellcolor[rgb]{ .98,  .827,  .722}\underline{15.74}  & \cellcolor[rgb]{ .98,  .831,  .729}\underline{27.93}  & \cellcolor[rgb]{ .98,  .827,  .722}\underline{24.94}  & \cellcolor[rgb]{ .98,  .827,  .722}\underline{12.78}  & \cellcolor[rgb]{ .98,  .827,  .718}\underline{26.63}  & \cellcolor[rgb]{ .98,  .82,  .71}\underline{23.25}  & \cellcolor[rgb]{ .98,  .82,  .71}\underline{48.84}  \\
				&   & TransCF & \cellcolor[rgb]{ .984,  .973,  .945}13.30  & \cellcolor[rgb]{ 1,  .984,  .976}6.61  & \cellcolor[rgb]{ 1,  .984,  .973}13.58  & \cellcolor[rgb]{ .992,  .918,  .871}15.26  & \cellcolor[rgb]{ .984,  .933,  .886}7.09  & \cellcolor[rgb]{ .984,  .918,  .863}15.89  & \cellcolor[rgb]{ .996,  .953,  .922}11.08  & \cellcolor[rgb]{ 1,  .98,  .969}26.29  \\
				&   & LRML & \cellcolor[rgb]{ .984,  .957,  .922}14.91  & \cellcolor[rgb]{ 1,  .973,  .953}7.48  & \cellcolor[rgb]{ .996,  .965,  .941}15.43  & \cellcolor[rgb]{ .992,  .906,  .851}16.49  & \cellcolor[rgb]{ .984,  .918,  .859}8.06  & \cellcolor[rgb]{ .984,  .906,  .839}17.51  & \cellcolor[rgb]{ .996,  .941,  .902}12.24  & \cellcolor[rgb]{ .996,  .941,  .902}31.89  \\
				& & AdaCML & \cellcolor[rgb]{ .98,  .875,  .792}23.02  & \cellcolor[rgb]{ .988,  .871,  .792}13.19  & \cellcolor[rgb]{ .988,  .882,  .812}23.38  & \cellcolor[rgb]{ .984,  .859,  .773}22.35  & \cellcolor[rgb]{ .98,  .859,  .769}11.31  & \cellcolor[rgb]{ .98,  .863,  .773}23.23  & \cellcolor[rgb]{ .984,  .851,  .757}19.88  & \cellcolor[rgb]{ .988,  .871,  .792}42.03  \\
   & & HLR   & \cellcolor[rgb]{ .984,  .902,  .835}20.30  & \cellcolor[rgb]{ .988,  .898,  .835}11.65  & \cellcolor[rgb]{ .992,  .91,  .851}20.96  & \cellcolor[rgb]{ .988,  .882,  .808}19.79  & \cellcolor[rgb]{ .98,  .886,  .812}9.88  & \cellcolor[rgb]{ .98,  .882,  .804}20.94  & \cellcolor[rgb]{ .988,  .882,  .808}17.06  & \cellcolor[rgb]{ .988,  .89,  .824}39.26  \\
				
				\cmidrule{2-11}      & \multirow{2}[2]{*}{Ours} & DPCML1 & \cellcolor[rgb]{ .98,  .847,  .753}25.39  & \cellcolor[rgb]{ .984,  .847,  .757}14.84  & \cellcolor[rgb]{ .984,  .847,  .757}26.56  & \cellcolor[rgb]{ .98,  .835,  .733}23.88  & \cellcolor[rgb]{ .98,  .839,  .737}12.11  & \cellcolor[rgb]{ .98,  .831,  .729}25.25  & \cellcolor[rgb]{ .98,  .835,  .733}22.26  & \cellcolor[rgb]{ .98,  .835,  .733}46.79  \\
				
				&   & DPCML2 & \cellcolor[rgb]{ .976,  .8,  .678}\textbf{29.88} & \cellcolor[rgb]{ .976,  .8,  .678}\textbf{17.13} & \cellcolor[rgb]{ .976,  .8,  .678}\textbf{31.22} & \cellcolor[rgb]{ .976,  .8,  .678}\textbf{28.70} & \cellcolor[rgb]{ .976,  .8,  .678}\textbf{14.51} & \cellcolor[rgb]{ .976,  .8,  .678}\textbf{30.56} & \cellcolor[rgb]{ .976,  .8,  .678}\textbf{24.10} & \cellcolor[rgb]{ .976,  .8,  .678}\textbf{51.95} \\
				\bottomrule
			\end{tabular}%
		}
	\end{table*}%
	
	\subsection{Experimental Setups}
	To begin with, we perform empirical experiments on several common recommendation benchmarks: \textit{MovieLens-1m}, \textit{Steam-200k}, \textit{CiteULike} and \textit{MovieLens-10m}. The detailed statistics in terms of these datasets are summarized in Tab.\ref{table1}. For the datasets with explicit feedbacks, we follow the previous works \cite{DBLP:conf/www/HeLZNHC17, tran2019improving} and transfer them into implicit feedback. Secondly, we evaluate the performance with five metrics, including \textbf{Precision} (P@$N$), \textbf{Recall} (R@$N$), \textbf{Normalized Discounted Cumulative Gain} (NDCG@$N$), \textbf{Mean Average Precision} (MAP), and \textbf{Mean Reciprocal Rank} (MRR). Moreover, we compared our proposed method with $14$ competitive competitors: a) Item-based CF method, \textbf{itemKNN} \cite{linden2003amazon}; b) MF-based algorithms, including the combination of MF and deep learning models and multi-vector MF-based approaches: \textbf{GMF}, \textbf{MLP}, \textbf{NeuMF} \cite{DBLP:conf/www/HeLZNHC17}, \textbf{M2F} \cite{DBLP:conf/eaamo/GuoKJG21} and \textbf{MGMF} \cite{DBLP:conf/eaamo/GuoKJG21}; c) CML-based methods, including \textbf{UniS} \cite{DBLP:conf/icdm/PanZCLLSY08}, \textbf{PopS} \cite{DBLP:conf/sigir/WuVSSR19}, \textbf{2stS} \cite{tran2019improving}, \textbf{HarS} \cite{hsieh2017collaborative,DBLP:journals/pr/GajicAG21}, \textbf{TransCF} \cite{DBLP:conf/icdm/ParkKXY18}, \textbf{LRML} \cite{DBLP:conf/www/TayTH18}, \textbf{AdaCML} \cite{DBLP:conf/dasfaa/ZhangZLXF0SC19} and \textbf{HLR} \cite{DBLP:conf/recsys/TranSHM21}.
	
	\subsection{Overall Performance} 
	The experimental results of all the involved competitors are shown in Tab.\ref{tab:addlabel} and Tab.\ref{results2} (in Appendix.\ref{suppa.4}). Consequently, we can draw the following conclusions: 1) In most cases, the best performance of CML-based methods consistently surpasses the best MF-based competitors. This suggests that it is necessary to develop CML-based RS algorithms. 2) Our proposed method consistently surpasses all the competitors significantly on all datasets, except the results for MAP and MRR on CiteULike. Even for the failure results, the performance is fairly competitive compared with the competitors. This shows the effectiveness of our proposed algorithm. 3) Compared with studies targeting joint accessibility (i.e., M2F and MGMF), our proposed method significantly outperforms M2F and MGMF on all benchmark datasets. This shows the advantage of the CML-based paradigm that deserves more attention along this direction in future work. 
	\begin{figure*}[!t]
		\centering
		\includegraphics[width=1.0\textwidth]{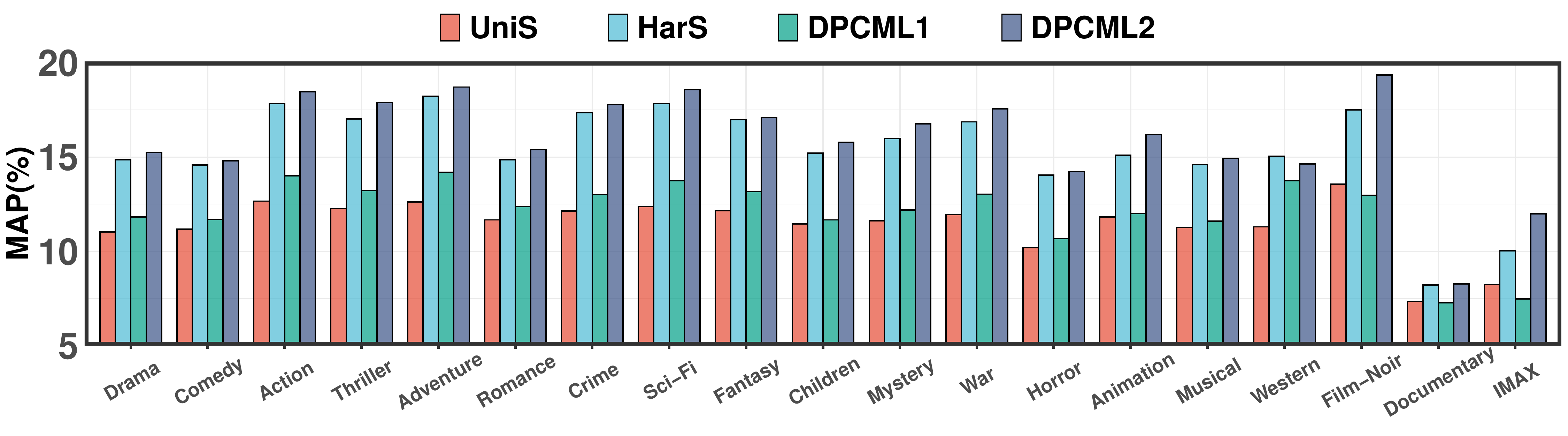}
		\caption{Fine-grained performance over each interest group on MovieLens-10m dataset. }
		\label{per_arrtribute_performance}
	\end{figure*}
	\subsection{Quantitative Analysis} \label{qqaa}
	\textbf{Fine-grained Performance Comparison.} Fig.\ref{per_arrtribute_performance} presents the MAP metric over each interest group (movie genre) on MovieLens-10m. We can observe that our proposed framework could not only significantly outperform their single-vector counterparts in the majority interests but also improve the performance of minority groups in most cases. Especially, compared with HarS, the performance improvement of DPCML2 on minority interests is sharp. This shows that DPCML could reasonably focus on potentially interesting items even with the imbalanced item distribution. 
	
		\begin{table}[!t]
  \centering
  \caption{The diversity performance comparison of CML-based algorithms on Steam-200k and MovieLens-1m datasets. Here a higher value implies more diverse recommendation results.}
  \scalebox{0.78}{
    \begin{tabular}{c|cccc}
    \toprule
    \multicolumn{5}{c}{Steam-200k} \\
    \midrule
    Method & MaxDiv@3 & MaxDiv@5 & MaxDiv@10 & MaxDiv@20 \\
    \midrule
    UniS & 1.354 & 4.750 & 23.520 & 117.927 \\
    HarS & 1.752 & 6.809 & 40.378 & 236.794 \\
    DPCML1 w/o DCRS & 1.643 & 5.857 & 30.425 & 155.193 \\
    DPCML1 & 1.822 & 6.713 & 34.727 & 179.065 \\
    DPCML2 w/o DCRS & 2.958 & 11.398 & 65.398 & 365.458 \\
    DPCML2 & 2.977 & 11.472 & 65.952 & 369.876 \\
    \midrule
    \multicolumn{5}{c}{MovieLens-1m} \\
    \midrule
    UniS & 1.739 & 6.142 & 30.127 & 140.095 \\
    HarS & 2.443 & 8.826 & 46.390 & 244.078 \\
    DPCML1 w/o DCRS & 1.623 & 5.857 & 29.500 & 140.057 \\
    DPCML1 & 1.744 & 6.195 & 30.755 & 145.615 \\
    DPCML2 w/o DCRS & 2.827 & 10.423 & 55.612 & 292.089 \\
    DPCML2 & 3.144 & 11.498 & 60.696 & 313.086 \\
    \bottomrule
    \end{tabular}%
    }
  \label{tab:diversity}%
\end{table}%
	
	\textbf{Recommendation Diversity Evaluation}. We test the performance of DPCML against CML-based competitors with \textit{max-sum diversification (MaxDiv)} \cite{10.1145/2213556.2213580}. The diversity results are shown in Tab.\ref{tab:diversity}. We observe that: a) For methods within the same negative sampling strategy (i.e., UniS and DPCML1, HarS and DPCML2), our proposed DPCML could achieve relatively higher max-sum values. This suggests the improvement of DPCML in terms of promoting recommendation diversity. b) In most cases (except for DPCML1 w/o DCRS on the MovieLens-1m dataset), DPCML outperforms other competitors even without regularization. c) Most importantly, equipped with the regularization term DCRS, DPCML could achieve better diversification results against w/o DCRS. 
	
\begin{figure}[!t]
	    \centering
	    \begin{minipage}[b]{0.43\columnwidth}
		\subfigure[DPCML1 (MRR)]{
			\includegraphics[width=0.46\columnwidth]{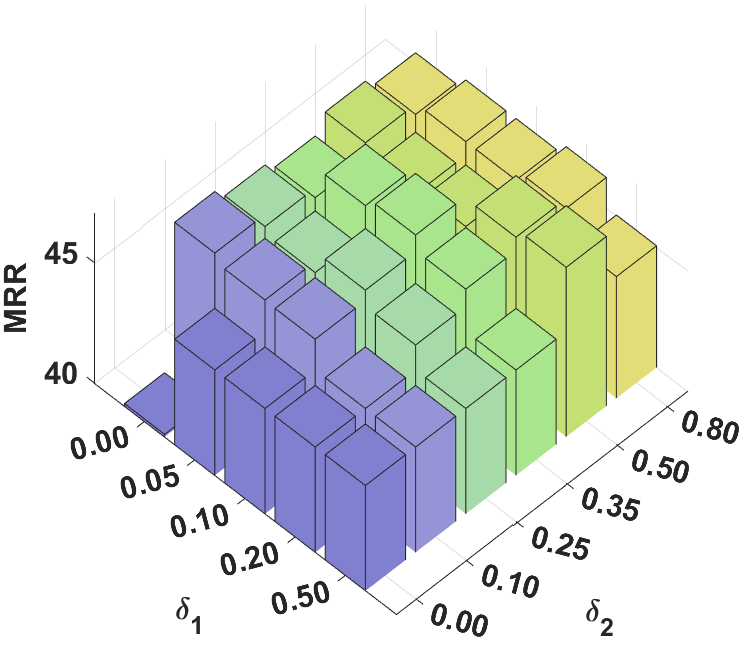}
		}
		\subfigure[DPCML2 (MRR)]{
			\includegraphics[width=0.46\columnwidth]{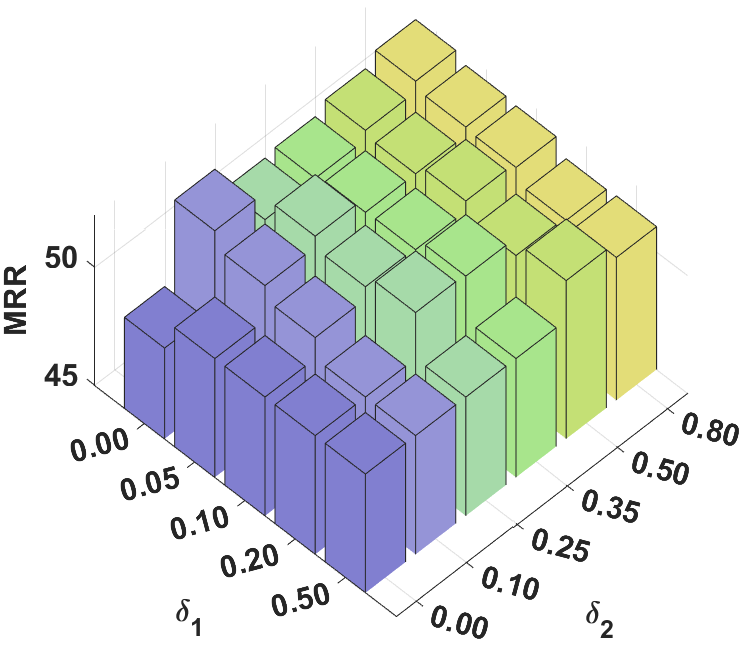}
		}
		\caption{Sensitivity analysis about $\delta_1$ and $\delta_2$ on Steam-200k datasets.}
		\label{sensitivity}
	\end{minipage}
    \hspace{0.1cm}
	\begin{minipage}[b]{0.43\columnwidth}
		\centering
		\subfigure[CiteULike]{
			\includegraphics[width=0.45\textwidth]{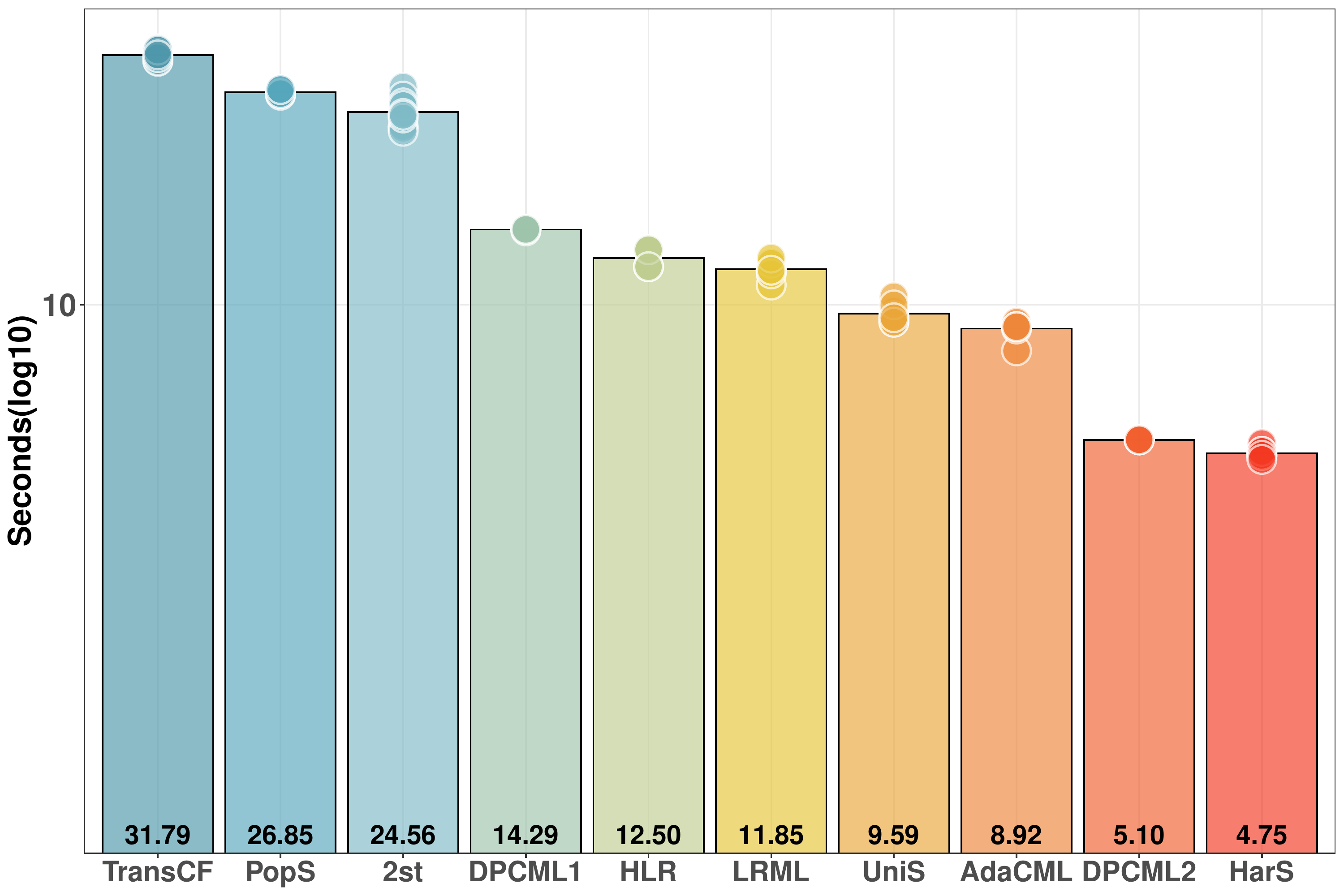}
			\label{ab.sub.citeulike}
		}
		\subfigure[MovieLens-10m]{
			\includegraphics[width=0.45\textwidth]{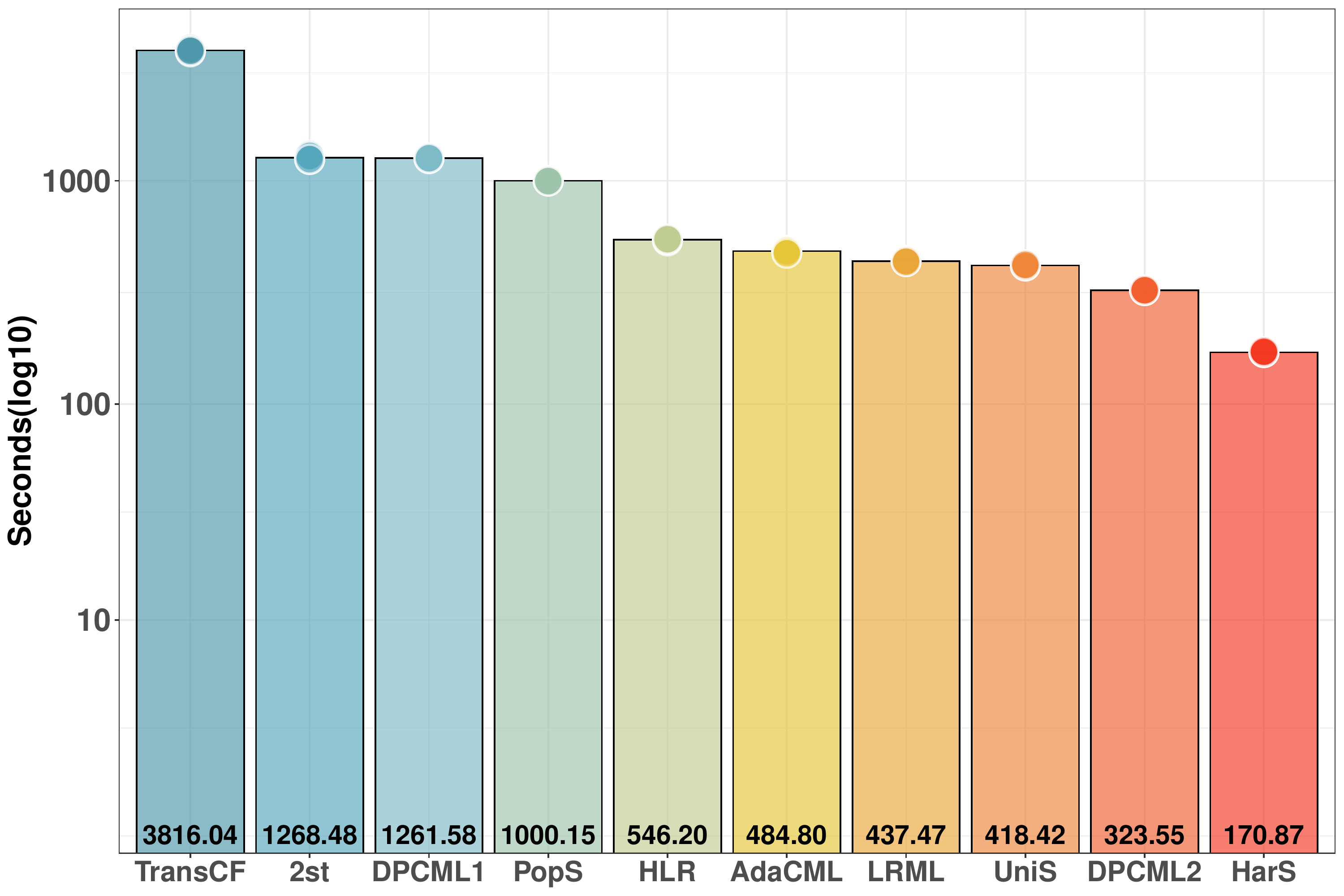}
			\label{ab.sub.ml-10m}
		}
		\caption{Training efficiency comparison among CML-based competitors.}
		\label{runtime}
	\end{minipage}
	\end{figure}

	\textbf{Effect of the Diversity Control Regularization.} Fig.\ref{sensitivity} illustrates a $3$D-barplot based on the results of grid search on Steam-200k. From the results, we can observe that the proposed regularization scheme could significantly boost performance on all metrics. Moreover, there would induce different performances with different diversity values. This suggests that controlling a proper diversity of the embeddings for the same user is essential to accommodate their preferences better.
	
\begin{figure}[!t]
	    \centering
	    \begin{minipage}[b]{0.43\columnwidth}
		\subfigure[DPCML1]{
		\includegraphics[width=0.46\columnwidth]{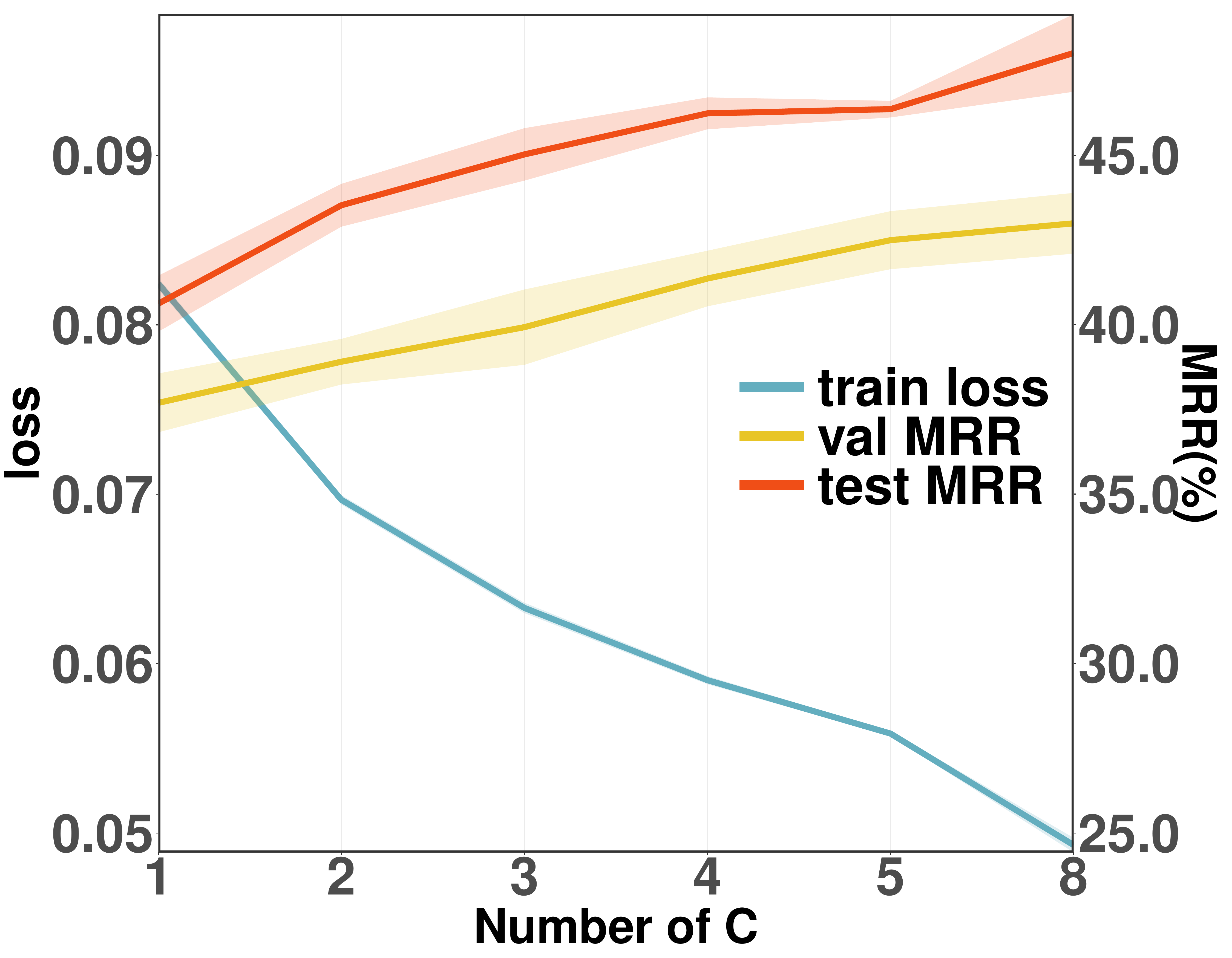}
		\label{DPCML1}
	}
	\subfigure[DPCML2]{
		\includegraphics[width=0.46\columnwidth]{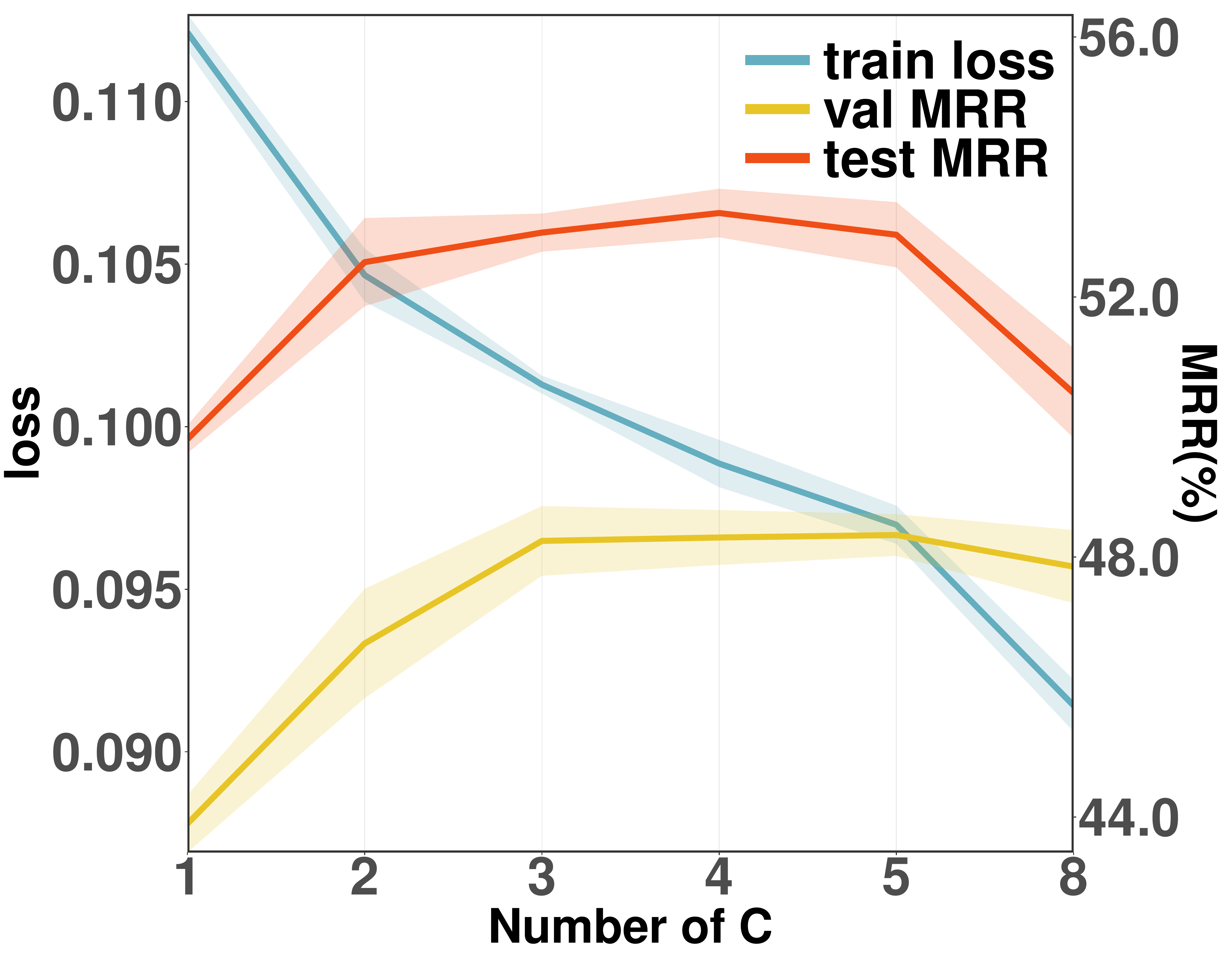}
		\label{DPCML2}
	}
	\caption{Empirical justification of Thm.\ref{them1}.}
	\label{just_thm1}
	\end{minipage}
	\hspace{0.1cm}
	\begin{minipage}[b]{0.43\columnwidth}
		\centering
		\subfigure[P@3]{
		\includegraphics[width=0.46\columnwidth]{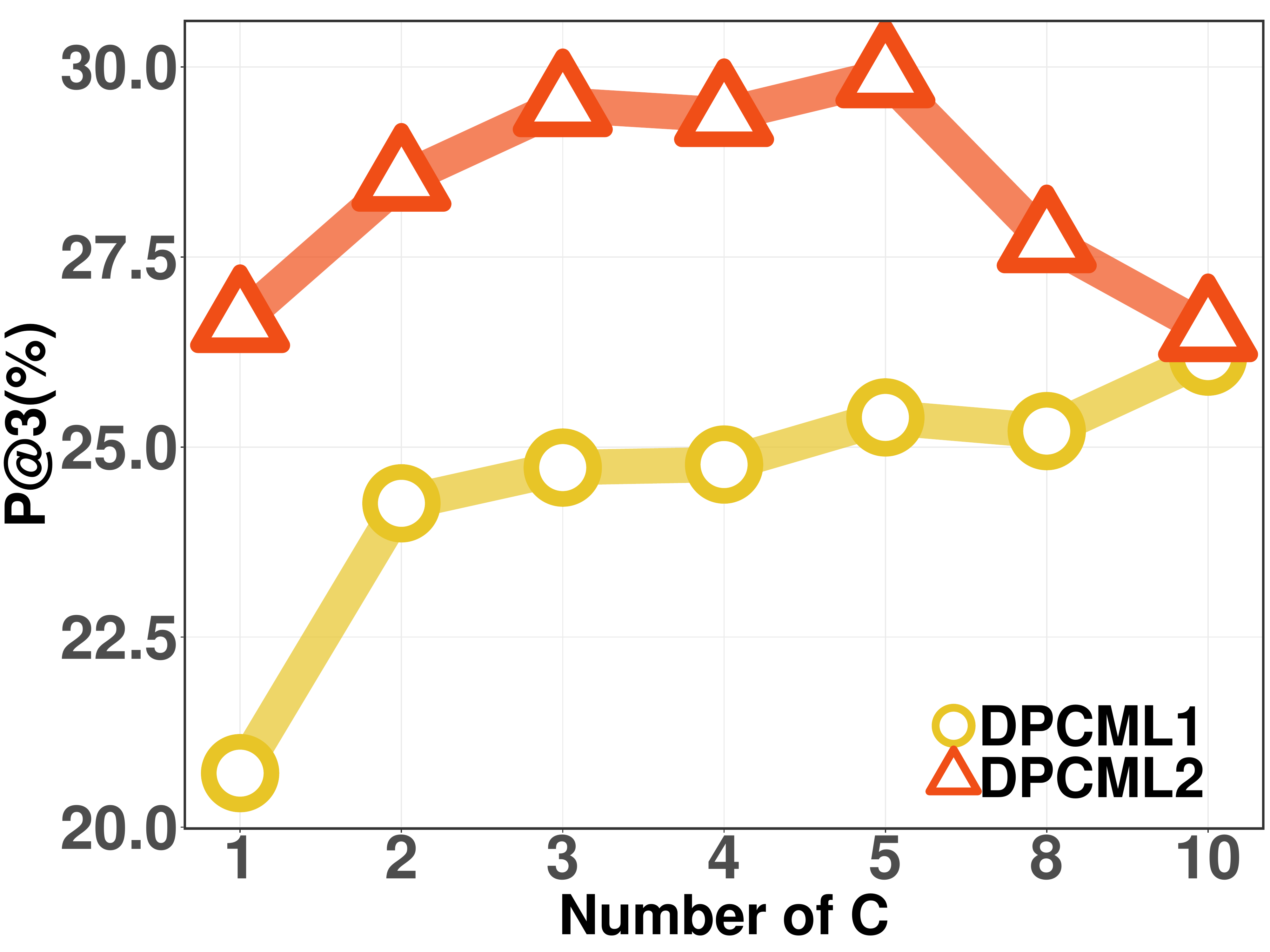}
		\label{effct_C:P@3}
	}
	\subfigure[P@5]{
		\includegraphics[width=0.46\columnwidth]{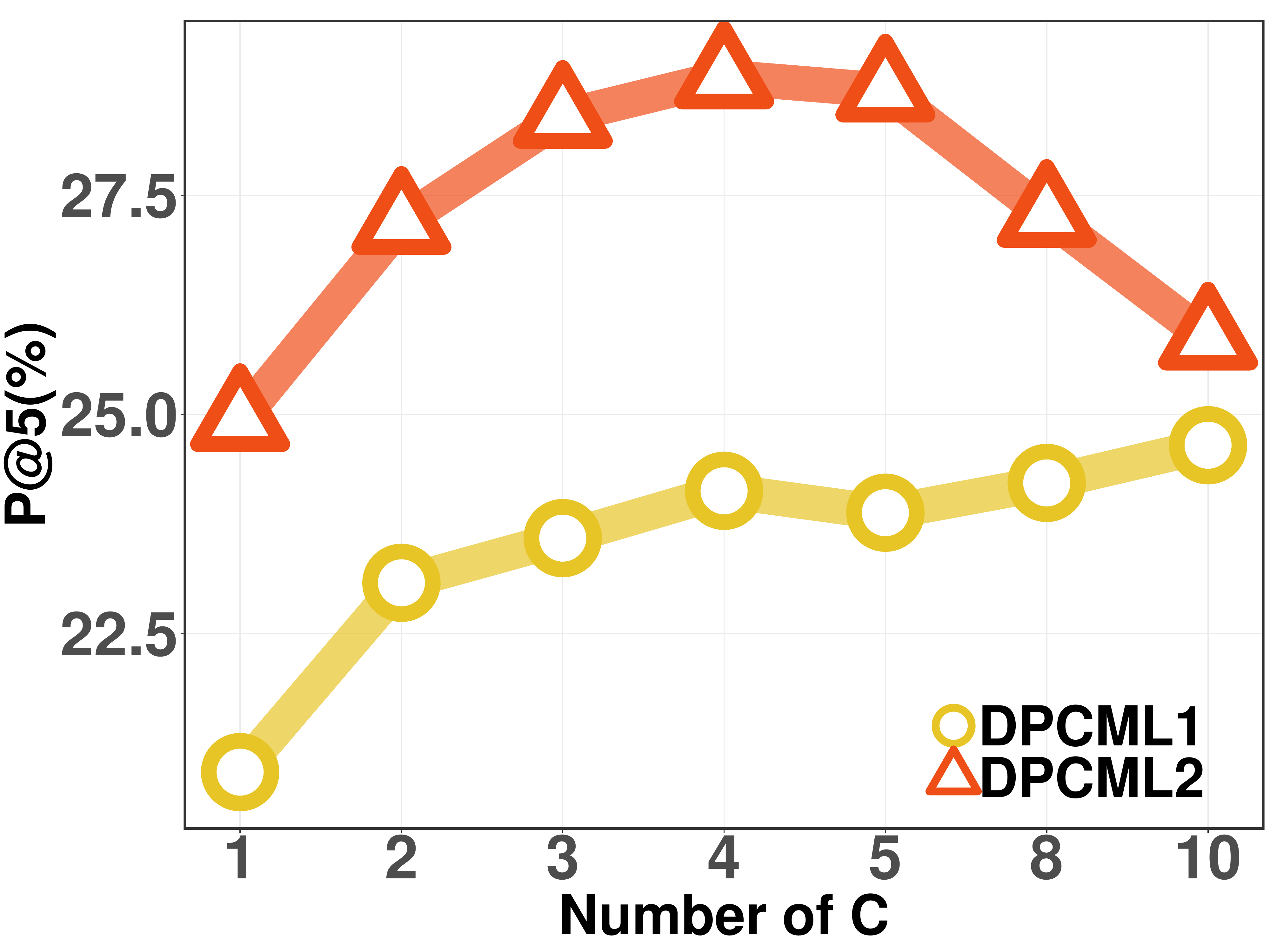}
		\label{effct_C:P@5}
	}
	\caption{Sensitive Analysis of different $C$.}
	\label{effect_of_C}
	\end{minipage}
	\end{figure}

 	\textbf{Empirical Justification of Corol.\ref{cor1}} Fig.\ref{just_thm1} shows the empirical results on Steam-200k dataset. Based on these results, we can see that, with the increase of $C$, the empirical risk (i.e., training loss) of DPCML ($C>1$) is significantly smaller than CML ($C=1$). In addition, DPCML could substantially improve the performance of the validation/test set. Thus, we can conclude that DPCML could induce a smaller generalization error than traditional CML. This is consistent with Corol.\ref{cor1}.

 	\textbf{Sensitive Analysis of $C$.} Fig.\ref{effect_of_C} demonstrates the performance of DPCML methods with different $C$ on Steam-200k dataset. We observe that a proper $C$ could significantly improve the performance. Besides, leveraging C too aggressively for DPCML2 may adversely hurt the performance since models optimized with hard samples are more likely to lead to the over-fitting problem with the increasing parameters.

    \textbf{Sensitivity analysis of $\eta$.} We investigate the sensitivity of $\eta \in \{0, 1, 3, 5, 10, 20, 30\}$ for recommendation results on the Steam-200k dataset. The experimental results are listed in Tab.\ref{tab:sen_dpcml1} and Tab.\ref{tab:sen_dpcml2}. We can conclude that a proper $\eta$ (roughly $10$) could significantly improve the performance, suggesting the essential role of the proposed diversity control regularization scheme.
	
	\textbf{Training Efficiency.} Since DPCML includes multiple user representations, it will inevitably introduce extra complexity to the overall optimization. We further investigate the efficiency of our proposed algorithm, presented in Fig.\ref{runtime}. This trend suggests that our proposed algorithm could achieve competitive performance with acceptable efficiency. 
	
	\textbf{Ablation Studies of DCRS.} In order to show the effectiveness of our proposed DCRS, we compare its performance with the its three variants: \textbf{a)} $\textbf{w/o DCRS}$. \textbf{b)} $\textbf{DCRS}-\delta_1$ and \textbf{c)} $\textbf{DCRS}-\delta_2$. Please refer to Appendix.\ref{ab_stu} to the details of each variant. The empircal results on Steam-200k dataset are provided in Tab.\ref{tab:DPCML1} and Tab.\ref{tab:DPCML2}. From the above results, we can see that: In most cases, only employing one of the two terms of DCRS could still improve the recommendation performance. However, none of them could outperform our proposed method. This strengthens the effectiveness of our proposed regularization scheme.
	
	\textbf{DCRS for MF-based Systems.} We attempt to apply the proposed diversity control regularization scheme (DCRS) for M2F \cite{DBLP:conf/recsys/WestonWY13,DBLP:conf/eaamo/GuoKJG21}. In addition, we further explore the effectiveness of DCRS for the general framework of joint accessibility (GFJA, Eq.(\ref{eq3123}) in the main paper). The experimental results are summarized in Tab.\ref{tab:reg_for_MF}. The overall performance suggests the superiority of our proposed method against the current multi-vector-based competitors.
	\section{Conclusion \& Future Remarks}
	This paper pays attention to developing an effective CML-based algorithm when users have multiple categories of interests. First, we point out that the current CML framework might induce preference bias, especially when the item category distribution is imbalanced. To this end, we propose a novel algorithm called DPCML. The key idea is to include multiple representations for each user in the model design. Meanwhile, a novel \textit{diversity control regularization} scheme is specifically tailored to serve our purpose better. To see the generalization ability of DPCML on unseen test data, we also provide high probability upper bounds for the generalization error. Finally, the experiments over a range of benchmark datasets speak to the efficacy of DPCML. However, following the paradigm of CML, a possible limitation of DPCML is that it generally applies to implicit feedback but not explicit feedback since CML only cares about the relative preference ranking instead of concrete magnitude. In the future, we will explore how to improve the recommendation diversity based on explicit feedback.
	
	\section{Acknowledgements}
	This work was supported in part by the National Key R\&D Program of China under Grant 2018AAA0102000, in part by National Natural Science Foundation of China: U21B2038, 61931008, 62025604, 61971016, 6212200758 and 61976202, in part by the Fundamental Research Funds for the Central Universities, in part by Youth Innovation Promotion Association CAS, in part by the Strategic Priority Research Program of Chinese Academy of Sciences, Grant No.XDB28000000, in part by the China National Postdoctoral Program for Innovative Talents under Grant BX2021298, and in part by China Postdoctoral Science Foundation under Grant 2022M713101.

	\clearpage
	
	\bibliography{neurips_2022}
	\bibliographystyle{neurips_2022}

	\clearpage
\section*{Checklist}

\begin{enumerate}

\item For all authors...
\begin{enumerate}
  \item Do the main claims made in the abstract and introduction accurately reflect the paper's contributions and scope?
    \answerYes{}
  \item Did you describe the limitations of your work?
    \answerNA{} 
  \item Did you discuss any potential negative societal impacts of your work?
    \answerNo{There are no immediate societal impacts associated with our research.}
  \item Have you read the ethics review guidelines and ensured that your paper conforms to them?
    \answerYes{}
\end{enumerate}

\item If you are including theoretical results...
\begin{enumerate}
  \item Did you state the full set of assumptions of all theoretical results?
    \answerYes{}
        \item Did you include complete proofs of all theoretical results?
    \answerYes{See the supplementary materials.}
\end{enumerate}

\item If you ran experiments...
\begin{enumerate}
  \item Did you include the code, data, and instructions needed to reproduce the main experimental results (either in the supplemental material or as a URL)?
    \answerYes{}
  \item Did you specify all the training details (e.g., data splits, hyperparameters, how they were chosen)?
    \answerYes{See Sec.\ref{app:train} in the Appendix.}
        \item Did you report error bars (e.g., with respect to the random seed after running experiments multiple times)?
    \answerNA{}
        \item Did you include the total amount of compute and the type of resources used (e.g., type of GPUs, internal cluster, or cloud provider)?
    \answerNA{}
\end{enumerate}

\item If you are using existing assets (e.g., code, data, models) or curating/releasing new assets...
\begin{enumerate}
  \item If your work uses existing assets, did you cite the creators?
    \answerYes{}
  \item Did you mention the license of the assets?
    \answerYes{See Sec.\ref{exp_supp}}
  \item Did you include any new assets either in the supplemental material or as a URL? \answerNA{} 
  
  \item Did you discuss whether and how consent was obtained from people whose data you're using/curating?
    \answerNo{} We use open-source software and datasets. 
  \item Did you discuss whether the data you are using/curating contains personally identifiable information or offensive content?
    \answerNA{}
\end{enumerate}

\item If you used crowdsourcing or conducted research with human subjects...
\begin{enumerate}
  \item Did you include the full text of instructions given to participants and screenshots, if applicable?
    \answerNA{}
  \item Did you describe any potential participant risks, with links to Institutional Review Board (IRB) approvals, if applicable?
    \answerNA{}
  \item Did you include the estimated hourly wage paid to participants and the total amount spent on participant compensation?
    \answerNA{}
\end{enumerate}
\end{enumerate}
	\newpage
	\appendix
	\onecolumn
	
	 \section*{\textcolor{blue}{\Large{Contents}}}
	 \startcontents[sections]
	 \printcontents[sections]{l}{1}{\setcounter{tocdepth}{2}}
	 
	 \clearpage
	 \section{Prior Arts} \label{rel_work}
	In this section, we briefly review the closely related studies along with our main topic.
	
	\subsection{One-Class Collaborative Filtering}
	
	In many real-world applications, the vast majority of interactions are implicitly expressed by users’ behaviors, e.g., downloads of movies, clicks of products and browses of news. In order to develop RS from such implicit feedback, researchers usually formulate the recommendation task as the \textit{One-Class Collaborative Filtering} (OCCF) problem \cite{DBLP:conf/icdm/PanZCLLSY08, DBLP:journals/www/YaoTYXZSL19, DBLP:conf/icml/HeckelR17,DBLP:journals/kbs/ZhangR21,DBLP:conf/sigir/LeeKJPY21, DBLP:journals/ijon/ZhangR21}.

	
	\noindent \textbf{Matrix Factorization (MF) based Algorithm}. Over the past decades, the Matrix Factorization (MF)-based algorithms are one of the most classical OCCF solutions \cite{DBLP:conf/ijcai/0001DWTTC18,DBLP:conf/icml/ZhengTDZ16,DBLP:conf/uai/RendleFGS09, DBLP:conf/aaai/ChenL019}. The key idea of MF is to factorize the user-item interaction into an inner product between user and item embeddings, respectively. For example, \cite{DBLP:conf/sigir/HeZKC16} proposes an item-oriented MF method with implicit feedback. Neural Collaborative Filtering (NCF) \cite{DBLP:conf/www/HeLZNHC17} develops a general framework that unifies the MF and the neural networks together, and then regards the recommendation task as a regression problem. 
	
	
	\noindent \textbf{Collaborative Metric Learning based Algorithm}. As mentioned in the Sec.\ref{intro}, \textit{Collaborative Metric Learning} (CML) \cite{hsieh2017collaborative} presents a very early trial to replace the inner product scoring function with Euclidean distance.
	Ever since CML \cite{hsieh2017collaborative} has shown great power in RS, there raises a new direction to develop recommendation algorithms. Typically, the existing methods fall into two camps. The first camp pursues model-based improvement over the original CML \cite{DBLP:conf/icdm/ParkKXY18, DBLP:conf/www/TayTH18}, with a goal to improve the formulation of the latent embeddings. 
	Representative work includes \cite{DBLP:conf/icdm/ParkKXY18,DBLP:conf/www/TayTH18,DBLP:conf/dasfaa/ZhangZLXF0SC19,DBLP:conf/recsys/TranSHM21}. 
	Moreover, since CML employs a negative sampling strategy \cite{DBLP:conf/kdd/YangDZYZT20,DBLP:conf/wsdm/RendleF14} to mitigate the high computational burden of pairwise learning, another camp attempts to improve the effectiveness of the negative sampling process \cite{DBLP:conf/acml/CanevetF14}, such as uniform sampling \cite{hsieh2017collaborative}, popularity-based sampling \cite{DBLP:conf/sigir/WuVSSR19}, two-stage negative sampling \cite{tran2019improving} and Hard negative sampling \cite{DBLP:journals/pr/GajicAG21,DBLP:conf/ijcai/DingQ00J19}.
	
	This paper falls into the first camp. We focus on a challenging scenario where a user has multiple categories of interests. Unfortunately, in this case, the current literature equipped with unique user representation might induce preference bias, especially when the item category distribution is imbalanced. 
	
	\subsection{Recommendation against Joint Accessibility} \label{Sec.2.2}
	Recently, some studies \cite{DBLP:conf/eaamo/GuoKJG21, DBLP:conf/icml/CurmeiDR21, DBLP:conf/fat/DeanRR20} have pointed out a \textit{joint accessibility} problem in the recommendation, which determines the opportunities of users to discover interesting content. More precisely, joint accessibility measures whether an item candidate with size $K$ could be jointly accessed by a user in a Top-$K$ recommendation \cite{DBLP:conf/eaamo/GuoKJG21}. In other words, joint accessibility also somewhat captures a fundamental requirement of content diversity. If there are sufficient preference records of a target user, he/she should be able to be recommended any combination of $K$ items that he/she may be interested in. In this direction, noteworthy is the work present in \cite{DBLP:conf/eaamo/GuoKJG21}, which provides the theoretically necessary and sufficient conditions to meet the joint accessibility. Subsequently, \cite{DBLP:conf/eaamo/GuoKJG21} proposes an alternative MF-based model (M2F) to improve joint accessibility. Formally, with respect to each user, it assigns $m$ feature vectors to users, and thus the predicted score of each item is defined as $
	s(j) = \max\limits_{i \in [m]} \boldsymbol{u}_i^{\top} \boldsymbol{v}_j$, where $\boldsymbol{u}_i, i \in [m]$ is the $i$-th user latent vector; $\boldsymbol{v}_j$ is the item feature and $[m] = \{1, \dots, m\}$. 
	Finally, M2F adopts the least square \cite{DBLP:conf/recsys/TakacsT12} loss to recover the missing values in the user-item matrix.
	
	The existing line of such work merely focuses on the MF-based algorithms, while we take a further step to explore the problem under the context of CML. It is also interesting to note that, under mild conditions, we could see that M2F is a particular case of our method (shown in Sec.\ref{sec.3.6}). In this sense, we generalize the original idea of joint accessibility.
	 \newpage
	\section{Generalization Bounds and its proofs} 
	\label{supp.sec.b}
	\subsection{Preliminary Lemmas}
	In this section, we first briefly review some preparatory knowledge for the proof.
		\begin{defi} [Bounded Difference Property]\label{def:bdp}
			Given a group of independent random variables $X_1, X_2, \cdots, X_n$ where $X_t \in \mathbb{X}, \forall t$, $f(X_1,X_2,\cdots, X_n)$ is satisfied with the bounded difference property, if there exists some non-negative constants $c_1, c_2,\cdots, c_n$, such that: 
			\begin{equation}
				\sup_{x_1,x_2,\cdots,x_n, x'_t} \left|f(x_1,\cdots,x_n) - f(x_1,\cdots, x_{t-1},x'_t,\cdots,x_n)\right| \le c_t, ~ \forall t,  1 \le t \le n.
			\end{equation}
		\end{defi}
		Hereafter, if any function $f$ holds the Bounded Difference Property, the following Mcdiarmid's inequality is always satisfied.
		\begin{lem}[Mcdiarmid's Inequality \cite{mc}] \label{lem:mc} Assume we have $n$ independent random variables $X_1, X_2, \dots, X_n$ that all of them are chosen from the set $\mathcal{X}$. For a function $f: \mathcal{X} \rightarrow \mathbb{R}$, $\forall t, 1 \le t \le n$, if the following inequality holds:
			\[
			\sup_{x_1,x_2,\cdots,x_n, x'_t} \left|f(x_1,\cdots,x_n) - f(x_1,\cdots, x_{t-1},x'_t,\cdots,x_n)\right| \le c_t, ~ \forall t,  1 \le t \le n.
			\]with $\boldsymbol{x} \neq \boldsymbol{x}'$, then for all $\epsilon >0$, we have
			\[\mathbb{P}[ \mathbb{E}(f) - f \ge \epsilon ] \le \exp\left(\dfrac{-2\epsilon^2}{\sum_{t=1}^nc_t^2} \right), \]
			\[
			\mathbb{P}[ f - \mathbb{E}(f) \ge \epsilon ] \le \exp\left(\dfrac{-2\epsilon^2}{\sum_{t=1}^nc_t^2} \right).
			\]
		\end{lem}
		\begin{lem}[Union bound/Boole’s inequality] \label{lem46} Given the countable or finite set of events $E_i$, the probability that at least one event happens is less than or equal to the sum of all probabilities of the events happened individually, i.e.,
			\begin{equation}
				\begin{aligned}
					\mathbb{P}\left[\mathop{\cup}\limits_{i} E_i\right] &\le \sum_{i} \mathbb{P}\left[E_i\right]
				\end{aligned}
			\end{equation}
			
		\end{lem}
		\begin{lem} [$\phi$-Lipschitz Continuous] \label{lem2} Given a set $\mathcal{X}$ and a function $f: \mathcal{X} \rightarrow \mathbb{R}$, if $f$ is continuously differentiable on $\mathcal{X}$ such that, $\forall x, y \in \mathcal{X}$, the following condition holds with a real constant $\phi$:
			\[
			\left\|f(x) - f(y)\right\| \le \phi \left\|x - y\right\|.
			\]
			Thereafter, $f$ is said to be a $\phi$-Lipschitz continuous function.
		\end{lem}
	
	\subsection{Key Lemmas}
		\begin{rdef2}
			[$\epsilon$-Covering] \label{rdef2} \cite{ledoux1991probability} Let $(\mathcal{F}, \rho)$ be a (pesudo) metric space, and $\mathcal{G} \subseteq \mathcal{F}$. $\{f_1, \dots, f_n\}$ is said to be an $\epsilon$-covering of $\mathcal{G}$ if $\mathcal{G} \subseteq \mathop{\cup}\limits_{i=1}^n \mathcal{B}(f_i, \epsilon)$, i.e., $\forall g \in \mathcal{G}$, $\exists i$ such that $\rho(g, f_i) \le \epsilon$.
		\end{rdef2}
		\begin{rdef3} [Covering Number] \label{rdef3} \cite{ledoux1991probability} According to the notations in Def.\ref{rdef2}, the covering number of $\mathcal{G}$ with radius $\epsilon$ is defined as:
			\begin{equation} \nonumber
				\mathcal{N}(\epsilon;\mathcal{G}, \rho) = \min\{n: \exists \epsilon-covering \ over \ \mathcal{G} \ with \ size \ n\}
			\end{equation}
		\end{rdef3}
		\begin{rassu1}
			[Basic Assumptions] \label{rassu1} We assume that all the embeddings of users and items are chosen from the following embedding hypothesis space:
			\begin{equation}
				\mathcal{H}_R = \left\{\bmg: \bmg \in \mathbb{R}^d, \|\bmg\| \le r\right\},
			\end{equation}
			where $\boldsymbol{g}^c_{u_i} \in \mathcal{H}_R, u_i \in \mcu, c \in [C]$ and $\boldsymbol{g}_{v_j} \in \mathcal{H}_R, v_j \in \mci$.
		\end{rassu1}
		\begin{rlem1} \label{covering_rlem} \cite{DBLP:conf/iclr/LongS20,DBLP:conf/icml/LiL21, DBLP:journals/jc/Zhou02}
			The covering number of the hypothesis class $\mathcal{H}_R$ has the following upper bound: 
			\begin{equation}
				\log \mathcal{N}(\epsilon;\mathcal{H}_R, \rho) \le d \log \left(\frac{3r}{\epsilon}\right),
			\end{equation}
			where $d$ is the dimension of embedding space. 
		\end{rlem1}
	
	In what follows, we will present the key lemmas to derive the upper bounds.
	
		\begin{lem} \label{lem8} Let $\varepsilon$ be the generalization error between $\hat{\mathcal{L}}_{\mcd}(\bmg)$ and $\expe[\hat{\mathcal{L}}_{\mcd}(\bmg)]$. Then by constructing an $\sigma$-covering $\{\bmg_1, \bmg_2, \dots, \bmg_n\}$ of $\mathcal{H}_R$ with $\sigma = \frac{\varepsilon}{16r(4 + \eta)}$, the following inequality holds
			\begin{equation}
				\begin{aligned} \label{eqlem8}
					\mathbb{P} \left[ \sup\limits_{\bmg \in \mathcal{B}(\bmg_l, \sigma)}\left|\hat{\mathcal{L}}_{\mcd}(\bmg) - \expe [\hat{\mathcal{L}}_{\mcd}(\bmg)]\right|\le \varepsilon \right] &\ge \mathbb{P}\left[\left|\hat{\mathcal{L}}_{\mcd}(\bmg_l) - \expe [\hat{\mathcal{L}}_{\mcd}(\bmg_l)]\right| \le \frac{\varepsilon}{2}\right], \ \ \forall l \in [n],
				\end{aligned}
			\end{equation}

			
		\end{lem}
	\begin{proof} Assume there exists an $\sigma$-covering $\{\bmg_1, \bmg_2, \dots, \bmg_n\}$ of $\mathcal{H}_R$. To prove (\ref{eqlem8}), we turn to prove the following inequality:
		\begin{equation}
			\begin{aligned} \label{eq122}
				\left||\hat{\mathcal{L}}_{\mcd}(\bmg) - \expe [\hat{\mathcal{L}}_{\mcd}(\bmg)]| - |\hat{\mathcal{L}}_{\mcd}(\bmg_l) - \expe [\hat{\mathcal{L}}_{\mcd}(\bmg_l)]|\right| &\le \frac{\varepsilon}{2}, \ \forall l \in [n].
			\end{aligned}
		\end{equation}
		
		Note that, we have
		\begin{equation}
			\begin{aligned}
				\left||\hat{\mathcal{L}}_{\mcd}(\bmg) - \expe [\hat{\mathcal{L}}_{\mcd}(\bmg)]| - |\hat{\mathcal{L}}_{\mcd}(\bmg_l) - \expe [\hat{\mathcal{L}}_{\mcd}(\bmg_l)]|\right|
				&\overset{\textcolor{orange}{(**)}}{\le} \left|\hat{\mathcal{L}}_{\mcd}(\bmg) - \expe [\hat{\mathcal{L}}_{\mcd}(\bmg)] - \left(\hat{\mathcal{L}}_{\mcd}(\bmg_l) - \expe [\hat{\mathcal{L}}_{\mcd}(\bmg_l)]\right)\right| \\
				&\overset{\textcolor{orange}{(*)}}{\le} \left|\hat{\mathcal{L}}_{\mcd}(\bmg) - \hat{\mathcal{L}}_{\mcd}(\bmg_l)\right| + \left|\expe [\hat{\mathcal{L}}_{\mcd}(\bmg_l)] - \expe [\hat{\mathcal{L}}_{\mcd}(\bmg)]\right|,
			\end{aligned}
		\end{equation}
		where $\textcolor{orange}{(*)}$ and $\textcolor{orange}{(**)}$ follows the facts $|x + y| \le |x| + |y|$ and $||x| - |y|| \le |x - y|$, respectively.
		
		Then, to achieve (\ref{eq122}), we only need to show that the following inequation holds: 
		\begin{equation}
			\begin{aligned}\label{eq177}
				\left|\hat{\mathcal{L}}_{\mcd}(\bmg) - \hat{\mathcal{L}}_{\mcd}(\bmg_l)\right| &\le \frac{\varepsilon}{4}, \ \ \forall l \in [n].\\
			\end{aligned}
		\end{equation}
		
		Recall that 
		\begin{equation} \label{eq15}
			\hat{\mcl}_{\mcd}(\bmg) = \hat{\mathcal{R}}_{\mcd, \bmg} + \eta \cdot \hat{\Omega}_{\mcd, \bmg},
		\end{equation}
		where 
		\begin{equation}
			\begin{aligned}
				\hat{\mathcal{R}}_{\mcd, \bmg} &= \frac{1}{|\mcu|} \sum_{u_i \in \mcu} \frac{1}{n_i^+n_i^-} \sum_{j=1}^{n_i^+} \sum_{k=1}^{n_i^-} \ell^{(i)}_g(v_j^+, v_k^-), \\
				\ell^{(i)}_g(v_j^+, v_k^-) &= \max (0, \lambda + s(u_i, v_j^+) - s(u_i, v_k^-)), \\
				s(u_i, v_j) &= \min\limits_{c \in [C]} \|\boldsymbol{g}_{u_i}^c - \boldsymbol{g}_{v_j}\|^2, \forall \ v_j, v_j \in \mci
			\end{aligned}
		\end{equation}
		and 
		\begin{equation}
			\begin{aligned}
				\hat{\Omega}_{\mcd, \bmg} &= \frac{1}{|\mcu|} \sum_{u_i \in \mcu} \left(\max\left(0, \delta_1 - \delta_{\bmg, u_i}) + \max(0, \delta_{\bmg, u_i} - \delta_2\right)\right), \\
				\delta_{\bmg, u_i} &= \frac{1}{2C(C-1)} \sum_{c_{1}, c_2 \in C} \|\boldsymbol{g}_{u_i}^{c_1} - \boldsymbol{g}_{u_i}^{c_2}\|^2.
			\end{aligned}
		\end{equation}
		
		Let us define some intermediate variables:
		\begin{equation}
			\begin{aligned}
				\hat{\mathcal{R}}_{\mathcal{D}_{u_i}, \bmg} &= \frac{1}{n_i^+n_i^-} \sum_{j=1}^{n_i^+}\sum_{k=1}^{n_i^-}\ell^{(i)}_g(v_j^+, v_k^-), \\
				\hat{\Omega}_{\mcd_{u_i}, \bmg} &= \max\left(0, \delta_1 - \delta_{\bmg, u_i}\right) + \max\left(0, \delta_{\bmg, u_i} - \delta_2\right), \\
				\Delta_{\bmg, \bmg_l}(c_1,c_2) &= \left(\|\boldsymbol{g}_{u_i}^{c_1} - \boldsymbol{g}_{v_j^+}\|^2 - \|\tilde{\boldsymbol{g}}_{u_i}^{c_2} - \tilde{\boldsymbol{g}}_{v_j^+}\|^2\right).
			\end{aligned}
		\end{equation}
		
		In this sense, we have 
		\begin{equation}
			\begin{aligned}
				\left|\hat{\mathcal{L}}_{\mcd}(\bmg) - \hat{\mathcal{L}}_{\mcd}(\bmg_l)\right| &= \left|\hat{\mathcal{R}}_{\mcd, \bmg} + \eta \cdot \hat{\Omega}_{\mcd, \bmg} -\hat{\mathcal{R}}_{\mcd, \bmg_l} - \eta \cdot \hat{\Omega}_{\mcd, \bmg_l} \right| \\
				& \le \underbrace{\left|\hat{\mathcal{R}}_{\mcd, \bmg} - \hat{\mathcal{R}}_{\mcd, \bmg_l}\right|}_{\textcolor{orange}{(1)}} + \underbrace{\eta\left|\Omega_{\mcd, \bmg} - \cdot \Omega_{\mcd, \bmg_l}\right|}_{\textcolor{orange}{(2)}} \\
			\end{aligned}
		\end{equation}
		
		Subsequently, in terms of $\textcolor{orange}{(1)}$, we first consider a specific user $u_i$ with her/his corresponding interaction records $\mathcal{D}_{u_i}$. We have
		\begin{equation}
			\begin{aligned}
				\left|\hat{\mathcal{R}}_{\mcd_{u_i}, \bmg} - \hat{\mathcal{R}}_{\mcd_{u_i}, \bmg_l}\right| &= \left|\frac{1}{n_i^+n_i^-} \sum_{j=1}^{n_i^+}\sum_{k=1}^{n_i^-}\ell^{(i)}_{\bmg}(v_j^+, v_k^-) - \frac{1}{n_i^+n_i^-} \sum_{j=1}^{n_i^+}\sum_{k=1}^{n_i^-}\ell^{(i)}_{\bmg_l}(v_j^+, v_k^-) \right| \\
				& \le \frac{1}{n_i^+n_i^-}  \sum_{j=1}^{n_i^+}\sum_{k=1}^{n_i^-} \left|\ell^{(i)}_{\bmg}(v_j^+, v_k^-) - \ell^{(i)}_{\bmg_l}(v_j^+, v_k^-) \right| \\
				& \overset{\textcolor{orange}{(a)}}{\le} \frac{1}{n_i^+n_i^-} \sum_{j=1}^{n_i^+}\sum_{k=1}^{n_i^-} \left|s(u_i, v_j^+) - s(u_i, v_k^-) - \tilde{s}_l(u_i, v_j^+) + \tilde{s}_l(u_i, v_k^-)\right| \\
				& \overset{\textcolor{orange}{(*)}}{\le} \frac{1}{n_i^+n_i^-} \sum_{j=1}^{n_i^+}\sum_{k=1}^{n_i^-} \left(\left|s(u_i, v_j^+)  - \tilde{s}_l(u_i, v_j^+) \right| +\left|\tilde{s}_l(u_i, v_k^-) - s(u_i, v_k^-) \right|\right) \\
			\end{aligned}
		\end{equation}
		
		where \textcolor{orange}{(a)} follows the Lem.\ref{lem2} and $\ell^{(i)}_{\bmg}$ is apparently a $1$-Lipschitz continuous function. 
		
		In terms of $\left|s(u_i, v_j^+)  - \tilde{s}_l(u_i, v_j^+) \right|$, the following equation holds:
		\begin{equation}
			\begin{aligned}
				\left|s(u_i, v_j^+)  - \tilde{s}_l(u_i, v_j^+) \right| &= \left|\min\limits_{c_1 \in [C]} \|\boldsymbol{g}_{u_i}^{c_1} - \boldsymbol{g}_{v_j^+}\|^2 - \min\limits_{c_2 \in [C]} \|\tilde{\boldsymbol{g}}_{u_i}^{c_2} - \tilde{\boldsymbol{g}}_{v_j^+}\|^2 \right| \\
				&= \left|\min\limits_{c_1 \in [C]} \max\limits_{c_2 \in [C]} \Delta_{\bmg, \bmg_l}(c_1,c_2)\right| \\
				&= \max \left\{\min\limits_{c_1 \in [C]} \max\limits_{c_2 \in [C]} \Delta_{\bmg, \bmg_l}(c_1,c_2), \max\limits_{c_1 \in [C]} \min\limits_{c_2 \in [C]} \Delta_{\bmg_l, \bmg}(c_2,c_1)\right\}
			\end{aligned}
		\end{equation}
		
		Moreover, we have 
		\begin{equation}
			\begin{aligned}
				\ \ \ \ & \min\limits_{c_1 \in [C]} \max\limits_{c_2 \in [C]} \Delta_{\bmg, \bmg_l}(c_1,c_2) \\
				&\le \max\limits_{c_1 = c2, c_1,c_2 \in [C]} \Delta_{\bmg, \bmg_l}(c_1,c_2) \\
				&\le \max\limits_{c_1 = c2, c_1,c_2 \in [C]} \left|\Delta_{\bmg, \bmg_l}(c_1,c_2) \right|\\
				&= \max\limits_{c \in [C]} \left|\left(\|\boldsymbol{g}_{u_i}^{c} - \boldsymbol{g}_{v_j^+}\| +\|\tilde{\boldsymbol{g}}_{u_i}^{c} - \tilde{\boldsymbol{g}}_{v_j^+}\| \right)\left(\|\boldsymbol{g}_{u_i}^{c} - \boldsymbol{g}_{v_j^+}\| -\|\tilde{\boldsymbol{g}}_{u_i}^{c} - \tilde{\boldsymbol{g}}_{v_j^+}\|\right)\right| \\
				& \overset{\textcolor{orange}{(**)}}{\le} \max\limits_{c \in [C]}  \left(\|\boldsymbol{g}_{u_i}^{c} - \boldsymbol{g}_{v_j^+}\| +\|\tilde{\boldsymbol{g}}_{u_i}^{c} - \tilde{\boldsymbol{g}}_{v_j^+}\| \right)\left(\|\boldsymbol{g}_{u_i}^{c} - \boldsymbol{g}_{v_j^+} -\tilde{\boldsymbol{g}}_{u_i}^{c} + \tilde{\boldsymbol{g}}_{v_j^+}\|\right) \\
				& \le 4r\left(\max\limits_{c\in [C]} \|\boldsymbol{g}_{u_i}^{c} - \tilde{\boldsymbol{g}}_{u_i}^{c}\| +  \|\tilde{\boldsymbol{g}}_{v_j^+} - \boldsymbol{g}_{v_j^+}\|\right) \\
				& \le 8r \sigma
			\end{aligned}
		\end{equation}
		
		where $\textcolor{orange}{(**)}$ follows the fact $||x| - |y|| \le |x - y|$.
		
		Similarly, we have
		\begin{equation}
			\begin{aligned}
				\left|\tilde{s}_l(u_i, v_k^-) - s(u_i, v_k^-) \right| &\le 4r\left(\max\limits_{c\in [C]} \|\boldsymbol{g}_{u_i}^{c} - \tilde{\boldsymbol{g}}_{u_i}^{c}\| +  \|\tilde{\boldsymbol{g}}_{v_k^-} - \boldsymbol{g}_{v_k^-}\|\right)  \\
				& \le 8r \sigma
			\end{aligned}
		\end{equation}

		Thus, we have
		
		\begin{equation}
			\begin{aligned}
				\left|\hat{\mathcal{R}}_{\mcd, g} - \hat{\mathcal{R}}_{\mcd, \tilde{g}_l}\right| = 
				\left|\hat{\mathcal{R}}_{\mcd_{u_i}, \bmg} - \hat{\mathcal{R}}_{\mcd_{u_i}, \bmg_l}\right| & \le 16r \sigma
			\end{aligned}
		\end{equation}
		
		Therefore, for all users, we also have
		\begin{equation} \label{proof1}
			\left|\hat{\mathcal{R}}_{\mcd, \bmg} - \hat{\mathcal{R}}_{\mcd, \bmg_l}\right| \le 16r \sigma.
		\end{equation}
		{\noindent} \rule[-5pt]{17.5cm}{0.05em}
		
		With respect to $\textcolor{orange}{(2)}$, we also first consider a specific user $u_i$, i.e.,
		\begin{equation}
			\begin{aligned}
				& \eta\left|\hat{\Omega}_{\mcd_{u_i}, \bmg} - \hat{\Omega}_{\mcd_{u_i}, \bmg_l}\right| \overset{\textcolor{orange}{(a)}}{\le} 2\eta \left|\delta_{\bmg, u_i} - \delta_{\bmg_l, u_i}\right| \\
				&= \frac{\eta}{C(C-1)} \left|\sum_{c_{1}, c_2 \in C} \left\|\boldsymbol{g}_{u_i}^{c_1} - \boldsymbol{g}_{u_i}^{c_2}\right\|^2 - \sum_{c_{1}, c_2 \in C} \left\|\tilde{\boldsymbol{g}}_{u_i}^{c_1} - \tilde{\boldsymbol{g}}_{u_i}^{c_2}\right\|^2\right| \\
				& \le \frac{\eta}{C(C-1)}\left|\sum_{c_{1}, c_2 \in C} \left(\left\|\boldsymbol{g}_{u_i}^{c_1} - \boldsymbol{g}_{u_i}^{c_2}\right\| + \left\|\tilde{\boldsymbol{g}}_{u_i}^{c_1} - \tilde{\boldsymbol{g}}_{u_i}^{c_2}\right\|\right)\left(\left\|\boldsymbol{g}_{u_i}^{c_1} - \boldsymbol{g}_{u_i}^{c_2}\right\| - \left\|\tilde{\boldsymbol{g}}_{u_i}^{c_1} - \tilde{\boldsymbol{g}}_{u_i}^{c_2}\right\|\right) \right| \\
				& \le \frac{\eta}{C(C-1)}\sum_{c_{1}, c_2 \in C}\left|  \left(\left\|\boldsymbol{g}_{u_i}^{c_1} - \boldsymbol{g}_{u_i}^{c_2}\right\| + \left\|\tilde{\boldsymbol{g}}_{u_i}^{c_1} - \tilde{\boldsymbol{g}}_{u_i}^{c_2}\right\|\right)\left(\left\|\boldsymbol{g}_{u_i}^{c_1} - \boldsymbol{g}_{u_i}^{c_2}\right\| - \left\|\tilde{\boldsymbol{g}}_{u_i}^{c_1} - \tilde{\boldsymbol{g}}_{u_i}^{c_2}\right\|\right)\right| \\
				& \overset{\textcolor{orange}{(**)}}{\le} \frac{4\eta r }{C(C-1)}\sum_{c_{1}, c_2 \in C}\left(\|\boldsymbol{g}_{u_i}^{c_1} -\tilde{\boldsymbol{g}}_{u_i}^{c_1} \| + \|\tilde{\boldsymbol{g}}_{u_i}^{c_2} - \boldsymbol{g}_{u_i}^{c_2}\|\right) \\
				& \le 4\eta r \left(\max\limits_{c \in [C]} \|\boldsymbol{g}_{u_i}^{c} -\tilde{\boldsymbol{g}}_{u_i}^{c}\|\right) \\
				& \le 4\eta r \sigma
			\end{aligned}
		\end{equation}
		where \textcolor{orange}{(a)} follows the Lem.\ref{lem2} and $\textcolor{orange}{(**)}$ follows $||x| - |y|| \le |x - y|$.
		
		
		In like wise, we have 
		\begin{equation}
			\begin{aligned} \label{proof2}
				\eta\left|\hat{\Omega}_{\mcd, \bmg} -  \hat{\Omega}_{\mcd, \bmg_l}\right|\le 4\eta r \sigma.
			\end{aligned}
		\end{equation}
		
		Finally, based on (\ref{proof1}) and (\ref{proof2}), we have
		\begin{equation}
			\left|\hat{\mathcal{L}}_{\mcd}(\bmg) - \hat{\mathcal{L}}_{\mcd}(\bmg_l)\right| \le 4r\sigma(4 + \eta).
		\end{equation}
		Based on this, by further choosing $\sigma = \frac{\varepsilon}{16r(4 + \eta)}$, we could construct the covering number $\mathcal{N}_1$ and $\mathcal{N}_2$ with respect to users and items, respectively, i.e., 
		\begin{equation}
			\begin{aligned}
				\mathcal{N}_1\left(\frac{\varepsilon}{16r(4 + \eta)}, \mathcal{H}_R,\rho_1\right), \ \ \rho_1 &= \max\limits_{c \in [C]} \|\boldsymbol{g}_{u_i}^{c} -\tilde{\boldsymbol{g}}_{u_i}^{c}\|, \ \ \forall u_i \in \mcu, \\
				\mathcal{N}_2\left(\frac{\varepsilon}{16r(4 + \eta)},\mathcal{H}_R, \rho_2\right), \ \ \rho_2 &=  \|\tilde{\boldsymbol{g}}_{v_j} - \boldsymbol{g}_{v_j}\|, \ \  \forall v_j \in \mci,
			\end{aligned}
		\end{equation}
		
		such that the following inequality holds:
		\[
		\left|\hat{\mathcal{L}}_{\mcd}(\bmg) - \hat{\mathcal{L}}_{\mcd}(\bmg_l)\right| \le \frac{\varepsilon}{4}.
		\]
		This completed the proof.
	\end{proof}
	
	{\noindent} \rule[-5pt]{17.5cm}{0.05em} \\

		\begin{lem} [Bounded Difference Property of DPCML] \label{lem9} Let $\mcd$ and $\mcd'$ be two independent datasets where exactly one instance is different instead of a term. We conclude that $\hat{\mathcal{L}}_{\mcd}(\bmg)$ satisfies the bounded difference property (Lem.\ref{def:bdp}).
		\end{lem}
	
	\begin{proof} We need to seek the upper bound of
		\[
		\sup\limits_{\bmg \in \mathcal{H}_R}\left|\hat{\mcl}_{\mathcal{D}'}(\bmg) - \hat{\mcl}_{\mcd}(\bmg)\right|.
		\]
		
		To achieve this, notice that, such difference between $\mathcal{D}$ and $\mathcal{D}'$  could be caused by either the user side or the item side. Therefore, we have the following three possible cases:
		\begin{itemize}
			\item \textbf{Case 1:} Only one user is different, i.e., 
			\begin{equation}
				\mcd = \mathop{\cup}\limits_{u_i \in \mathcal{U}} \ \mathcal{D}_{u_i}, \ \ \ \, \mathcal{D}' = \left(\mcd \backslash \mathcal{D}_{u_t}\right) \cup \mathcal{D}_{u'_t}, \ \ \forall t, t = 1, 2, \dots, |\mcu|.
			\end{equation}
			Under this circumstance, we have
			\begin{equation}
				\begin{aligned}\label{eq17}
					\sup\limits_{\bmg \in \mathcal{H}_R}\left|\hat{\mcl}_{\mathcal{D}'}(\bmg) - \hat{\mcl}_{\mcd}(\bmg)\right| &\overset{\textcolor{orange}{(b)}}{\le} \underbrace{\sup\limits_{\bmg \in \mathcal{H}_R} \left|\hat{\mcr}_{\mcd, \bmg} - \hat{\mcr}_{\mcd', \bmg} \right|}_{\textcolor{orange}{(3)}} + \underbrace{\sup\limits_{\bmg \in \mathcal{H}_R} \left|\eta \hat{\Omega}_{\mcd, \bmg} - \eta \hat{\Omega}_{\mathcal{D}', \bmg}\right|}_{\textcolor{orange}{(4)}} \\
				\end{aligned}  
			\end{equation}
			where $\textcolor{orange}{(b)}$ is achieved by the inequality: $\sup (x + y) \le \sup (x) + \sup (y)$.
			
			Based on (\ref{eq17}), in what follows, we will show the upper bound of term $\textcolor{orange}{(3)}$ and $\textcolor{orange}{(4)}$, respectively.
			
			At first, we define some intermediate variables:
			\begin{equation}\nonumber
				\begin{aligned}
					\hat{\mathcal{R}}_{\mathcal{D}_{u_i}, \bmg} &= \frac{1}{n_i^+n_i^-} \sum_{j=1}^{n_i^+}\sum_{k=1}^{n_i^-}\ell^{(i)}_{\bmg}(v_j^+, v_k^-), \\
					\phi_{\bmg}(c_1, c_2) &= \|\boldsymbol{g}^{c_1}_{u_t} - \boldsymbol{g}_{v_j^+}\|^2 - \|\boldsymbol{g}^{c_2}_{u'_t} - \boldsymbol{g}_{v_j^+}\|^2, \forall c_1, c_2, c_1, c_2 \in\ [C]\\
				\end{aligned}
			\end{equation}
			Then, with respect to term $\textcolor{orange}{(3)}$, we have 
			\begin{equation}
				\begin{aligned} \label{eq333}
					&\sup\limits_{\bmg \in \mathcal{H}_R} \left|\hat{\mcr}_{\mcd, \bmg} - \hat{\mcr}_{\mcd', \bmg} \right| = \frac{1}{|\mcu|}\sup\limits_{\bmg \in \mathcal{H}_R} \left|\hat{\mcr}_{\mcd_{u_t}, \bmg} - \hat{\mcr}_{\mcd_{u'_t}, \bmg} \right| \\
					=& \frac{1}{|\mcu|} \sup\limits_{\bmg \in \mathcal{H}_R} \left|\frac{1}{n_i^+n_i^-} \sum_{j=1}^{n_i^+}\sum_{k=1}^{n_i^-}\ell^{(t)}_{\bmg}(v_j^+, v_k^-) - \frac{1}{n_i^+n_i^-} \sum_{j=1}^{n_i^+}\sum_{k=1}^{n_i^-}\ell^{(t')}_{\bmg}(v_j^+, v_k^-)\right| \\
					\le& \frac{1}{|\mcu|} \frac{1}{n_i^+n_i^-} \sup\limits_{\bmg \in \mathcal{H}_R} \sum_{j=1}^{n_i^+}\sum_{k=1}^{n_i^-} \left|\ell^{(t)}_{\bmg}(v_j^+, v_k^-) - \ell^{(t')}_{\bmg}(v_j^+, v_k^-)\right| \\
					\overset{\textcolor{orange}{(b)}}{\le}& \frac{1}{|\mcu|} \frac{1}{n_i^+n_i^-} \sum_{j=1}^{n_i^+}\sum_{k=1}^{n_i^-} \sup\limits_{\bmg \in \mathcal{H}_R} \left|\ell^{(t)}_{\bmg}(v_j^+, v_k^-) - \ell^{(t')}_{\bmg}(v_j^+, v_k^-)\right| \\
					\overset{\textcolor{orange}{(a)}}{\le}& \frac{1}{|\mcu|} \frac{1}{n_i^+n_i^-}  \sum_{j=1}^{n_i^+}\sum_{k=1}^{n_i^-} \sup\limits_{\bmg \in \mathcal{H}_R} \left|s(u_t, v_j^+) - s(u_t, v_k^-) - \left(s(u_t', v_j^+) - s(u_t', v_k^-)\right) \right| \\
					\overset{\textcolor{orange}{(*)}}{\le}& \frac{1}{|\mcu|} \frac{1}{n_i^+n_i^-}  \sum_{j=1}^{n_i^+}\sum_{k=1}^{n_i^-} \sup\limits_{g \in \mathcal{H}_R} \left(
					\left|s(u_t, v_j^+) - s(u_t', v_j^+)\right| + \left|s(u_t', v_k^-) - s(u_t, v_k^-)\right|\right) \\
					\overset{\textcolor{orange}{(b)}}{\le}& \frac{1}{|\mcu|} \frac{1}{n_i^+n_i^-}  \sum_{j=1}^{n_i^+}\sum_{k=1}^{n_i^-} \left(\sup\limits_{\bmg \in \mathcal{H}_R} 
					\left|s(u_t, v_j^+) - s(u_t', v_j^+)\right| + \sup\limits_{\bmg \in \mathcal{H}_R} 
					\left|s(u_t', v_k^-) - s(u_t, v_k^-)\right| \right)
				\end{aligned}
			\end{equation}
			
			For $\sup\limits_{\bmg \in \mathcal{H}_R} 
			\left|s(u_t, v_j^+) - s(u_t', v_j^+)\right|$, the following results hold:
			\begin{equation}
				\begin{aligned} \label{eq322}
					\sup\limits_{\bmg \in \mathcal{H}_R} \left|s(u_t, v_j^+) - s(u_t', v_j^+)\right| &=
					\sup\limits_{\bmg \in \mathcal{H}_R}  \left|\min\limits_{c_1 \in [C]} \|\boldsymbol{g}^{c_1}_{u_t} - \boldsymbol{g}_{v_j^+}\|^2 - \min\limits_{c_2 \in [C]} \|\boldsymbol{g}^{c_2}_{u'_t} - \boldsymbol{g}_{v_j^+}\|^2\right| \\
					& \le \sup\limits_{\bmg \in \mathcal{H}_R}  \left|\min\limits_{c_1 \in [C]} \max\limits_{c_2 \in [C]} \left(\|\boldsymbol{g}^{c_1}_{u_t} - \boldsymbol{g}_{v_j^+}\|^2 - \|\boldsymbol{g}^{c_2}_{u'_t} - \boldsymbol{g}_{v_j^+}\|^2 \right)\right| \\
					& \le \max \left\{\min\limits_{c_1 \in [C]} \max\limits_{c_2 \in [C]} \phi_{\bmg}(c_1, c_2), \max\limits_{c_1 \in [C]} \min\limits_{c_2 \in [C]} \phi_{\bmg}(c_2, c_1)\right\}\
				\end{aligned}
			\end{equation}
			
			According to (\ref{eq322}), we can go a step further:
			\begin{equation}
				\begin{aligned} \label{eq355}
					\min\limits_{c_1 \in [C]} \max\limits_{c_2 \in [C]} \phi_{\bmg}(c_1, c_2) & \le \max\limits_{c_1 = c_2, c1,c2 \in [C]} \phi_{\bmg}(c_1, c_1) \\
					& \le \max\limits_{c \in [C]} \left|\phi_{\bmg}(c, c)\right| \\
					&= \max\limits_{c \in [C]} \left|\|\boldsymbol{g}^{c}_{u_t} - \boldsymbol{g}_{v_j^+}\|^2 - \|\boldsymbol{g}^{c}_{u'_t} - \boldsymbol{g}_{v_j^+}\|^2\right| \\
					&= \max\limits_{c \in [C]} \left|\left(\|\boldsymbol{g}^{c}_{u_t} - \boldsymbol{g}_{v_j^+}\| + \|\boldsymbol{g}^{c}_{u'_t} - \boldsymbol{g}_{v_j^+}\| \right)\left(\|\boldsymbol{g}^{c}_{u_t} - \boldsymbol{g}_{v_j^+}\| - \|\boldsymbol{g}^{c}_{u'_t} - \boldsymbol{g}_{v_j^+}\| \right) \right| \\
					& \overset{\textcolor{orange}{(**)}}{\le} 4 r \max\limits_{c \in [C]} \|\boldsymbol{g}^{c}_{u_t} -\boldsymbol{g}^{c}_{u'_t} \| \\
					& \le 8r^2
				\end{aligned}
			\end{equation}
			
			Based on the result of (\ref{eq355}), we have the following result for (\ref{eq333})
			
			\begin{equation}
				\begin{aligned}
					\frac{1}{|\mcu|}\sup\limits_{\bmg \in \mathcal{H}_R} \left|\hat{\mcr}_{\mcd_{u_t}, \bmg} - \hat{\mcr}_{\mcd_{u'_t}, \bmg} \right| & \le \frac{16r^2}{|\mcu|}
				\end{aligned}
			\end{equation}
			
			
			With respect to $\textcolor{orange}{(4)}$, recall that, we have
			\[
			\hat{\Omega}_{\mcd, \bmg} = \frac{1}{|\mcu|} \sum_{u_i \in \mcu} \psi_{\bmg}(u_{i}),
			\]
			where 
			\begin{equation}\nonumber
				\begin{aligned}
					\psi_{\bmg}(u_{i}) &= \max\left(0, \delta_1 - \delta_{\bmg, u_i}\right) + \max\left(0, \delta_{\bmg, u_i} - \delta_2\right), \\
					\delta_{\bmg, u_i} &= \frac{1}{2C(C-1)} \sum_{c_{1}, c_2 \in C} \|\boldsymbol{g}_{u_i}^{c_1} - \boldsymbol{g}_{u_i}^{c_2}\|^2, \\
				\end{aligned}
			\end{equation}
			Moreover, let us define some intermediate variables:
			\begin{equation}\nonumber
				\begin{aligned}
					\psi_{\bmg, \delta_1}(u_{i}, u_j) &= \max\left(0, \delta_1 - \delta_{\bmg, u_i}\right) - \max\left(0, \delta_1 - \delta_{\bmg, u_j}\right), \\
					\psi_{\bmg, \delta_2}(u_{i}, u_j) &= \max\left(0, \delta_{\bmg, u_i} - \delta_2\right) - \max\left(0, \delta_{\bmg, u_j} - \delta_2\right).
				\end{aligned}
			\end{equation}
			In this sense, in terms of (\ref{eq17}), the following result holds:
			\begin{equation}
				\begin{aligned}\label{eq19}
					\sup\limits_{\bmg \in \mathcal{H}_R}\left|\hat{\mcl}_{\mathcal{D}'}(\bmg) - \hat{\mcl}_{\mcd}(\bmg)\right| & = \eta \cdot \left|\frac{1}{|\mcu|}\psi_{\bmg}(u_{t}) - \frac{1}{|\mcu|}\psi_{\bmg}(u'_{t})\right| \\
					&= \frac{\eta}{|\mcu|} \cdot \left|\psi_{\bmg}(u_{t}) - \psi_{\bmg}(u'_{t})\right| \\
					& \overset{\textcolor{orange}{(*)}}{\le} \frac{\eta}{|\mcu|} \cdot \left(\left|\psi_{\bmg, \delta_1}(u_t, u'_t)\right| + \left|\psi_{\bmg, \delta_2}(u_t, u'_t)\right|\right) \\
					& \overset{\textcolor{orange}{(a)}}{\le} \frac{2\eta}{|\mcu|} \cdot \left|\delta_{\bmg, u_t} - \delta_{\bmg, u'_t}\right| \\
					& = \frac{\eta}{C(C-1)|\mcu|} \left|\sum_{c_{1}, c_2 \in C} \left( \|\boldsymbol{g}_{u_t}^{c_1} - \boldsymbol{g}_{u_t}^{c_2}\|^2 - \|\boldsymbol{g}_{u'_t}^{c_1} - \boldsymbol{g}_{u'_t}^{c_2}\|^2\right)\right| \\
					& \overset{\textcolor{orange}{(*)}}{\le} \frac{\eta}{C(C-1)|\mcu|}\sum_{c_{1}, c_2 \in C} \left|\|\boldsymbol{g}_{u_t}^{c_1} - \boldsymbol{g}_{u_t}^{c_2}\|^2 - \|\boldsymbol{g}_{u'_t}^{c_1} - \boldsymbol{g}_{u'_t}^{c_2}\|^2\right| \\
					& \le \frac{4r^2\eta}{|\mcu|}
				\end{aligned}
			\end{equation}
			where $\textcolor{orange}{(*)}$ achieves via the inequality $|x + y| \le |x| + |y|$ and $\textcolor{orange}{(a)}$ follows the Lem.\ref{lem2}.
			
			Finally, in this case, we have
			\begin{equation}
				\begin{aligned}
					\sup\limits_{\bmg \in \mathcal{H}_R}\left|\hat{\mcl}_{\mathcal{D}'}(\bmg) - \hat{\mcl}_{\mcd}(\bmg)\right| \le \frac{16r^2 + 4r^2\eta}{|\mcu|}.
				\end{aligned}
			\end{equation}
			
			\item \textbf{Case 2:} \label{SFCML:case1} Only one positive item is different. In this case, we consider such difference occurs in the positive item $v_{t_1}^+$ with respect to a specific user $u_i$ and there are $|\mcu|$ cases for all users. Mathematically, we have
			\begin{equation}
				\begin{aligned}
					\mathcal{D}_{u_i} = \{v_j^+\}_{j=1}^{n_i^+} \cup \{v_k^-\}_{k=1}^{n_i^-}, \ \ \ \ \mathcal{D}_{u_i}' = (\mathcal{D}_{u_i} \backslash	\{v_{t_1}^+\}) \cup \{\tilde{v}_{t_1}^{+}\}, \label{SFCML:eq26}
				\end{aligned}
			\end{equation}
			where $\forall t_1, t_1 = 1, 2, \dots, n_i^+$ and $n_i^+ + n_i^- = |\mci|$.
			Then, it is obvious that in this case only the first term in (\ref{eq15}) contributes to the upper bound. According to this observation, the upper bound could be simplified as follows:
			\begin{equation}
				\begin{aligned}
					\sup\limits_{\bmg \in \mathcal{H}_R}\left|\hat{\mcl}_{\mathcal{D}'}(\bmg) - \hat{\mcl}_{\mcd}(\bmg)\right| 
					&= \sup\limits_{\bmg \in \mathcal{H}_R} \left|\hat{\mathcal{R}}_{\mcd}(\bmg) - \hat{\mathcal{R}}_{\mathcal{D}'}(\bmg)\right|	\\
					&= \sup\limits_{\bmg \in \mathcal{H}_R} \left|\hat{\mathcal{R}}_{\mathcal{D}_{u_i}}(\bmg) - \hat{\mathcal{R}}_{\mathcal{D}'_{u_i}}(\bmg)\right|,\\
					\label{SFCML:eq28}
				\end{aligned}
			\end{equation}
			where again we denote
			\[
			\hat{\mathcal{R}}_{\mathcal{D}_{u_i}}(\bmg) = \frac{1}{|\mcu|} \cdot \sum_{j=1}^{n_i^+} \sum_{k=1}^{n_i^-} \ell^{(i)}_{\bmg}(v_j^+, v_k^-),
			\]
			and 
			\[
			\ell^{(i)}_{\bmg}(v_j^+, v_k^-) = \max (0, \lambda + s(u_i, v_j^+) - s(u_i, v_k^-)).
			\]
			Let
			\begin{equation} \label{eqqq}
				\Delta_{\bmg}(c_1, c_2) = \|\boldsymbol{g}_{u_i}^{c_1} - \boldsymbol{g}_{v_{t_1}^+}\|^2 -\|\boldsymbol{g}_{u_i}^{c_2} - \boldsymbol{g}_{\tilde{v}_{t_1}^+}\|^2.
			\end{equation}
			
			Then, since $v_j^+$ and $\tilde{v}_j^{+}$ are different in this case, we have
			\begin{equation}
				\begin{aligned}
					\sup\limits_{\bmg \in \mathcal{H}_R}\left|\hat{\mcl}_{\mathcal{D}'}(\bmg) - \hat{\mcl}_{\mcd}(\bmg)\right| 
					& = \frac{1}{|\mcu|}\sup\limits_{\bmg \in \mathcal{H}_R} \left| \frac{1}{n_i^+n_i^-} \sum_{k=1}^{n_i^-} \ell^{(i)}_{\bmg}(v_{t_1}^+, v_k^-) -  \frac{1}{n_i^+n_i^-} \sum_{k=1}^{n_i^-} \ell^{(i)}_{\bmg}(\tilde{v}_{t_1}^{+}, v_k^-)\right|\\ 
					&  \overset{\textcolor{orange}{(*)}}{\le} \frac{1}{|\mcu|n_i^+n_i^-}  \sup\limits_{\bmg \in \mathcal{H}_R} \sum_{k=1}^{n_i^-} \left|\ell^{(i)}_{\bmg}(v_{t_1}^+, v_k^-) - \ell^{(i)}_{\bmg}(\tilde{v}_{t_1}^{+}, v_k^-) \right| \\
					& \overset{\textcolor{orange}{(b)}}{\le} \frac{1}{|\mcu|n_i^+n_i^-}  \sum_{k=1}^{n_i^-} \left(\sup\limits_{\bmg \in \mathcal{H}_R} \left|\ell^{(i)}_{\bmg}(v_{t_1}^+, v_k^-) - \ell^{(i)}_{\bmg}(\tilde{v}_{t_1}^{+}, v_k^-) \right|\right) \\
					& \overset{\textcolor{orange}{(a)
					}}{\le} \frac{1}{|\mcu|n_i^+n_i^-}   \sum_{k=1}^{n_i^-}\sup\limits_{\bmg \in \mathcal{H}_R}\left|s(u_i, v_{t_1}^+)-s(u_i, \tilde{v}_{t_1}^{+}) \right| \\
					&= \frac{1}{|\mcu|n_i^+n_i^-}  \sum_{k=1}^{n_i^-} \sup\limits_{\bmg \in \mathcal{H}_R} \left|\min\limits_{c_1 \in [C]} \|\boldsymbol{g}_{u_i}^{c_1} - \boldsymbol{g}_{v_{t_1}^+}\|^2 - \min\limits_{c_2 \in [C]} \|\boldsymbol{g}_{u_i}^{c_2} - \boldsymbol{g}_{\tilde{v}_{t_1}^+}\|^2\right| \\
					&= \frac{1}{|\mcu|n_i^+n_i^-}  \sum_{k=1}^{n_i^-} \sup\limits_{\bmg \in \mathcal{H}_R} \left| \min\limits_{c_1 \in [C]} \max\limits_{c_2 \in [C]} \left(\|\boldsymbol{g}_{u_i}^{c_1} - \boldsymbol{g}_{v_{t_1}^+}\|^2 -\|\boldsymbol{g}_{u_i}^{c_2} - \boldsymbol{g}_{\tilde{v}_{t_1}^+}\|^2 \right)\right| \\
					\label{SFCML:thoe3:1}
				\end{aligned}
			\end{equation}
			
			According to (\ref{eqqq}), we have
			\begin{equation}
				\begin{aligned}\label{eq11}
					\sup\limits_{\bmg \in \mathcal{H}_R}\left|\hat{\mcl}_{\mathcal{D}'}(\bmg) - \hat{\mcl}_{\mcd}(\bmg)\right| 
					& \le \frac{1}{|\mcu|n_i^+n_i^-}  \sum_{k=1}^{n_i^-} \sup\limits_{\bmg \in \mathcal{H}_R} \left|\min\limits_{c_1 \in [C]} \max\limits_{c_2 \in [C]} \Delta_{\bmg}(c_1, c_2)\right| \\
					&= \frac{1}{|\mcu|n_i^+n_i^-}  \sum_{k=1}^{n_i^-}\left|  \max\left\{\min\limits_{c_1 \in [C]} \max\limits_{c_2 \in [C]} \Delta_{\bmg}(c_1, c_2), \max\limits_{c_1 \in [C]} \min\limits_{c_2 \in [C]} \Delta_{\bmg}(c_2, c_1)\right\} \right| \\
				\end{aligned}
			\end{equation}
			
			It is easy to show that, 
			\begin{equation}
				\begin{aligned}\label{eq12}
					\min\limits_{c_1 \in [C]} \max\limits_{c_2 \in [C]} \Delta_{\bmg}(c_1, c_2) & \le \max\limits_{c_1 = c_2, c_1, c_2 \in [C]} \Delta_{\bmg}(c_1, c_2) \\
					& \le \max\limits_{c_1 = c_2, c_1, c_2 \in [C]} \left| \Delta_{\bmg}(c_1, c_2)\right| \\
					&= \max\limits_{c \in [C]}\left|\|\boldsymbol{g}_{u_i}^{c} - \boldsymbol{g}_{v_{t_1}^+}\|^2 -\|\boldsymbol{g}_{u_i}^{c} - \boldsymbol{g}_{\tilde{v}_{t_1}^+}\|^2\right|\\
					&= \max\limits_{c\in [C]}\left| \left(\|\boldsymbol{g}_{u_i}^{c} - \boldsymbol{g}_{v_{t_1}^+}\| + \|\boldsymbol{g}_{u_i}^{c} - \boldsymbol{g}_{\tilde{v}_{t_1}^+}\|\right) \left(\|\boldsymbol{g}_{u_i}^{c} - \boldsymbol{g}_{v_{t_1}^+}\| - \|\boldsymbol{g}_{u_i}^{c} - \boldsymbol{g}_{\tilde{v}_{t_1}^+}\|\right) \right|\\
					& \overset{\textcolor{orange}{(**)}}{\le} \left(\|\boldsymbol{g}_{u_i}^{c} - \boldsymbol{g}_{v_{t_1}^+}\| + \|\boldsymbol{g}_{u_i}^{c} - \boldsymbol{g}_{\tilde{v}_{t_1}^+}\|\right) \left(\|\boldsymbol{g}_{u_i}^{c} - \boldsymbol{g}_{v_{t_1}^+} -\boldsymbol{g}_{u_i}^{c} +  \boldsymbol{g}_{\tilde{v}_{t_1}^+}\|\right) \\
					& \le 8r^2
				\end{aligned}
			\end{equation}
			In the same way, we also have 
			\begin{equation}\label{eq13}
				\max\limits_{c_1 \in [C]} \min\limits_{c_2 \in [C]} \Delta_{\bmg}(c_2, c_1) \le 8r^2
			\end{equation}
			
			Therefore, applying (\ref{eq12}) and (\ref{eq13}) to (\ref{eq11}), we have
			\begin{equation}
				\begin{aligned}\label{eq14}
					\sup\limits_{\bmg \in \mathcal{H}_R}\left|\hat{\mcl}_{\mathcal{D}'}(\bmg) - \hat{\mcl}_{\mcd}(\bmg)\right|  \le \frac{8r^2}{|\mcu|n_i^+}
				\end{aligned}
			\end{equation}
			\item \textbf{Case 3:} \label{SFCML:case2} Only one negative item is different. In this case, we assume such difference occurs in the negative item $v_{t_2}^-$ with respect to a specific user $u_i$, and there are also $|\mcu|$ cases for all users. Mathematically, we have
			\begin{equation}
				\mathcal{D}_i = \{v_j^+\}_{j=1}^{n_i^+} \cup \{v_k^-\}_{k=1}^{n_i^-}, \ \ \ \ \mathcal{D}_i' = (\mathcal{D}_i \backslash \{v_{t_2}^-\}) \cup \{\tilde{v}_{t_2}^{-}\}. \label{SFCML:eq27}
			\end{equation}
			where $\forall t_2, t_2 = 1, 2, \dots, n_i^-$.
			
			Similarly, if $v_k^-$ and $\tilde{v}_k^{-}$ are different, we can also hold
			\begin{equation}
				\sup\limits_{\bmg \in \mathcal{H}_R}\left|\hat{\mcl}_{\mathcal{D}'}(\bmg) - \hat{\mcl}_{\mcd}(\bmg)\right|  \le \frac{8r^2}{|\mcu|n_i^-} \label{SFCML:thoe3:2}
			\end{equation}
		\end{itemize}
		
		Finally, taking all above three cases into account, one can conclude that $\hat{\mathcal{L}}_{\mcd}(\bmg)$ is satisfied with the bounded difference property (Lem.\ref{def:bdp}). 
		
		This completed the proof.
	\end{proof}
	
		\begin{lem} \label{lem6} Equipped with Lem.\ref{lem8} and Lem.\ref{lem9}, the following inequality holds:
			\begin{equation}\nonumber
				\begin{aligned}
					\mathbb{P}\left[\left|\hat{\mathcal{L}}_{\mcd}(\bmg_l) - \expe [\hat{\mathcal{L}}_{\mcd}(\bmg_l)]\right| \ge \frac{\varepsilon}{2}\right] & \le 2 \exp \left(\frac{- \varepsilon^2 \tilde{N}}{2}\right),
				\end{aligned}
			\end{equation}
			where 
			\[
			\tilde{N} = \left(4r^2\sqrt{\left(\frac{(4 + \eta)^2}{|\mcu|} + \frac{2}{|\mcu|^2} \sum_{u_i \in \mcu} \left(\frac{1}{n_i^+} + \frac{1}{n_i^-}\right)\right)}\right)^{-2}
			\]
		\end{lem}
	\begin{proof}
		The proof could be easily achieved by applying Lem.\ref{lem:mc} on top of Lem.\ref{lem8} and Lem.\ref{lem9}. 
	\end{proof}
	
	{\noindent} \rule[-10pt]{17.5cm}{0.05em}	
	
	\subsection{Proof of the Main Result}
	\subsubsection{Proof of Thm.\ref{them1}}
		\begin{rthm1} [Generalization Upper Bound of DPCML] \label{rthem1} Let $\expe [\hat{\mathcal{L}}_{\mcd}(\bmg)]$ be the population risk of $\hat{\mathcal{L}}_{\mcd}(\bmg)$. Then, $\forall \ \bmg, \bmg \in \mathcal{H}_R$, with high probability, the following inequation holds:
			\begin{equation}
				\begin{aligned}
					\left| \hat{\mathcal{L}}_{\mcd}(\bmg) - \expe [\hat{\mathcal{L}}_{\mcd}(\bmg)]\right| \le \sqrt{\frac{2d\log \left(3r \tilde{N}\right)}{\tilde{N}}},
				\end{aligned}
			\end{equation}
			where we have
			\begin{equation} \nonumber
				\tilde{N} = \left(4r^2\sqrt{\left(\frac{(4 + \eta)^2}{|\mcu|} + \frac{2}{|\mcu|^2} \sum_{u_i \in \mcu} \left(\frac{1}{n_i^+} + \frac{1}{n_i^-}\right)\right)}\right)^{-2}.
			\end{equation}
		\end{rthm1}
	
	\begin{proof} \textcolor{blue}{Step 1}. In order to obtain the generalization bound, we need to first figure out the following probability:
		
		\[
		\mathbb{P}\left[\sup_{\bmg \in \mathcal{H}_R} \left| \hat{\mathcal{L}}_{\mcd}(\bmg) - \expe [\hat{\mathcal{L}}_{\mcd}(\bmg)]\right|\ge \varepsilon\right],
		\]
		where $\varepsilon$ is the generalization error and usually a very small value.
		
		%
		Denote the covering number of $\sigma$-covering in Lem.\ref{lem8} as $\mathcal{N}_3(\sigma; \mathcal{H}_R, \rho_3)$. Then, according to Def.\ref{rdef2}, Def.\ref{rdef3}, Lem.\ref{lem46} and Lem.\ref{lem8}, we have
		\begin{equation}
			\begin{aligned}\label{eq10}
				\mathbb{P}\left[\sup_{\bmg \in \mathcal{H}_R} \left| \hat{\mathcal{L}}_{\mcd}(\bmg) - \expe [\hat{\mathcal{L}}_{\mcd}(\bmg)]\right| \ge \varepsilon \right] 
				&\le \mathbb{P}\left[\sup\limits_{\bmg \in \mathop{\cup}\limits_{l=1}^{\mathcal{N}_3} \mathcal{B}(\bmg_l, \sigma)}  \left|\hat{\mathcal{L}}_{\mcd}(\bmg) - \expe [\hat{\mathcal{L}}_{\mcd}(\bmg)]\right| \ge \varepsilon\right]\\
				& \overset{\textcolor{orange}{Lem.\ref{lem46}}}{\le} \sum_{l=1}^{\mathcal{N}_3} \mathbb{P} \left[ \sup\limits_{\bmg \in \mathcal{B}(\bmg_l, \sigma)}\left|\hat{\mathcal{L}}_{\mcd}(\bmg) - \expe [\hat{\mathcal{L}}_{\mcd}(\bmg)]\right|\ge \varepsilon \right] \\
				& \overset{\textcolor{orange}{Lem.\ref{lem8}}}{\le} \sum_{l=1}^{\mathcal{N}_3} \mathbb{P}\left[\left|\hat{\mathcal{L}}_{\mcd}(\bmg_l) - \expe [\hat{\mathcal{L}}_{\mcd}(\bmg_l)]\right| \ge \frac{\varepsilon}{2}\right] \\				
			\end{aligned}
		\end{equation}
		where, without the loss of generality, we denote the covering number as $\mathcal{N}_3$ for short.
		
		Note that, from Lem.\ref{lem8} we have $\sigma = \frac{\varepsilon}{16r(4 + \eta)}$, and 
		\[
		\mathcal{N}_3(\sigma; \mathcal{H}_R, \rho_3) \le \mathcal{N}_1\left(\frac{\varepsilon}{16r(4 + \eta)}, \mathcal{H}_R,\rho_1\right) \cdot \mathcal{N}_2\left(\frac{\varepsilon}{16r(4 + \eta)}, \mathcal{H}_R,\rho_2\right).
		\]
		Therefore, we further have
		
		\begin{equation}
			\begin{aligned}
				\mathbb{P}\left[\sup_{\bmg \in \mathcal{H}_R} \left| \hat{\mathcal{L}}_{\mcd}(\bmg) - \expe [\hat{\mathcal{L}}_{\mcd}(\bmg)]\right| \ge \varepsilon \right] & \le \mathcal{N}_1 \cdot \mathcal{N}_2 \cdot \mathbb{P}\left[\left|\hat{\mathcal{L}}_{\mcd}(\bmg_l) - \expe [\hat{\mathcal{L}}_{\mcd}(\bmg_l)]\right| \ge \frac{\varepsilon}{2}\right]
			\end{aligned}
		\end{equation}
		
		\textcolor{blue}{\textbf{Step 2}}. Now, according to Lem.\ref{lem6}, we have 
		
		\begin{equation}
			\begin{aligned}
				\mathbb{P}\left[\sup_{\bmg \in \mathcal{H}_R} \left| \hat{\mathcal{L}}_{\mcd}(\bmg) - \expe [\hat{\mathcal{L}}_{\mcd}(\bmg)]\right| \ge \varepsilon \right] & \le 2 \mathcal{N}_1 \cdot \mathcal{N}_2 \cdot \exp \left(\frac{- \varepsilon^2 \tilde{N}}{2}\right).
			\end{aligned}
		\end{equation}
		
		Then with Lem.\ref{covering_lem} and by further choosing 
		\begin{equation} \nonumber
			\begin{aligned}
				\varepsilon = \sqrt{\frac{2d}{\tilde{N}} \log \left(3r \tilde{N}\right)},
			\end{aligned}
		\end{equation}
		we have:
		\begin{equation}
			\begin{aligned}
				\mathbb{P}\left[\sup_{\bmg \in \mathcal{H}_R} \left| \hat{\mathcal{L}}_{\mcd}(\bmg) - \expe [\hat{\mathcal{L}}_{\mcd}(\bmg)]\right|\ge \sqrt{\frac{2d\log \left(3r \tilde{N}\right)}{\tilde{N}}}\right] &\le 2\left(\frac{3rB^2}{2d\log\left(3r \tilde{N}\right)}\right)^d,
			\end{aligned}
		\end{equation}
		where again 
		\[
		\tilde{N} = \left(4r^2\sqrt{\left(\frac{(4 + \eta)^2}{|\mcu|} + \frac{2}{|\mcu|^2} \sum_{u_i \in \mcu} \left(\frac{1}{n_i^+} + \frac{1}{n_i^-}\right)\right)}\right)^{-2},
		\] 
		and 
		\begin{equation}
			\begin{aligned}
				B &= 16r(4 + \eta)
			\end{aligned}
		\end{equation}
		
		Therefore, we can conclude that, with high probability, 
		\begin{equation}
			\begin{aligned}
				\left| \hat{\mathcal{L}}_{\mcd}(\bmg) - \expe [\hat{\mathcal{L}}_{\mcd}(\bmg)]\right| \le \sqrt{\frac{2d\log \left(3r \tilde{N}\right)}{\tilde{N}}}, \ \ \forall \ \bmg, \bmg \in \mathcal{H}_R.
			\end{aligned}
		\end{equation}
		
		This completed the proof.
	\end{proof}
	
	\subsubsection{Proof of Corol.\ref{cor1}}
	\begin{rcor1} \label{rcor1} On the top of Thm.\ref{them1}, DPCML could enjoy a smaller generalization error than CML.
	\end{rcor1}
	\begin{proof} For simplication of notations, let $\mathcal{X}_{=1}$ and $\mathcal{X}_{>1}$ be the feasible regions of CML ($C=1$) and DPCML ($C > 1$), and $\hat{\mathcal{L}}_{=1}(\boldsymbol{g})$ and $\hat{\mathcal{L}}_{>1}(\boldsymbol{g})$ be the empirical risks of CML ($C=1$) and DPCML ($C > 1$), respectively. Then, since DPCML leverages $\min\limits_{c \in C} \|\boldsymbol{g}_{u_i}^c - \boldsymbol{g}_{v_j}\|^2$ as the distance, which can be regarded as a minimum of multiple single version CML, it is easy to know that the feasible solution of CML is also included in DPCML, i.e., $\mathcal{X}_{=1}\subseteq \mathcal{X}_{>1}$. Therefore, we can conclude that $\hat{\mathcal{L}}_{>1}(\boldsymbol{g}) \le \hat{\mathcal{L}}_{=1}(\boldsymbol{g})$. Denote $\Delta = \sqrt{\frac{2d\log \left(3r \tilde{N}\right)}{\tilde{N}}}$ as the residuals between $\mathbb{E}[\hat{\mathcal{L}}(\boldsymbol{g})]$ and $\hat{\mathcal{L}}(\boldsymbol{g})$. Moreover, we have $\Delta_{\text{DPCML}} = \Theta(\Delta_{\text{CML}})$ since $\Delta$ in our bound does not depend on $C$. This is consistent with the over-parameterization phenomenon \cite{DBLP:journals/corr/abs-2109-02355, DBLP:conf/iclr/NakkiranKBYBS20}. According to Thm.\ref{them1}, we see that $\mathbb{E}[\hat{\mathcal{L}_*}(\boldsymbol{g})] \le \hat{\mathcal{L}_*}(\boldsymbol{g}) + \Delta$, where $*$ represents $=1$ or $>1$. Therefore, we can conclude that DPCML could enjoy a smaller generalization error than the traditional CML. We also empirically demonstrate this in the experiment Sec.\ref{qqaa}.
	\end{proof}
	
	\section{Experiments} 
	\label{exp_supp}
	\subsection{Dataset}
	We perform empirical experiments on several public and real-world benchmark datasets, including:
	\begin{itemize}[leftmargin=*]
		\item \textbf{MovieLens}\footnote{\url{https://grouplens.org/datasets/movielens/}} - One of the most popular benchmark datasets with many versions. Specifically, it includes explicit user-item ratings ranging from 1 to 5 and movie types in terms of various movies. We adopt \textbf{MovieLens-1m}\footnote{\url{https://grouplens.org/datasets/movielens/1m/}} and \textbf{MovieLens-10m}\footnote{\url{https://grouplens.org/datasets/movielens/10m/}} here to evaluate the performance. To obtain the implicit preference feedback, if the score of item $v_j$ rated by user $u_i$ is no less than 4, we regard item $v_j$ as a positive item for user $u_i$ following the
		previous and successful research \cite{DBLP:conf/www/HeLZNHC17}. 
		
		\item \textbf{CiteULike}\footnote{\url{http://www.citeulike.org/faq/data.adp}} \cite{DBLP:conf/ijcai/WangCL13} - An implicit feedback dataset that includes the preferences of users toward different articles. There are two configurations of CiteULike collected from CiteULike and Google Scholar. Following \cite{hsieh2017collaborative}, we adopt \textbf{CiteULike-T} here to evaluate the performance.
		\item \textbf{Steam-200k\footnote{\url{https://www.kaggle.com/tamber/steam-video-games}}} - This dataset is collected from the Steam which is the world's most popular PC gaming hub. The observed behaviors of users include 'purchase' and 'play' signals. In order to obtain the implicit feedback, if a user has purchased a game as well as the playing hours $play > 0$, we treat this game as a positive item. 
	\end{itemize}
	The detailed statistics in terms of these datasets are summarized in Tab.\ref{rtable1}.
	\begin{table*}[!h]
		\centering
		\setlength{\abovecaptionskip}{6pt}    
		\setlength{\belowcaptionskip}{15pt}    
		\setlength{\tabcolsep}{9pt}
		\caption{Basic Information of the Datasets. \%Density is defined as $\frac{\#Ratings}{\#Users \times \#Items} \times 100\% $.}	
		\label{rtable1}
		\begin{tabular}{c|ccccc}
			\toprule
			Datasets & MovieLens-1M & Steam-200k & CiteULike-T & MovieLens-10M\\
			\midrule
			Domain   & Movie & Game & Paper & Movie \\
			\#Users   & 6,034  & 3,757 & 5,219 & 69,167\\
			\#Items   & 3,953 & 5,113 & 25,975 & 10,019\\
			\#Ratings   & 575,271  & 115,139 & 125,580 & 5,003,437\\
			\%Density  & 2.4118\% & 0.5994\% & 0.0926\%  & 0.7220\%\\
			\bottomrule
		\end{tabular}
	\end{table*}
	
	\subsection{Evaluation Metrics}
	In some typical recommendation systems, users often care about the top-$N$ items in recommendation lists, so the most relevant items should be ranked first as much as possible. Motivated by this, we evaluate the performance of competitors and our algorithm with the following extensively adopted metrics, including:
	\begin{itemize}
		\item \textbf{Precision} (P@$N$) counts the proportion that the ground-truth items are among the Top-$N$ recommended list.
		\[
		\text{P@}N = \frac{1}{|\mcu|} \sum_{u_i \in \mathcal{U}}\frac{|\mathcal{D}_{u_i}^+ \cap I^{u_i}_N|}{N}
		\]
		where again $\mathcal{D}_{u_i}^+$ is the set of ground-truth items of user $u_i$; $I^{u_i}_N$ is the top-$N$ recommendation list for user $u_i$; and $|\cdot|$ means the size of set.
		\item \textbf{Recall} (R@$N$) is defined as the number of the ground-truth items in top-$N$ recommendation list divided by the amount of totally ground-truth items. This reflects the ability of model to find the relevant items.
		\[
		\text{R@}N = \frac{1}{|\mcu|} \sum_{u_i \in \mathcal{U}} \frac{|\mathcal{D}_{u_i}^+ \cap I^{u_i}_N|}{|\mathcal{D}_{u_i}^+|}
		\]
		\item \textbf{Normalized Discounted Cumulative Gain} (NDCG@$N$) counts the ground-truth items in the top-$N$ recommendation list with a position weighting strategy, i.e., assigning a larger value on top items than bottom ones.
		\[
		\text{NDCG@}N = \frac{1}{|\mcu|}\sum_{u_i \in \mathcal{U}} \frac{\text{DCG}_{u_i}\text{@}N}{\text{IDCG}_{u_i}\text{@}N}
		\]
		Specifically, the $\text{DCG}_{u_i}\text{@}N$ and $\text{IDCG}_{u_i}\text{@}N$ are defined as:
		\[
		\text{DCG}_{u_i}\text{@}N = \sum_{j = 1} ^ {N}\frac{1 \cdot \mathbb{I}(I^{u_i}_{N, j} \in \mathcal{D}_{u_i}^+)}{\log_2(j + 1)},
		\]
		\[
		\text{IDCG}_{u_i}\text{@}N = \sum_{k = 1}^{\min(N,|\mathcal{D}_{u_i}^+|)} \frac{1}{\log_2(k + 1)},
		\]
		where $I^{u_i}_{N, j}$ respresents the $j$-th item in the top-$N$ recommendation list; $\mathbb{I}(\cdot)$ is an indicator function that returns $1$ if the statement is true and returns $0$, otherwise.
		\item \textbf{Mean Average Precision} (MAP) is an extension of Average Precision(AP). AP is the average of precision values at all positions where ground-truth items are found.
		\[
		\text{AP}_{u_i} = \frac{1}{|\mathcal{D}_{u_i}^+|}\sum_{j = 1}^{|\hat{I}_{u_i}|}\frac{|\mathcal{D}_{u_i}^+ \cap \hat{I}_{u_i, 1: j}| \cdot \mathbb{I}(j \in \mathcal{D}_{u_i}^+)}{rank_j^{u_i}}
		\] 
		\[
		\text{MAP} = \frac{1}{|\mcu|} \sum_{u_i \in \mathcal{U}} \text{AP}_{u_i}
		\]
		where different from $I^{u_i}_N$, $\hat{I}_{u_i}$ is the recommendation rankings in terms of all items for user $u_i$; $\hat{I}_{u_i, 1:j}$ represents the top-$j$ recommendation list for user $u_i$; and $rank_j^{u_i}$ means the ranking of item $j$ in $\hat{I}_{u_i}$.
		\item \textbf{Mean Reciprocal Rank} (MRR) takes the rank of each recommended item into account. It is the average of reciprocal ranks of the desired item:
		\[
		\text{MRR} = \frac{1}{|\mcu|} \sum_{u_i\in \mathcal{U}} \sum_{j = 1}^{|\hat{I}_{u_i}|} \frac{1}{rank_j^{u_i}} \cdot \mathbb{I}(\hat{I}_{u_i, j} \in \mathcal{D}_{u_i}^+)
		\]
	\end{itemize}
	
	Note that, for all the above metrics, the higher the metric is, the better the algorithm achieves.
	
	\subsection{Competitors}
	The involved competitors roughly fall into three groups here, including:
	
	\textbf{1) Item-based collaborative filtering algorithm.} 
	\begin{itemize}[leftmargin=*]
		\item \textbf{itemKNN} \cite{linden2003amazon} is designed on the criterion of the k-nearest neighborhood (KNN), which directly considers the similarity (such as cosine similarity) between the candidate and the previously interacted items to make the recommendations.
	\end{itemize}
	%
	
	\textbf{2) MF-based algorithms including the combination of MF and deep learning network and multi-vector MF-based methods.}
	
	\begin{itemize}[leftmargin=*]
		\item \textbf{Generalized Matrix Factorization} (GMF) adopts a linear kernel to capture the preference of users such that it is more expressive than the traditional MF algorithms.
		\item \textbf{Multi-Layer Perceptron} (MLP) leverages a multi-layer perceptron endowed with reasonable flexibility and non-linearity to model the users' preference toward items.
		\item \textbf{Neural network-based Collaborative Filtering} (NeuMF)  \footnote{\url{https://github.com/guoyang9/NCF}} \cite{DBLP:conf/www/HeLZNHC17} is a seminal and competitive deep learning based recommendation framework. Specifically, NCF integrates the GMF and MLP algorithms and makes recommendation via regarding the recommendation task as a regression problem.
		\item \textbf{Multi-vector MF} (M2F) \cite{DBLP:conf/eaamo/GuoKJG21} is a state-of-the-art MF-based recommendation algorithm, which models the diversity preference of users by assigning them multiple embeddings in the dot-product space. This could be regarded as a competitive baseline to figure out the superiority of our proposed algorithm.
		\item \textbf{Multi-vector GMF} (MGMF). Consider that the original algorithm \cite{DBLP:conf/eaamo/GuoKJG21} might be specifically tailored for the explicit feedback rather than the implicit signals, we further apply a multiple set of users' representations to GMF \cite{DBLP:conf/www/HeLZNHC17}.
	\end{itemize}
	\textbf{3) CML-based recommendation competitors.}
	\begin{itemize}[leftmargin=*]
		\item \textbf{Uniform Negative Sampling} (UniS) \cite{DBLP:conf/icdm/PanZCLLSY08} in terms of each user, uniformly sample $S$ items from unobserved interactions as negative instances to optimize the pairwise ranking loss.
		\item \textbf{Popularity-based Negative Sampling} (PopS) \cite{DBLP:conf/sigir/WuVSSR19} samples $S$ negative candidates from unobserved interactions based on their popularity/frequencies. 
		\item  \textbf{Two-Stage Negative Sampling} (2stS) \footnotemark \cite{tran2019improving} adopts a two-stage sampling strategy. 1) A candidate set of items are sampled based on their popularity; 2) according to their inner product values with anchors (positive items), the most informative samples are selected from this candidate.
		\item \textbf{Hard Negative Sampling} (HarS)  \footnote{\url{https://github.com/changun/CollMetric}} is similar to the negative sample mining process broadly used in metric learning \cite{DBLP:journals/pr/GajicAG21,DBLP:conf/ijcai/DingQ00J19}.
		It can be divided into two steps: a) uniformly sample $S$ candidates from unobserved items; b) select the hard (the most distant) item from the candidates as negative in the light of the distance between the aimed user and item candidates. 
		\item \textbf{Collaborative Translational Metric Learning (TransCF)} \cite{DBLP:conf/icdm/ParkKXY18} is a translation-based method. Specifically, such translation-based algorithms employ $\bm{d}(i, j) = ||\boldsymbol{g}_{u_i} + \boldsymbol{g}_{r_{ij}} - \boldsymbol{g}_{v_j}||^2$ as the distance/score between user $u_i$ and item $v_j$ instead of $||\boldsymbol{g}_{u_i} - \boldsymbol{g}_{v_j}||^2$, where $\boldsymbol{g}_{r_{ij}}$ is a specific translation vector for $u_i$ and $v_j$. In light of this, TransCF discovers such user–item translation vectors via the users' relationships with their neighbor items.
		\item \textbf{Latent Relational Metric Learning (LRML)} \cite{DBLP:conf/www/TayTH18} is also a translation-based CML method. As a whole, the key idea of LRML is similar to TransCF. The main difference is how to access the translation vectors effectively. Concretely, TransCF leverages the neighborhood information of users and items to acquire the translation vectors while LRML introduces an attention-based memory-augmented neural architecture to learn the exclusive and optimal translation vectors. 
		\item \textbf{AdaCML} \cite{DBLP:conf/dasfaa/ZhangZLXF0SC19} learns an adaptive user representation via a memory component and an attention mechanism to accurately model the implicit relationships of user-item pairs and users’ interests.
		\item \textbf{HLR} \cite{DBLP:conf/recsys/TranSHM21} is a state-of-the-art CML-based approach that employs the memory-based attention networks to hierarchically capture users' preferences from both latent user-item and item-item relations. 
		\end{itemize}
	
	\subsection{Implementation Details}  \label{app:train}
	We implement our model with PyTorch \footnote{\url{https://pytorch.org/}} \cite{paszke2017automatic} and employ \textit{Adam} \cite{DBLP:journals/corr/KingmaB14} as the optimizer. In terms of all benchmark datasets, each user interactions are divided into training/validation/test sets with a $0.6:0.2:0.2$ split ratio. According to this, to ensure that each user has at least one positive interaction in training/validation/test, users who have less than five interactions are filtered out from these datasets. We adopt grid search for all methods to select the best parameters based on the validation set and report the corresponding performance on the test set. Specifically, for all competitors, the batch size is set to $256$ and the learning rate is searched within $\{0.001, 0.003, 0.01\}$. The number of epochs is set as $100$. The dimension of embedding $d$ is fixed as $100$, and the margin $\lambda$ is searched within $\{1.0, 1.5, 2.0\}$. The number of user representations $C$ is tuned among $\{2, 3, 4, 5\}$. For the regularization term, $\eta$ is searched within $\{10, 20, 30\}$, $\delta_1 \in \{0.,0.05, 0.1, 0.2, 0.5\}$ and $\delta_2 \in \{0.1, 0.25, 0.35, 0.5, 0.8\}$. For our proposed algorithm, following \cite{DBLP:conf/www/HeLZNHC17,DBLP:journals/tmm/LiXJMCH21}, we also adopt the negative sampling strategy to avoid the heavy burden of pairwise learning. Here we employ two negative sampling strategies i.e., uniform \cite{DBLP:conf/icdm/PanZCLLSY08} and hard negative sampling \cite{DBLP:conf/iccv/HenriquesCCB13}, denoted as \textbf{DPCML1} and \textbf{DPCML2} respectively. To make sure a reasonable comparison, we set the sampling constant $S = 10$ for all methods. For the other parameters of baseline models, we follow their tuning strategies in the original papers. Finally, in terms of the top-$N$ recommendation, we evaluate the performance at $N \in \{3, 5\}$, respectively.
	
	\subsection{Overall Performance} \label{suppa.4}
	The experimental results of all the involved competitors are shown in Tab.\ref{tab:addlabel} and Tab.\ref{results2} . Consequently, we can draw the following conclusions: 1) In most cases, the best performance of CML-based methods consistently surpasses the best MF-based competitors. This suggests that it is necessary to develop CML-based RS algorithms. 2) Our proposed method consistently surpasses all the competitors significantly on all datasets, except the results for MAP and MRR on CiteULike. Even for the failure results, the performance is fairly competitive compared with the competitors. This shows the effectiveness of our proposed algorithm. 3) Compared with studies targeting joint accessibility (i.e., M2F and MGMF), our proposed method significantly outperforms M2F and MGMF on all benchmark datasets. This shows the advantage of the CML-based paradigm that deserves more attention along this direction in future work. 
	\begin{table*}[htbp]
		\centering
		\caption{Performance comparisons on CiteULike and MovieLens-10m datasets. The best and second-best are highlighted in bold and underlined, respectively.}
		\label{results2}
		\scalebox{0.8}{
			\begin{tabular}{c|c|c|cccccccc}
				\toprule
				& Type & Method & P@3 & R@3 & NDCG@3 & P@5 & R@5 & NDCG@5 & MAP & MRR \\
				\midrule
				\multirow{14}[8]{*}{CiteULike} & Item-based & itemKNN & 1.20  & 0.83  & 1.23  & 1.15  & 0.77  & \cellcolor[rgb]{ .984,  .992,  .976}1.16  & \cellcolor[rgb]{ .98,  .988,  .996}1.44  & 3.78  \\
				\cmidrule{2-11}      & \multirow{5}[2]{*}{MF-based} & GMF & \cellcolor[rgb]{ .98,  .988,  .996}1.86  & \cellcolor[rgb]{ .992,  .996,  1}0.96  & \cellcolor[rgb]{ .976,  .984,  .996}2.05  & \cellcolor[rgb]{ .976,  .984,  .996}2.15  & \cellcolor[rgb]{ .992,  .996,  1}0.97  & \cellcolor[rgb]{ .957,  .976,  .973}2.40  & \cellcolor[rgb]{ .984,  .992,  .996}1.34  & \cellcolor[rgb]{ .973,  .984,  .996}5.53  \\
				&   & MLP & \cellcolor[rgb]{ .973,  .984,  .996}2.06  & \cellcolor[rgb]{ .984,  .992,  .996}1.08  & \cellcolor[rgb]{ .969,  .98,  .992}2.22  & \cellcolor[rgb]{ .969,  .98,  .992}2.40  & \cellcolor[rgb]{ .98,  .988,  .996}1.16  & \cellcolor[rgb]{ .953,  .973,  .973}2.61  & \cellcolor[rgb]{ .976,  .984,  .996}1.52  & \cellcolor[rgb]{ .867,  .918,  .969}12.37  \\
				&   & NeuMF & \cellcolor[rgb]{ .973,  .984,  .996}2.06  & \cellcolor[rgb]{ .984,  .992,  .996}1.08  & \cellcolor[rgb]{ .969,  .98,  .992}2.21  & \cellcolor[rgb]{ .969,  .984,  .996}2.36  & \cellcolor[rgb]{ .98,  .988,  .996}1.16  & \cellcolor[rgb]{ .953,  .973,  .973}2.57  & \cellcolor[rgb]{ .973,  .984,  .996}1.54  & \cellcolor[rgb]{ .867,  .922,  .969}12.22  \\
				&   & M2F & \cellcolor[rgb]{ .984,  .992,  .996}1.76  & \cellcolor[rgb]{ .996,  1,  1}0.90  & \cellcolor[rgb]{ .976,  .988,  .996}1.97  & \cellcolor[rgb]{ .984,  .992,  .996}1.87  & \cellcolor[rgb]{ .992,  .996,  1}0.93  & \cellcolor[rgb]{ .961,  .98,  .976}2.18  & 0.93  & \cellcolor[rgb]{ .988,  .996,  1}4.53  \\
				&   & MGMF & \cellcolor[rgb]{ .965,  .98,  .992}2.31  & \cellcolor[rgb]{ .973,  .984,  .996}1.23  & \cellcolor[rgb]{ .961,  .976,  .992}2.48  & \cellcolor[rgb]{ .969,  .98,  .992}2.42  & \cellcolor[rgb]{ .984,  .992,  .996}1.12  & \cellcolor[rgb]{ .949,  .973,  .973}2.71  & \cellcolor[rgb]{ .976,  .984,  .996}1.51  & \cellcolor[rgb]{ .965,  .98,  .992}6.18  \\
				\cmidrule{2-11}      & \multirow{8}[2]{*}{CML-based} & UniS & \cellcolor[rgb]{ .788,  .875,  .949}7.34  & \cellcolor[rgb]{ .804,  .882,  .953}3.71  & \cellcolor[rgb]{ .796,  .878,  .953}7.48  & \cellcolor[rgb]{ .78,  .867,  .949}9.54  & \cellcolor[rgb]{ .78,  .867,  .949}5.13  & \cellcolor[rgb]{ .776,  .867,  .945}10.02  & \cellcolor[rgb]{ .792,  .875,  .953}5.59  & \cellcolor[rgb]{ .788,  .871,  .949}17.27  \\
				&   & PopS & \cellcolor[rgb]{ .855,  .914,  .969}5.41  & \cellcolor[rgb]{ .855,  .914,  .969}2.94  & \cellcolor[rgb]{ .851,  .91,  .965}5.77  & \cellcolor[rgb]{ .855,  .914,  .965}6.75  & \cellcolor[rgb]{ .855,  .914,  .969}3.62  & \cellcolor[rgb]{ .843,  .906,  .957}7.23  & \cellcolor[rgb]{ .835,  .902,  .961}4.61  & \cellcolor[rgb]{ .835,  .898,  .961}14.39  \\
				&   & 2stS & \cellcolor[rgb]{ .824,  .894,  .957}6.40  & \cellcolor[rgb]{ .827,  .898,  .961}3.35  & \cellcolor[rgb]{ .82,  .89,  .957}6.77  & \cellcolor[rgb]{ .816,  .886,  .957}8.27  & \cellcolor[rgb]{ .824,  .894,  .957}4.29  & \cellcolor[rgb]{ .804,  .882,  .949}8.81  & \cellcolor[rgb]{ .82,  .89,  .957}4.99  & \cellcolor[rgb]{ .812,  .886,  .957}15.87  \\
				&   & HarS & \cellcolor[rgb]{ .753,  .851,  .941}\underline{8.44}  & \cellcolor[rgb]{ .757,  .851,  .941}\underline{4.41}  & \cellcolor[rgb]{ .753,  .851,  .941}\underline{8.82}  & \cellcolor[rgb]{ .757,  .855,  .941}\underline{10.43}  & \cellcolor[rgb]{ .757,  .851,  .941}\underline{5.60}  & \cellcolor[rgb]{ .749,  .847,  .941}\underline{11.25}  & \cellcolor[rgb]{ .741,  .843,  .937}\textbf{6.67}  & \cellcolor[rgb]{ .741,  .843,  .937}\textbf{20.08}  \\
				&   & TransCF & \cellcolor[rgb]{ .843,  .906,  .965}5.79  & \cellcolor[rgb]{ .851,  .91,  .965}3.03  & \cellcolor[rgb]{ .843,  .906,  .965}6.09  & \cellcolor[rgb]{ .835,  .902,  .961}7.45  & \cellcolor[rgb]{ .839,  .906,  .965}3.93  & \cellcolor[rgb]{ .827,  .898,  .953}7.84  & \cellcolor[rgb]{ .839,  .902,  .961}4.54  & \cellcolor[rgb]{ .831,  .898,  .961}14.50  \\
				&   & LRML & \cellcolor[rgb]{ .957,  .973,  .992}2.52  & \cellcolor[rgb]{ .969,  .98,  .992}1.33  & \cellcolor[rgb]{ .957,  .976,  .992}2.58  & \cellcolor[rgb]{ .953,  .973,  .988}3.06  & \cellcolor[rgb]{ .957,  .976,  .992}1.64  & \cellcolor[rgb]{ .937,  .965,  .973}3.19  & \cellcolor[rgb]{ .957,  .976,  .992}1.91  & \cellcolor[rgb]{ .961,  .976,  .992}6.45  \\
				& &AdaCML & \cellcolor[rgb]{ .8,  .878,  .953}7.04  & \cellcolor[rgb]{ .8,  .878,  .953}3.75  & \cellcolor[rgb]{ .8,  .878,  .953}7.31  & \cellcolor[rgb]{ .804,  .882,  .953}8.70  & \cellcolor[rgb]{ .812,  .886,  .957}4.52  & \cellcolor[rgb]{ .796,  .878,  .949}9.18  & \cellcolor[rgb]{ .792,  .875,  .953}5.57  & \cellcolor[rgb]{ .788,  .871,  .949}17.31  \\
    & & HLR   & \cellcolor[rgb]{ .973,  .984,  .996}2.03  & \cellcolor[rgb]{ .984,  .992,  .996}1.08  & \cellcolor[rgb]{ .969,  .984,  .996}2.20  & \cellcolor[rgb]{ .973,  .984,  .996}2.25  & \cellcolor[rgb]{ .984,  .992,  .996}1.13  & \cellcolor[rgb]{ .953,  .973,  .973}2.52  & \cellcolor[rgb]{ .98,  .988,  .996}1.45  & \cellcolor[rgb]{ .969,  .98,  .992}5.86  \\
				\cmidrule{2-11}      & \multirow{2}[2]{*}{Ours} & DPCML1 & \cellcolor[rgb]{ .776,  .863,  .945}7.78  & \cellcolor[rgb]{ .78,  .867,  .949}4.04  & \cellcolor[rgb]{ .773,  .863,  .945}8.14  & \cellcolor[rgb]{ .769,  .859,  .945}10.03  & \cellcolor[rgb]{ .769,  .863,  .945}5.33  & \cellcolor[rgb]{ .761,  .859,  .941}10.64  & \cellcolor[rgb]{ .769,  .863,  .945}6.08  & \cellcolor[rgb]{ .765,  .859,  .945}18.75  \\
				&   & DPCML2 & \cellcolor[rgb]{ .741,  .843,  .937}\textbf{8.70}  & \cellcolor[rgb]{ .741,  .843,  .937}\textbf{4.59}  & \cellcolor[rgb]{ .741,  .843,  .937}\textbf{9.06}  & \cellcolor[rgb]{ .741,  .843,  .937}\textbf{10.96}  & \cellcolor[rgb]{ .741,  .843,  .937}\textbf{5.85}  & \cellcolor[rgb]{ .741,  .843,  .937}\textbf{11.47}  & \cellcolor[rgb]{ .753,  .851,  .941}\underline{6.44}  & \cellcolor[rgb]{ .745,  .847,  .941}\underline{19.96}  \\
				\midrule
				\multirow{14}[7]{*}{MovieLens-10m} & Item-based & itemKNN & \cellcolor[rgb]{ .914,  .953,  .886}11.44  & \cellcolor[rgb]{ .882,  .937,  .843}3.70  & \cellcolor[rgb]{ .914,  .953,  .886}11.78  & \cellcolor[rgb]{ .914,  .953,  .886}12.27  & \cellcolor[rgb]{ .89,  .941,  .855}4.93  & \cellcolor[rgb]{ .914,  .953,  .886}12.63  & \cellcolor[rgb]{ .89,  .941,  .855}8.25  & \cellcolor[rgb]{ .894,  .941,  .859}25.85  \\
				\cmidrule{2-11}      & \multirow{5}[2]{*}{MF-based} & GMF & \cellcolor[rgb]{ .875,  .933,  .831}13.55  & \cellcolor[rgb]{ .875,  .933,  .831}3.87  & \cellcolor[rgb]{ .871,  .933,  .831}13.91  & \cellcolor[rgb]{ .867,  .929,  .824}14.67  & \cellcolor[rgb]{ .871,  .929,  .827}5.41  & \cellcolor[rgb]{ .871,  .929,  .827}15.13  & \cellcolor[rgb]{ .875,  .933,  .831}9.14  & \cellcolor[rgb]{ .863,  .925,  .816}28.91  \\
				&   & MLP & \cellcolor[rgb]{ .839,  .914,  .788}15.27  & \cellcolor[rgb]{ .816,  .902,  .761}4.93  & \cellcolor[rgb]{ .843,  .914,  .792}15.46  & \cellcolor[rgb]{ .839,  .914,  .788}16.08  & \cellcolor[rgb]{ .824,  .906,  .765}6.53  & \cellcolor[rgb]{ .847,  .918,  .796}16.38  & \cellcolor[rgb]{ .804,  .894,  .741}12.77  & \cellcolor[rgb]{ .827,  .906,  .773}32.21  \\
				&   & NeuMF & \cellcolor[rgb]{ .839,  .914,  .788}15.19  & \cellcolor[rgb]{ .812,  .898,  .753}5.02  & \cellcolor[rgb]{ .847,  .918,  .796}15.27  & \cellcolor[rgb]{ .839,  .914,  .788}16.09  & \cellcolor[rgb]{ .816,  .902,  .757}6.65  & \cellcolor[rgb]{ .847,  .918,  .8}16.24  & \cellcolor[rgb]{ .804,  .894,  .741}12.76  & \cellcolor[rgb]{ .831,  .91,  .776}31.87  \\
				&   & M2F & 7.03  & 1.41  & 7.21  & 7.55  & 2.23  & 7.98  & 2.50  & 15.17  \\
				&   & MGMF & \cellcolor[rgb]{ .851,  .922,  .804}14.62  & \cellcolor[rgb]{ .851,  .922,  .804}4.26  & \cellcolor[rgb]{ .847,  .918,  .8}15.15  & \cellcolor[rgb]{ .851,  .922,  .804}15.53  & \cellcolor[rgb]{ .847,  .918,  .796}5.96  & \cellcolor[rgb]{ .847,  .918,  .8}16.26  & \cellcolor[rgb]{ .851,  .918,  .804}10.30  & \cellcolor[rgb]{ .839,  .914,  .788}31.07  \\
				\cmidrule{2-11}      & \multirow{8}[2]{*}{CML-based} & UniS & \cellcolor[rgb]{ .941,  .969,  .922}10.15  & \cellcolor[rgb]{ .925,  .961,  .902}2.84  & \cellcolor[rgb]{ .941,  .969,  .922}10.33  & \cellcolor[rgb]{ .933,  .965,  .91}11.19  & \cellcolor[rgb]{ .925,  .961,  .902}4.08  & \cellcolor[rgb]{ .937,  .969,  .918}11.38  & \cellcolor[rgb]{ .878,  .933,  .839}8.92  & \cellcolor[rgb]{ .91,  .953,  .878}24.24  \\
				&   & PopS & \cellcolor[rgb]{ .973,  .984,  .961}8.61  & \cellcolor[rgb]{ .914,  .953,  .886}3.06  & \cellcolor[rgb]{ .969,  .984,  .957}8.96  & \cellcolor[rgb]{ .988,  .992,  .98}8.34  & \cellcolor[rgb]{ .937,  .969,  .918}3.76  & \cellcolor[rgb]{ .984,  .992,  .98}8.84  & \cellcolor[rgb]{ .933,  .965,  .91}6.08  & \cellcolor[rgb]{ .941,  .969,  .925}20.97  \\
				&   & 2stS & \cellcolor[rgb]{ .808,  .898,  .749}16.47  & \cellcolor[rgb]{ .816,  .902,  .757}4.89  & \cellcolor[rgb]{ .812,  .898,  .749}16.72  & \cellcolor[rgb]{ .804,  .894,  .741}17.62  & \cellcolor[rgb]{ .804,  .894,  .741}6.87  & \cellcolor[rgb]{ .804,  .894,  .741}18.06  & \cellcolor[rgb]{ .796,  .89,  .733}12.89  & \cellcolor[rgb]{ .804,  .894,  .745}33.75  \\
				&   & HarS & \cellcolor[rgb]{ .804,  .894,  .741}\underline{17.00}  & \cellcolor[rgb]{ .816,  .902,  .757}4.97  & \cellcolor[rgb]{ .808,  .898,  .749}\underline{17.16}  & \cellcolor[rgb]{ .796,  .89,  .733}\underline{18.34}  & \cellcolor[rgb]{ .804,  .894,  .741}6.96  & \cellcolor[rgb]{ .804,  .894,  .737}\underline{18.70}  & \cellcolor[rgb]{ .796,  .89,  .729}13.14  & \cellcolor[rgb]{ .808,  .894,  .745}\underline{34.20}  \\
				&   & TransCF & \cellcolor[rgb]{ .922,  .961,  .898}11.00  & \cellcolor[rgb]{ .882,  .937,  .843}3.70  & \cellcolor[rgb]{ .929,  .965,  .906}10.91  & \cellcolor[rgb]{ .925,  .961,  .902}11.62  & \cellcolor[rgb]{ .89,  .941,  .851}4.94  & \cellcolor[rgb]{ .933,  .965,  .914}11.61  & \cellcolor[rgb]{ .894,  .945,  .863}7.99  & \cellcolor[rgb]{ .914,  .953,  .886}23.67  \\
				&   & LRML & \cellcolor[rgb]{ .871,  .929,  .827}13.72  & \cellcolor[rgb]{ .867,  .929,  .827}3.96  & \cellcolor[rgb]{ .871,  .929,  .831}13.98  & \cellcolor[rgb]{ .871,  .929,  .827}14.53  & \cellcolor[rgb]{ .863,  .925,  .816}5.58  & \cellcolor[rgb]{ .871,  .929,  .827}15.08  & \cellcolor[rgb]{ .875,  .933,  .835}8.99  & \cellcolor[rgb]{ .863,  .925,  .82}28.77  \\
			& &	AdaCML & \cellcolor[rgb]{ .867,  .929,  .824}13.65  & \cellcolor[rgb]{ .859,  .925,  .816}4.00   & \cellcolor[rgb]{ .871,  .929,  .827}13.82  & \cellcolor[rgb]{ .863,  .925,  .82}14.64  & \cellcolor[rgb]{ .859,  .925,  .812}5.52  & \cellcolor[rgb]{ .867,  .929,  .824}14.98  & \cellcolor[rgb]{ .835,  .91,  .78}11.13  & \cellcolor[rgb]{ .851,  .918,  .804}29.58  \\
    & & HLR   & \cellcolor[rgb]{ .835,  .914,  .784}15.13  & \cellcolor[rgb]{ .796,  .89,  .733}\underline{5.12}  & \cellcolor[rgb]{ .847,  .918,  .8}14.94  & \cellcolor[rgb]{ .827,  .906,  .773}16.40  & \cellcolor[rgb]{ .796,  .89,  .729}\underline{7.00}  & \cellcolor[rgb]{ .843,  .914,  .792}16.23  & \cellcolor[rgb]{ .792,  .886,  .722}\underline{13.40}  & \cellcolor[rgb]{ .827,  .906,  .773}31.66  \\
				\cmidrule{2-11}      & \multirow{2}[1]{*}{Ours} & DPCML1 & \cellcolor[rgb]{ .894,  .945,  .863}12.73  & \cellcolor[rgb]{ .886,  .941,  .851}3.82  & \cellcolor[rgb]{ .902,  .949,  .871}13.05  & \cellcolor[rgb]{ .89,  .941,  .855}13.12  & \cellcolor[rgb]{ .882,  .937,  .843}5.07  & \cellcolor[rgb]{ .898,  .945,  .863}13.72  & \cellcolor[rgb]{ .855,  .922,  .808}10.32  & \cellcolor[rgb]{ .875,  .933,  .835}28.65  \\
				&   & DPCML2 & \cellcolor[rgb]{ .776,  .878,  .706}\textbf{18.00} & \cellcolor[rgb]{ .776,  .878,  .706}\textbf{5.46} & \cellcolor[rgb]{ .776,  .878,  .706}\textbf{18.37} & \cellcolor[rgb]{ .776,  .878,  .706}\textbf{18.97} & \cellcolor[rgb]{ .776,  .878,  .706}\textbf{7.37} & \cellcolor[rgb]{ .776,  .878,  .706}\textbf{19.57} & \cellcolor[rgb]{ .776,  .878,  .706}\textbf{14.01} & \cellcolor[rgb]{ .776,  .878,  .706}\textbf{36.44} \\
				\bottomrule
			\end{tabular}%
		}
	\end{table*}%
		
	\subsection{Quantitative Analysis} \label{ext_qa}
	\subsubsection{Fine-grained Performance Comparison}
	At first, we expect to examine the performance improvements of our proposed method with respect to minority preferences. To do this, we show the fine-grained MAP metric over each interest group (movie genre) on MovieLens-10m. The empirical results are reported in Fig.\ref{per_arrtribute_performance}. Here the $x$-axis is organized with a descending order according to the item category distribution presented in Fig.\ref{ml-10m_freq}. We can observe that our proposed framework could not only significantly outperform their single-vector counterparts in the majority interests but also improve the performance of minority groups in most cases. Especially, compared with HarS, the performance improvement of DPCML2 on minority interests is sharp. This shows that DPCML could reasonably focus on potentially interesting items even with the imbalanced item distribution to alleviate the preference bias induced by the conventional CML-based studies. 
	\subsubsection{Recommendation Diversity Evaluation}
	Furthermore, to show the improvement of DPCML in promoting diversity, we evaluate the performance of DPCML against CML-based competitors with a metric called \textit{max-sum diversification (MaxDiv)} \cite{10.1145/2213556.2213580}, i.e.,
	\[
	\textit{MaxDiv}@N = \frac{1}{|\mathcal{U}|}\sum_{u_i \in \mathcal{U}} \sum_{\substack{v_i, v_j \in \mathcal{I}^N_{u_i}, \\ v_i \neq v_j}} s(v_i, v_j),
	\]
	where $s(v_i, v_j)=\|\boldsymbol{g}_{v_i} - \boldsymbol{g}_{v_j}\|^2$ is the square of Euclidean distance between item $v_i$ and $v_j$; $\mathcal{I}^N_{u_i}$ is the top-$N$ recommendation items for user $u_i$.
	
	Generally speaking, MaxDiv@$N$ measures the recommendation diversification by considering item-side similarity, where a high value implies that the recommendation results are relatively diverse. Then, we compare DPCML with the following competitors for a fair evaluation: a) \textbf{UniS} b) \textbf{HarS} c) \textbf{DPCML1 without (w/o) DCRS} and d) \textbf{DPCML2 without (w/o) DCRS}. The experiments are conducted on the Steam-200k and MovieLens-1m datasets with $N \in \{3, 5, 10, 20\}$. 

	The diversity results are shown in Tab.\ref{tab:diversity}. We observe that: a) For methods within the same negative sampling strategy (i.e., UniS and DPCML1, HarS and DPCML2), our proposed DPCML could achieve relatively higher max-sum values. This suggests the improvement of DPCML in terms of promoting recommendation diversity. b) In most cases (except for DPCML1 w/o DCRS on the MovieLens-1m dataset), DPCML outperforms other competitors even without regularization. c) Most importantly, equipped with the regularization term DCRS, DPCML could achieve better diversification results against w/o DCRS. This once again shows the rationality/importance of DCRS. 

	\subsubsection{Effect of the Diversity Control Regularization}
\begin{figure*}[]
		\centering
		\subfigure[DPCML1 (P@$3$)]{
			\includegraphics[width=0.23\columnwidth]{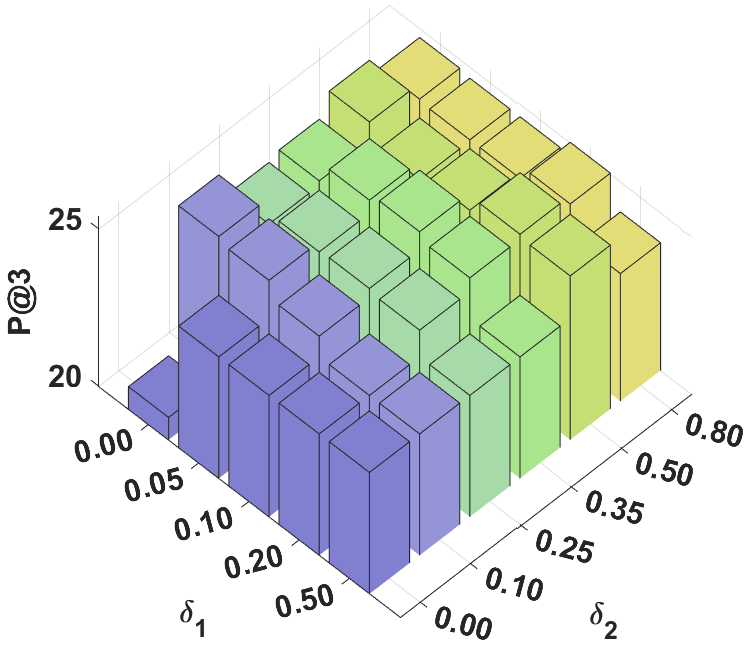}
		}
		\subfigure[DPCML1 (R@$3$)]{
			\includegraphics[width=0.23\columnwidth]{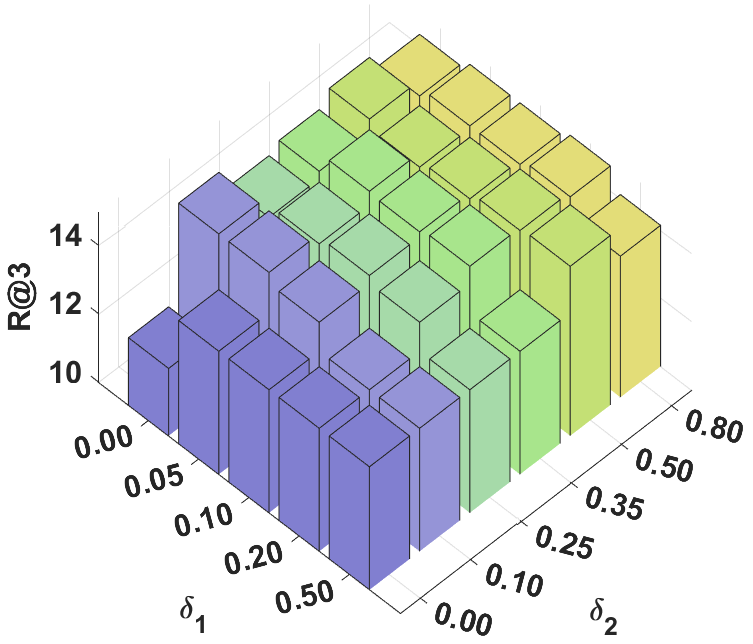}
		}
		\subfigure[DPCML1 (NDCG@$3$)]{
			\includegraphics[width=0.23\columnwidth]{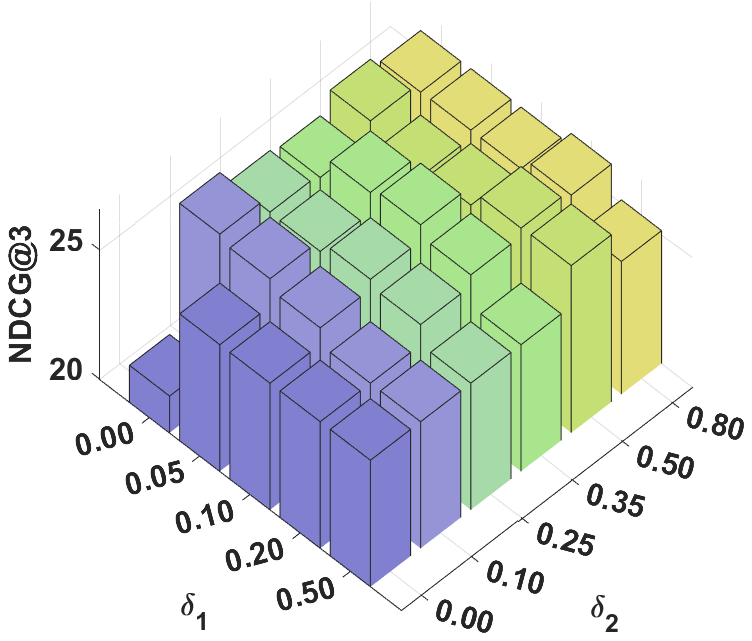}
		}
		\subfigure[DPCML1 (P@$5$)]{
			\includegraphics[width=0.23\columnwidth]{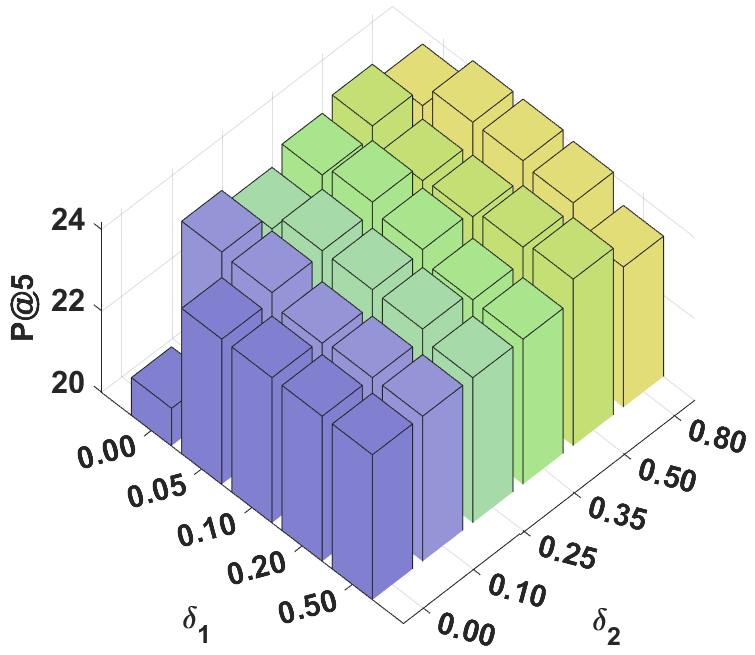}
		}
		\subfigure[DPCML1 (R@$5$)]{
			\includegraphics[width=0.23\columnwidth]{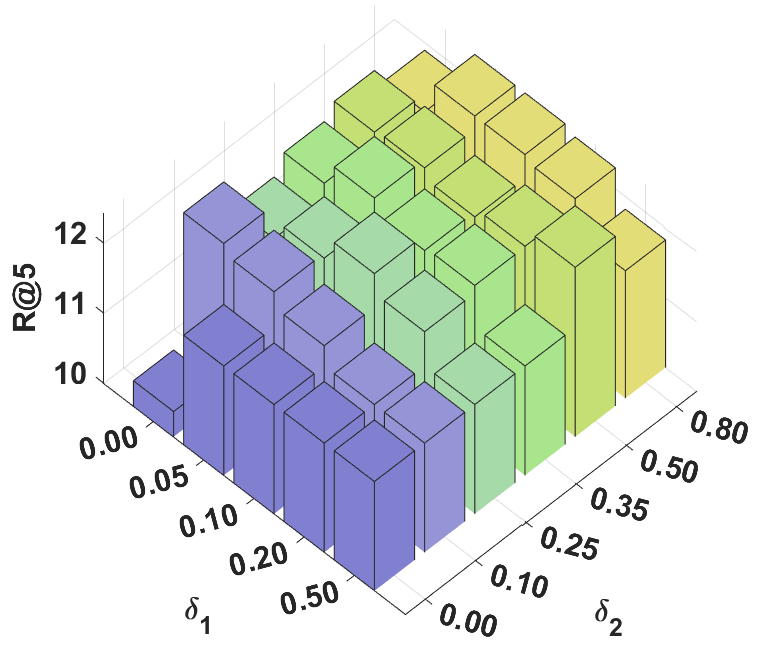}
		}
		\subfigure[DPCML1 (NDCG@$5$)]{
			\includegraphics[width=0.23\columnwidth]{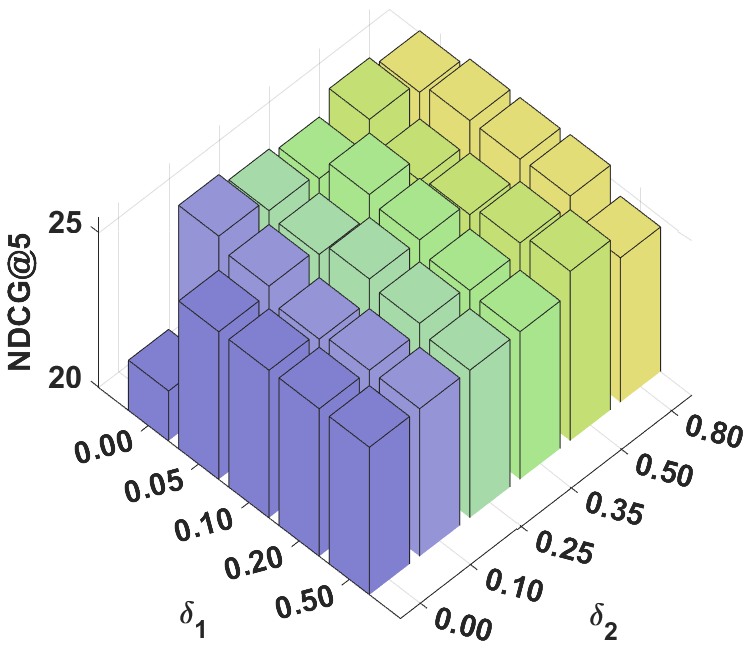}
		}
		\subfigure[DPCML1 (MAP)]{
			\includegraphics[width=0.23\columnwidth]{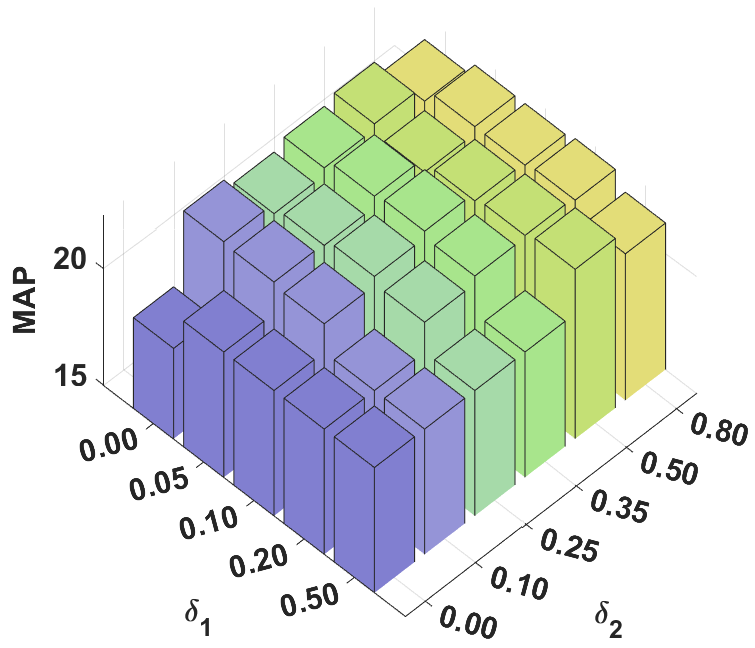}
		}
		\subfigure[DPCML1 (MRR)]{
			\includegraphics[width=0.23\columnwidth]{img/sensitivity/DPCML1_MRR.png}
		}
		\caption{Sensitivity against $\delta_1$ and $\delta_2$ for DPCML1 on Steam-200k datasets. The $x$- and $y$-axis stand for the value of $\delta_1$ and $\delta_2$ respectively, and the $z$-axis shows the performance.}
		\label{sensitivity_supp1}
	\end{figure*}

\begin{figure*}[]
		\centering
		
		\subfigure[DPCML2 (P@$3$)]{
			\includegraphics[width=0.23\columnwidth]{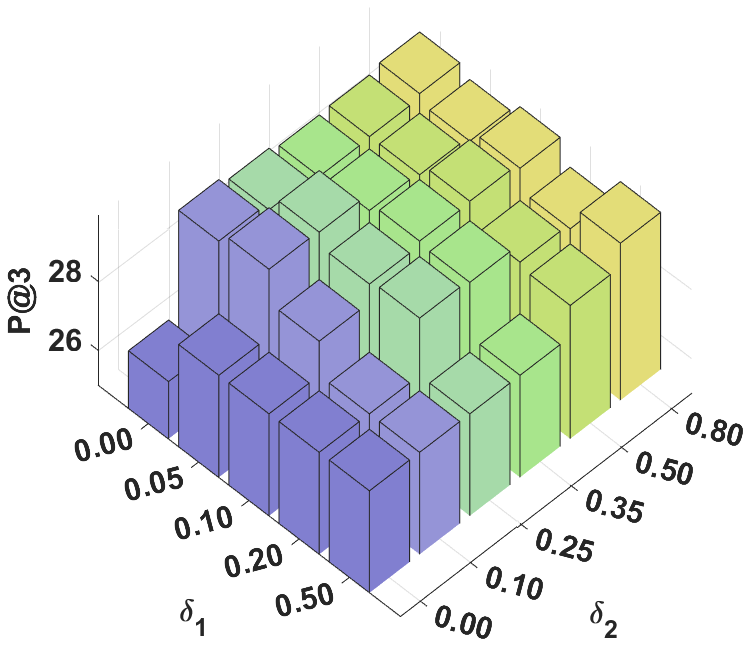}
		}
		\subfigure[DPCML2 (R@$3$)]{
			\includegraphics[width=0.23\columnwidth]{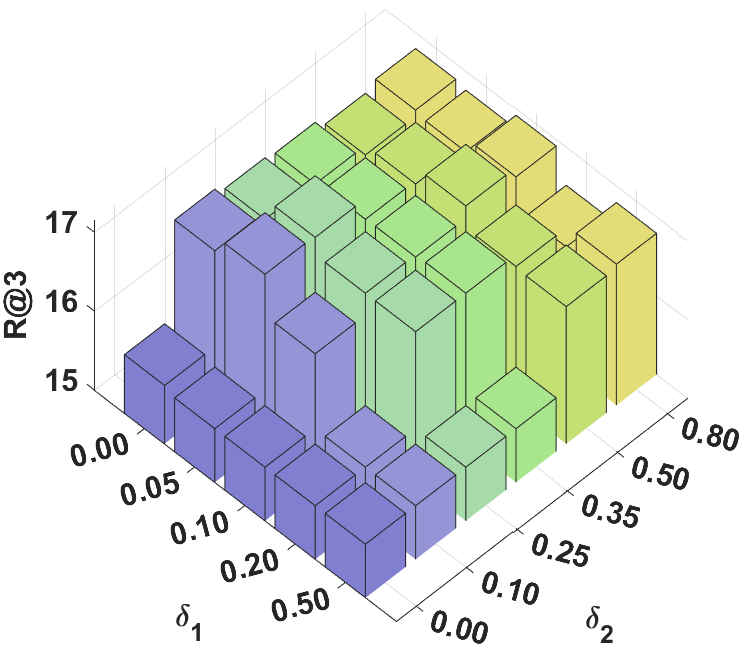}
		}
		\subfigure[DPCML2 (NDCG@$3$)]{
			\includegraphics[width=0.23\columnwidth]{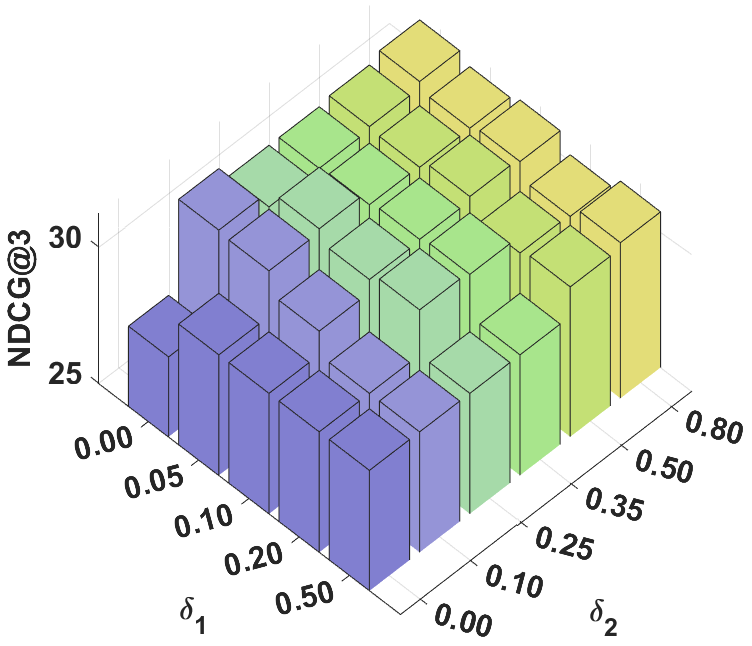}
		}
		\subfigure[DPCML2 (P@$5$)]{
			\includegraphics[width=0.23\columnwidth]{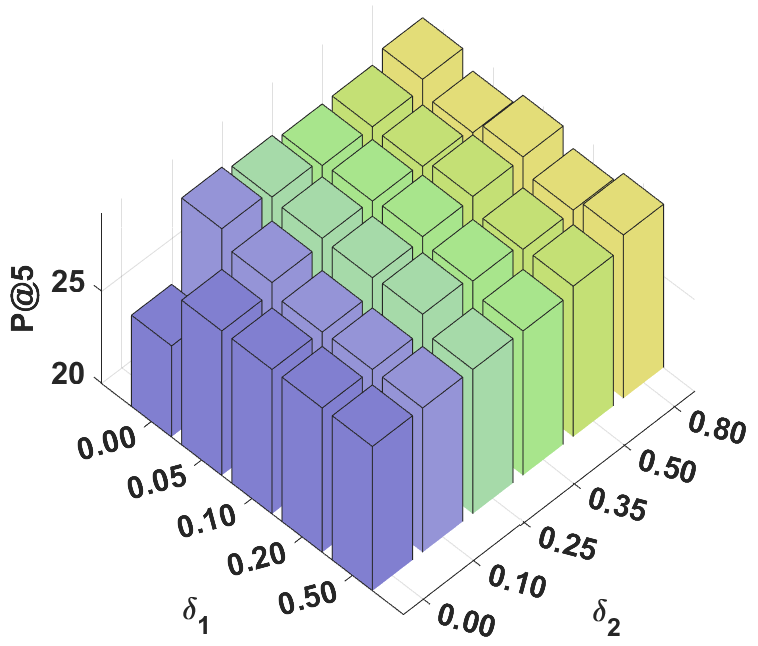}
		}
		\subfigure[DPCML2 (R@$5$)]{
			\includegraphics[width=0.23\columnwidth]{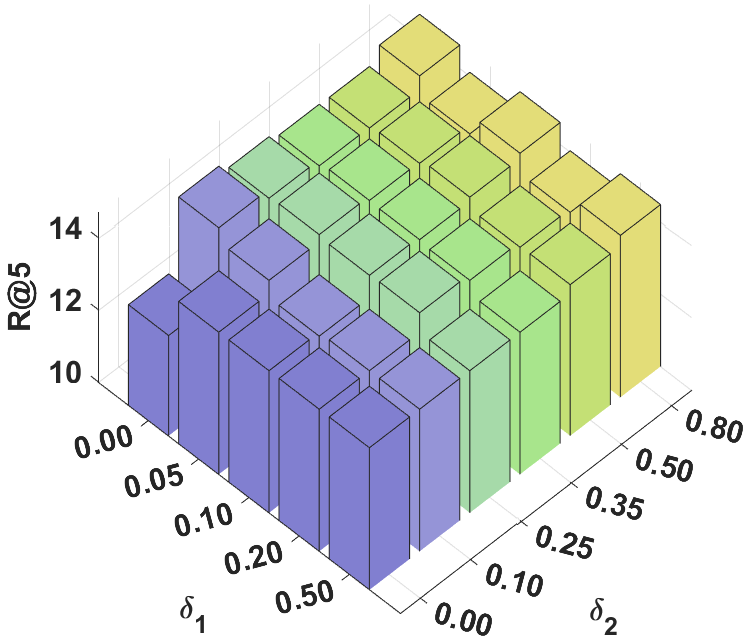}
		}
		\subfigure[DPCML2 (NDCG@$5$)]{
			\includegraphics[width=0.23\columnwidth]{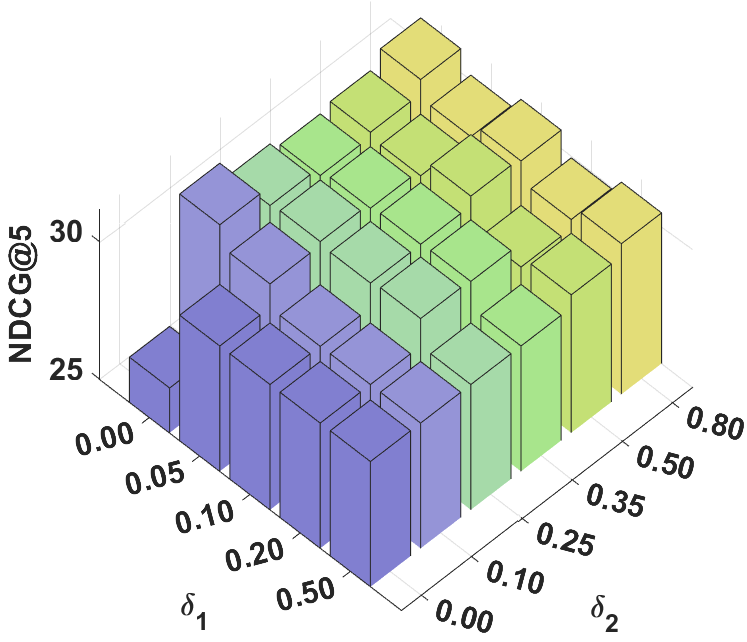}
		}
		\subfigure[DPCML2 (MAP)]{
			\includegraphics[width=0.23\columnwidth]{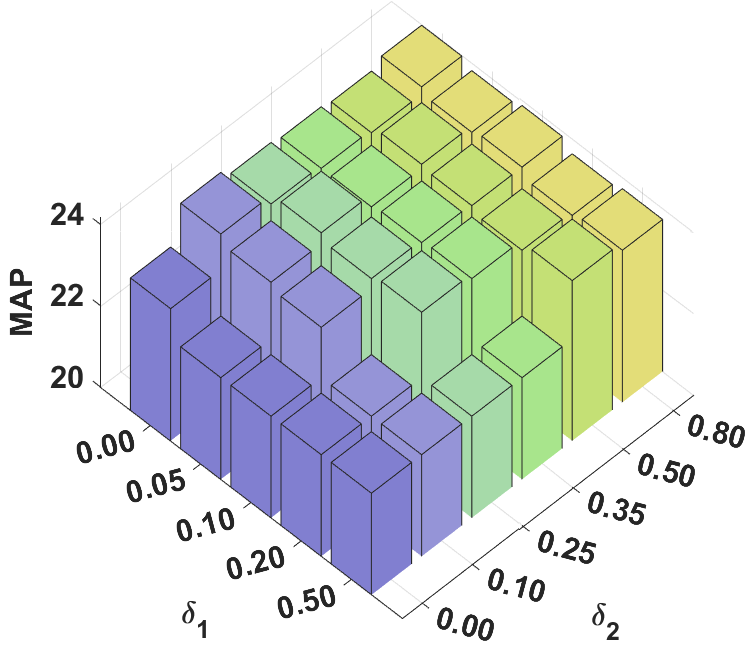}
		}
		\subfigure[DPCML2 (MRR)]{
			\includegraphics[width=0.23\columnwidth]{img/sensitivity/DPCML2_MRR.png}
		}
		\caption{Sensitivity against $\delta_1$ and $\delta_2$ for DPCML2 on Steam-200k datasets. The $x$- and $y$-axis stand for the value of $\delta_1$ and $\delta_2$ respectively, and the $z$-axis shows the performance.}
		\label{sensitivity_supp2}
	\end{figure*}

	Next, we study the effectiveness of our proposed diversity control regularization scheme. To do this, we analyze the influence of two main hyper-parameters, $\delta_1$ and $\delta_2$. We illustrate a $3$D-barplot based on the results of grid search on Steam-200k. The results are presented in Fig.\ref{sensitivity}, Fig.\ref{sensitivity_supp1} and Fig.\ref{sensitivity_supp2}. For a clear comparison, $\delta_1 = \delta_2 = 0$ represents the standard single-vector counterparts performances and $\delta_1 > \delta_2$ indicates the results of DPCML removing the diversity control regularization scheme. Moreover, we set the trade-off coefficient $\eta=10$ and the representation number $C=5$ here. From these results, we can observe that the proposed regularization scheme could significantly boost performance on all metrics, which demonstrates the effectiveness of the diversity control regularization term. In addition, one can see that there would induce different performances with different diversity values. This suggests that controlling a proper diversity of the embeddings for the same user is essential to accommodate their preferences better.
	
	\begin{figure}[!t]
	    \centering
		\subfigure[DPCML1]{
		\includegraphics[width=0.46\columnwidth]{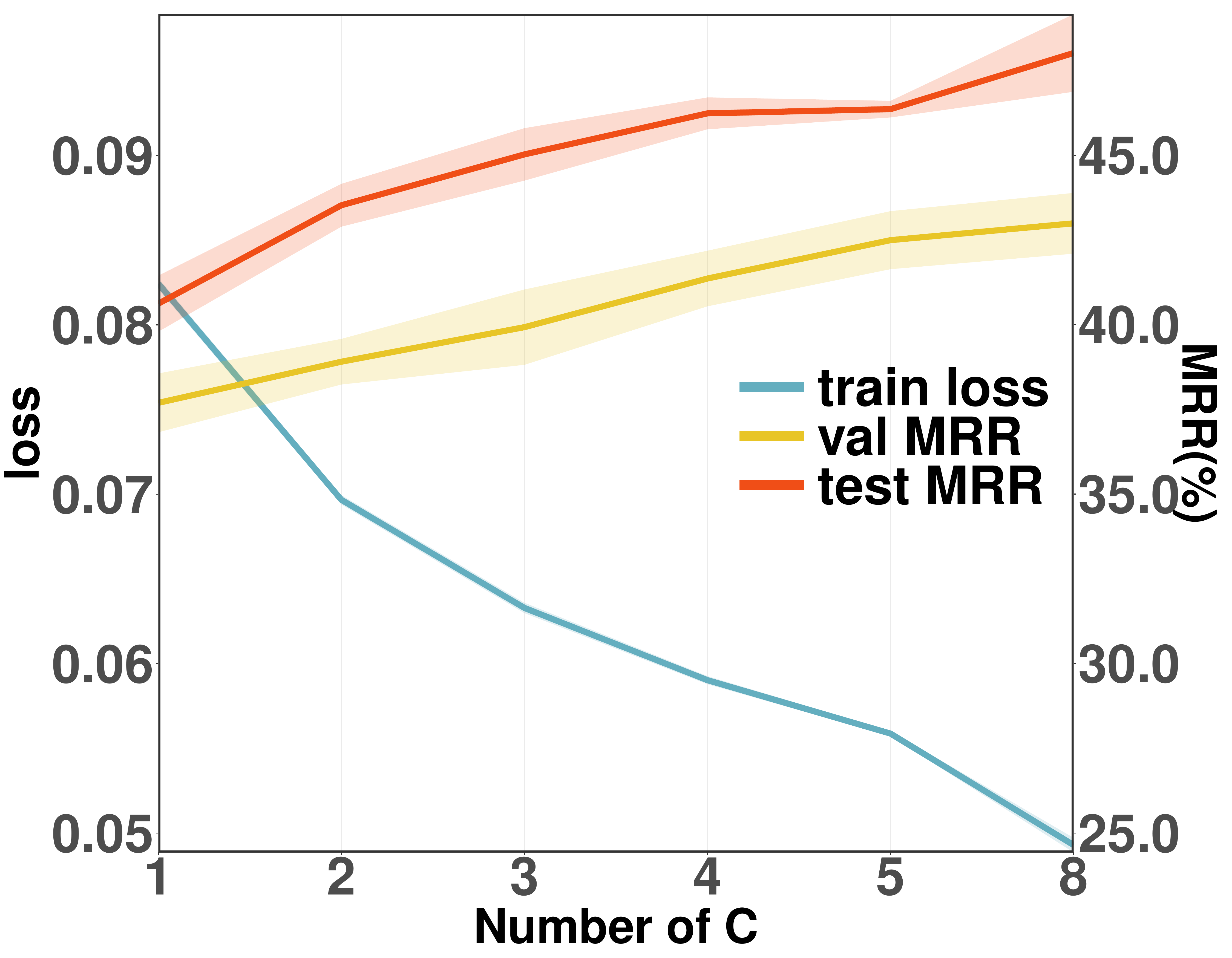}
		\label{supp:DPCML1}
	}
	\subfigure[DPCML2]{
		\includegraphics[width=0.46\columnwidth]{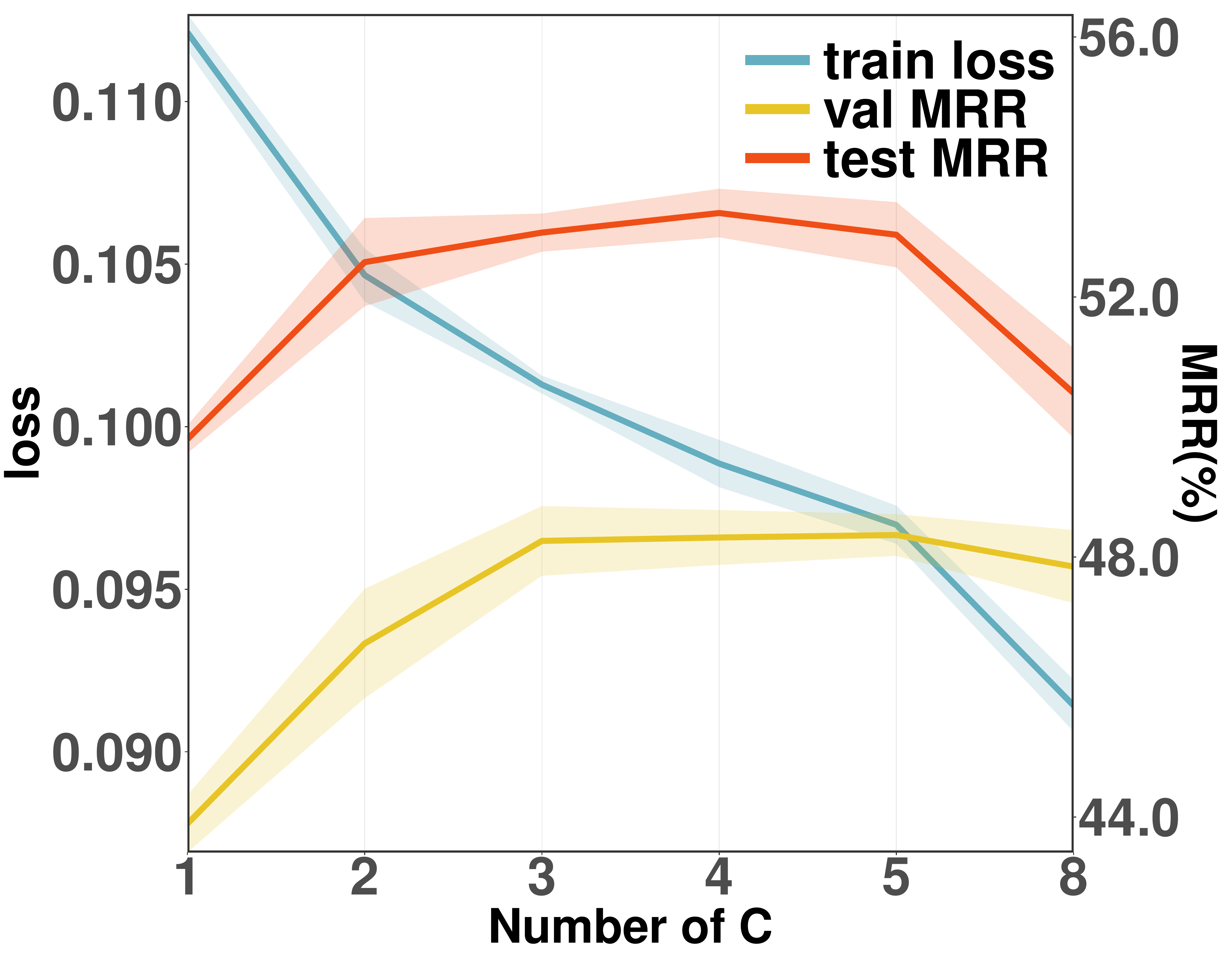}
		\label{supp:DPCML2}
	}
	\caption{Empirical justification of Thm.\ref{them1}.}
	\label{supp:just_thm1}
	\end{figure}
	
	\subsubsection{Empirical Justification of Corol.\ref{cor1} \label{exp:cor1}} 
	We conduct the empirical studies of Corol.\ref{cor1} on the Steam-200k dataset. Expressly, we set $C \in \{1, 2, 3, 4, 5, 8\}$ and record the results of training loss and validation/test MRR metric. Moreover, the experiments are repeated $5$ times. The empirical results are shown in Fig.\ref{supp:just_thm1}, where the shades represent the variance among $5$ experiments. Based on these results, we can see that, with the increase of $C$, the empirical risk (i.e., training loss) of DPCML ($C>1$) is significantly smaller than CML ($C=1$). In addition, DPCML could substantially improve the performance of the validation/test set. Thus, we can conclude that DPCML could induce a smaller generalization error than traditional CML. Overall, this experiment empirically suggests the correctness of Corol.\ref{cor1}.
    
    \begin{figure}
        \centering
		\subfigure[P@3]{
		\includegraphics[width=0.46\columnwidth]{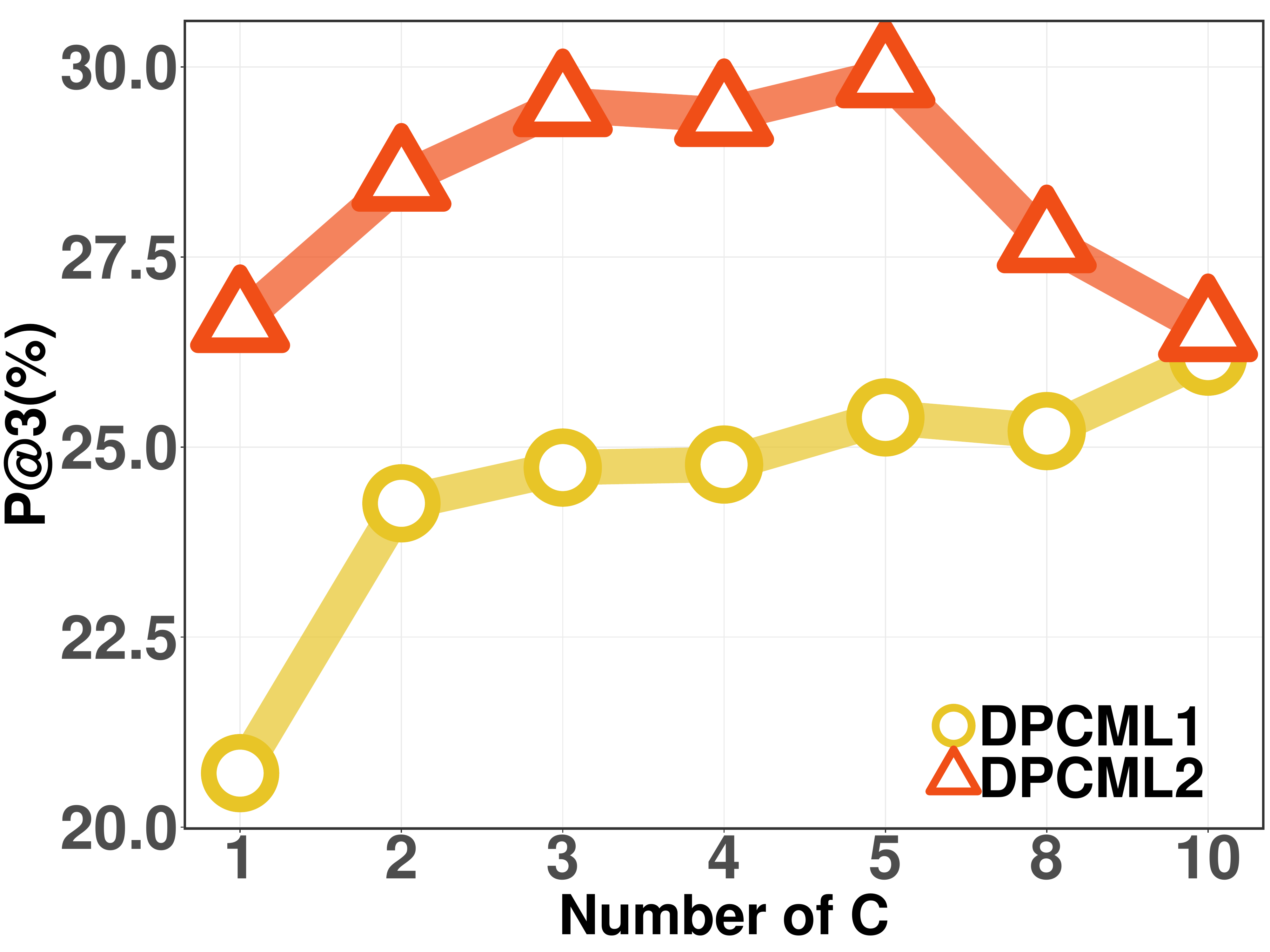}
		\label{supp:effct_C:P@3}
	}
	\subfigure[P@5]{
		\includegraphics[width=0.46\columnwidth]{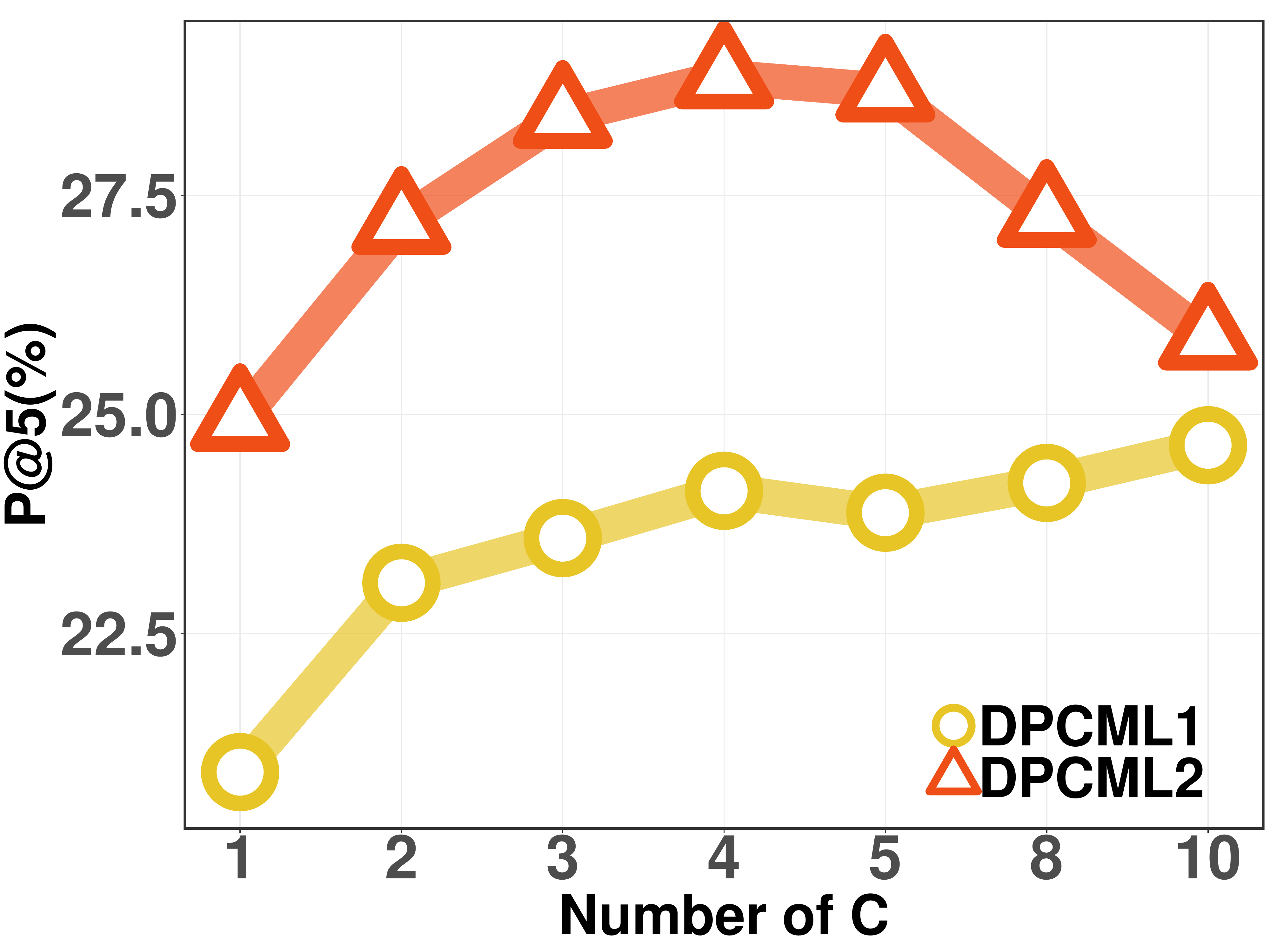}
		\label{supp:effct_C:P@5}
	}
	\caption{Sensitive Analysis of different $C$.}
	\label{supp:effect_of_C}
    \end{figure}
    
 	\subsubsection{Sensitive Analysis of $C$} 
 	Fig.\ref{supp:effect_of_C} demonstrates the P@$3$ and P@$5$ performance of DPCML methods with different $C$ on Steam-200k dataset. We observe that a proper $C$ could significantly improve the performance. Besides, leveraging C too aggressively for DPCML2 may adversely hurt the performance since models optimized with hard samples are more likely to lead to the over-fitting problem with the increasing parameters.
	
\begin{table}[htbp]
  \centering
  \caption{Sensitivity analysis for DPCML1 ($C=5$) on the Steam-200k dataset. }
    \begin{tabular}{c|cccccccc}
    \toprule
    $\eta$ & P@3 & R@3 & NDCG@3 & P@5 & R@5 & NDCG@5 & MAP & MRR \\
    \midrule
    1 & 25.04 & 14.65 & 26.01 & 24.60 & 12.55 & 25.81 & 21.65 & 45.55 \\
    3 & 24.67 & 14.43 & 25.50 & 23.88 & 12.25 & 24.96 & 21.56 & 44.73 \\
    5 & 25.24 & 14.91 & 26.65 & 23.80 & 12.17 & 25.34 & 22.17 & 47.23 \\
    10 & 25.39 & 14.84 & 26.56 & 23.88 & 12.11 & 25.25 & 22.26 & 46.79 \\
    20 & 24.60 & 14.34 & 25.79 & 24.03 & 12.05 & 25.17 & 21.87 & 46.20 \\
    30 & 25.23 & 14.69 & 26.19 & 24.25 & 12.08 & 25.58 & 21.94 & 46.00 \\
    \bottomrule
    \end{tabular}%
  \label{tab:sen_dpcml1}%
\end{table}%

\begin{table}[htbp]
  \centering
  \caption{Sensitivity analysis for DPCML2 ($C=5$) on the Steam-200k dataset.}
    \begin{tabular}{c|cccccccc}
    \toprule
    $\eta$ & P@3 & R@3 & NDCG@3 & P@5 & R@5 & NDCG@5 & MAP & MRR \\
    \midrule
    1 & 28.55 & 16.35 & 29.92 & 27.82 & 13.94 & 29.65 & 22.90 & 50.57 \\
    3 & 28.68 & 16.32 & 29.96 & 27.71 & 13.90 & 29.59 & 23.13 & 50.19 \\
    5 & 29.34 & 16.82 & 30.45 & 27.98 & 13.95 & 29.75 & 23.42 & 50.62 \\
    10 & 29.88 & 17.13 & 31.22 & 28.70 & 14.51 & 30.56 & 24.10 & 51.95 \\
    20 & 29.81 & 17.12 & 31.08 & 29.11 & 14.65 & 30.77 & 24.35 & 51.90 \\
    30 & 29.43 & 16.99 & 30.67 & 28.96 & 14.53 & 30.56 & 24.50 & 51.36 \\
    \bottomrule
    \end{tabular}%
  \label{tab:sen_dpcml2}%
\end{table}%

	\subsubsection{Sensitivity analysis of $\eta$}
	We investigate the sensitivity of $\eta \in \{0, 1, 3, 5, 10, 20, 30\}$ for recommendation results on the Steam-200k dataset. The experimental results are listed in Tab.\ref{tab:sen_dpcml1} and Tab.\ref{tab:sen_dpcml2} for DPCML1 and DPCML2, respectively. We can conclude that a proper $\eta$ (roughly $10$) could significantly improve the performance, suggesting the essential role of the proposed diversity control regularization scheme.
	
	\subsubsection{Training Efficiency}
	Since DPCML includes multiple user representations, it will inevitably introduce extra complexity to the overall optimization. We further investigate the training overheads of our proposed algorithm. Fig.\ref{runtime} shows the efficiency performance. Specifically, every method is conducted 10 epochs, and the average running time across 10 epochs is reported at the bottom of the boxplot. This trend suggests that our proposed algorithm could achieve competitive performance with acceptable efficiency.
	
	\subsubsection{Ablation Studies of Diversity Control Regularization Scheme (DCRS) \label{ab_stu}}
	In order to show the effectiveness of our proposed DCRS, we compare its performance with the following three variants of DCRS:
	\begin{itemize}[leftmargin=*]
	    \item $\textbf{w/o DCRS}$: This is a variant of our method where no regularization is adopted at all. 
	    \item $\textbf{DCRS}-\delta_1$: This is a variant of our method where the punishment on a \textbf{large} diversity is \textbf{removed}. In other words, we will use the following regularization term:
$$
\psi_{\boldsymbol{g}}(u_i) = \max(0, \delta_1 - \delta_{\boldsymbol{g}, u_i}).
$$ 
\item $\textbf{DCRS}-\delta_2$: This is a variant of our method where the punishment on a \textbf{small} diversity is \textbf{removed}. In other words, we will use the following regularization term:
$$
\psi_{\boldsymbol{g}}(u_i) = \max(0, \delta_{\boldsymbol{g}, u_i} - \delta_2).
$$
	\end{itemize}
	The empircal results on Steam-200k dataset are provided in Tab.\ref{tab:DPCML1} and Tab.\ref{tab:DPCML2}. From the above results, we can see that: In most cases, only employing one of the two terms of DCRS could still improve the recommendation performance. However, none of them could outperform our proposed method. This strengthens the effectiveness of our proposed regularization scheme.
	
	\begin{figure*}[!t]
		\centering
		
		\subfigure[MovieLens-1m]{
			\includegraphics[width=0.45\textwidth]{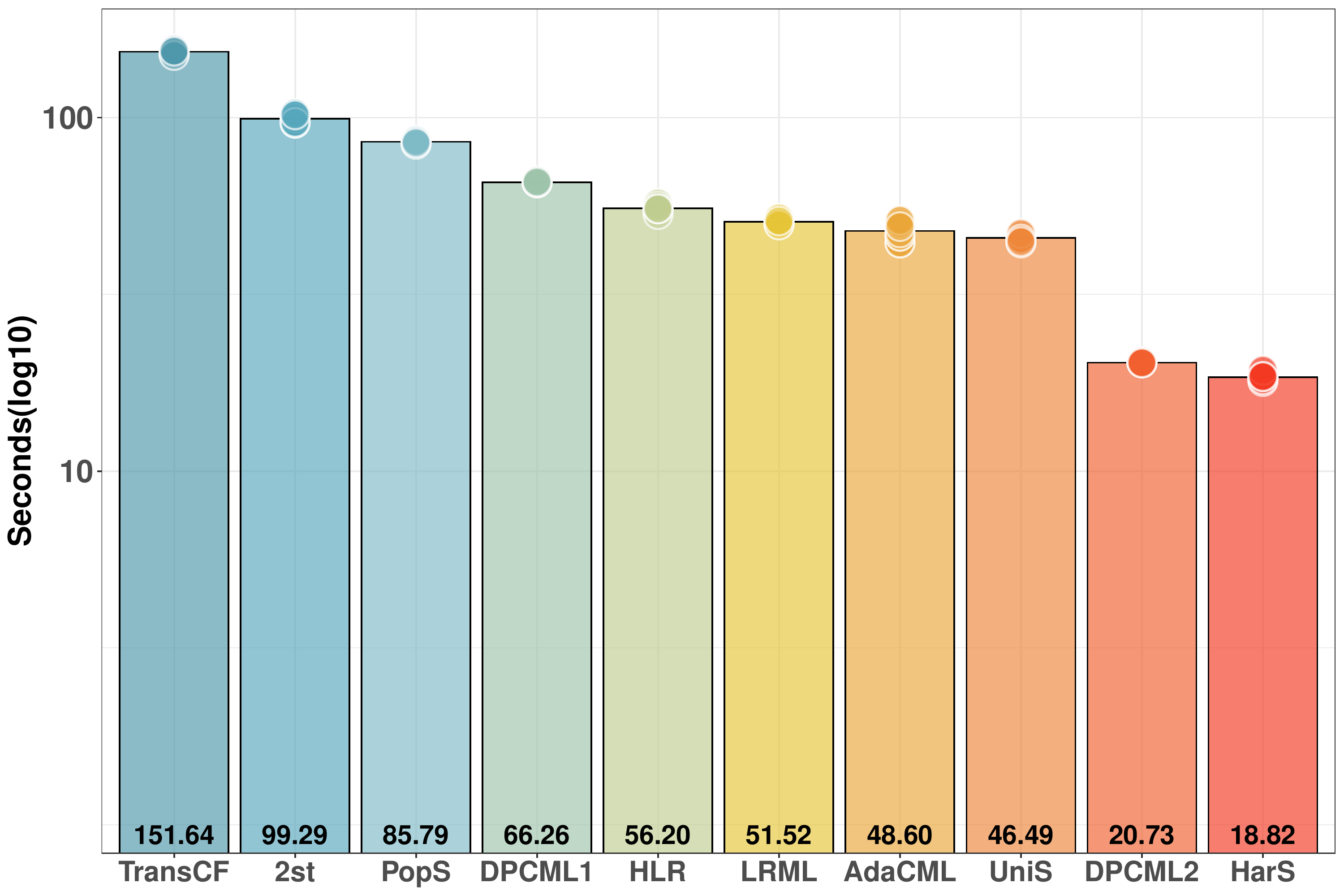}
			\label{ab.sub.ml-1m}
		}
		\subfigure[Steam-200k]{
			\includegraphics[width=0.45\textwidth]{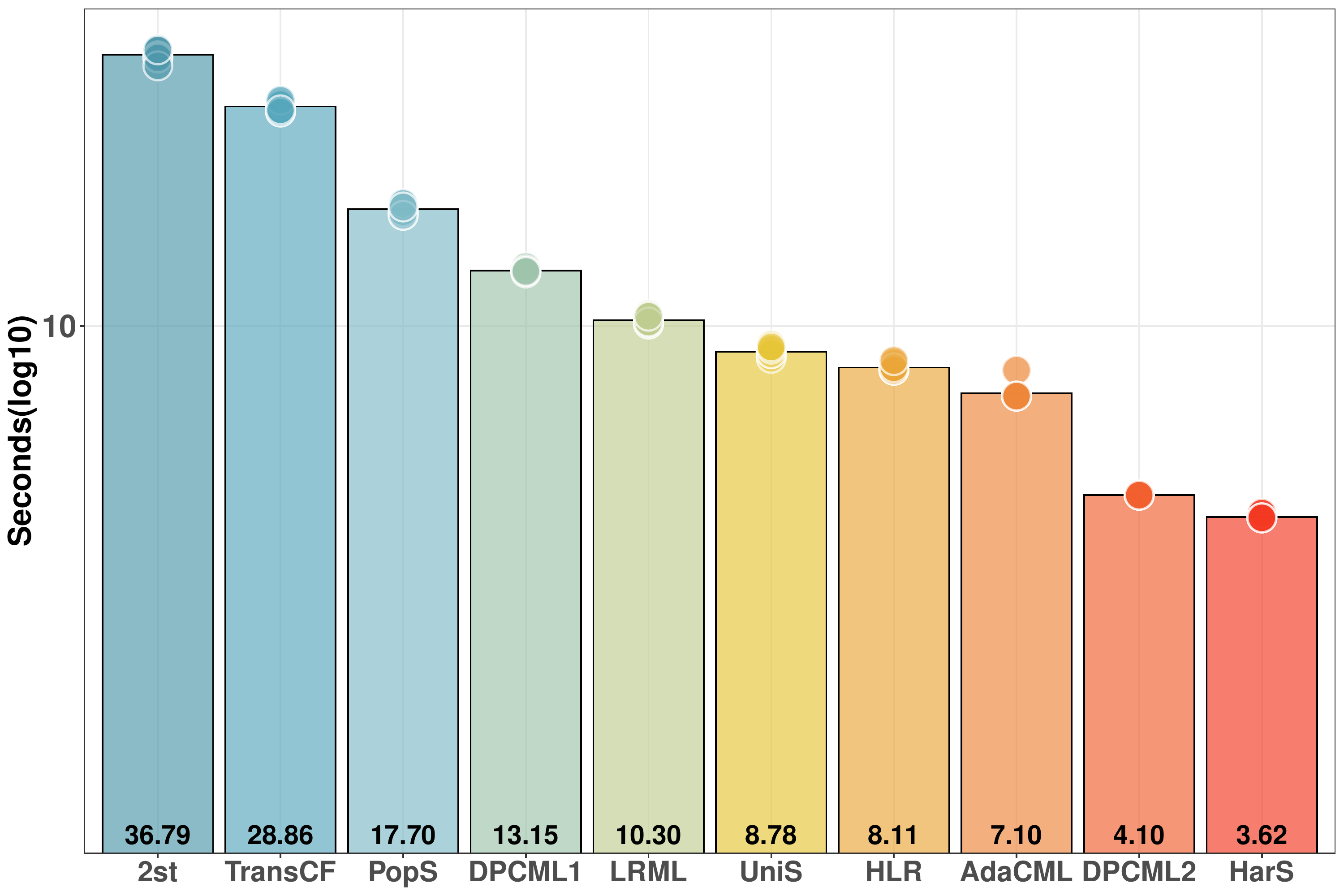}
			\label{ab.sub.200k}
		}
		\subfigure[CiteULike]{
			\includegraphics[width=0.45\textwidth]{img/training_time/CiteULike.pdf}
			\label{ab.sub.citeulike}
		}
		\subfigure[MovieLens-10m]{
			\includegraphics[width=0.45\textwidth]{img/training_time/ml-10m.pdf}
			\label{ab.sub.ml-10m}
		}
		\caption{Training efficiency comparison among CML-based competitors.}
		\label{app:runtime}
	\end{figure*}
	
\begin{table}[htbp]
  \centering
  \caption{Ablation studies of DPCML1 on Steam-200k dataset.}
    \begin{tabular}{c|cccccccc}
    \toprule
     Method & P@3 & R@3 & NDCG@3 & P@5 & R@5 & NDCG@5 & MAP & MRR \\
    \midrule
    w/o DCRS & 23.86 & 13.06 & 24.90 & 23.57 & 11.56 & 24.77 & 20.37 & 44.38 \\
    DCRS-$\delta_1$ & 24.28 & 14.38 & 25.61 & 22.48 & 11.35 & 24.13 & 21.02 & 45.76 \\
    DCRS-$\delta_2$ & 24.56 & 14.36 & 25.41 & 23.82 & 11.97 & 24.74 & 21.67 & 45.14 \\
    DCRS & \textbf{25.39} & \textbf{14.84} & \textbf{26.56} & \textbf{23.88} & \textbf{12.11} & \textbf{25.25} & \textbf{22.26} & \textbf{46.79} \\
    \bottomrule
    \end{tabular}%
  \label{tab:DPCML1}%
\end{table}%

\begin{table}[htbp]
  \centering
  \caption{Ablation studies of DPCML2 on Steam-200k dataset.}
    \begin{tabular}{c|cccccccc}
    \toprule
    Method & P@3 & R@3 & NDCG@3 & P@5 & R@5 & NDCG@5 & MAP & MRR \\
    \midrule
    w/o DCRS & 27.96 & 15.68 & 29.42 & 27.85 & 13.94 & 29.56 & 22.50 & 50.03 \\
    DCRS-$\delta_1$ & 28.28 & 16.05 & 29.60 & 27.25 & 13.75 & 29.17 & 22.63 & 50.00 \\
    DCRS-$\delta_2$ & 29.26 & 16.83 & 30.61 & 28.47 & 14.28 & 30.16 & 23.86 & 51.14 \\
    DCRS & \textbf{29.88} & \textbf{17.13} & \textbf{31.22} & \textbf{28.70} & \textbf{14.51} & \textbf{30.56} & \textbf{24.10} & \textbf{51.95} \\
    \bottomrule
    \end{tabular}%
  \label{tab:DPCML2}%
\end{table}%

\subsubsection{The Effectiveness of DCRS for MF-based Systems}
To see this, we attempt to apply the proposed diversity control regularization scheme (DCRS) for M2F \cite{DBLP:conf/recsys/WestonWY13,DBLP:conf/eaamo/GuoKJG21}. In addition, we further explore the effectiveness of DCRS for the general framework of joint accessibility (GFJA, Eq.(\ref{eq3123}) in the main paper). Here we also conduct a grid search to choose the best performance of M2F with DCRS on the Steam-200k and MovieLens-1m datasets, where the parameters space stays the same as DPCML. The experimental results are summarized in Tab.\ref{tab:reg_for_MF}. From the above results, we can draw the following observations: 1) The proposed DCRS does not work well for MF-based models. A possible reason here is that the metric space of MF-based and CML-based methods are intrinsically different. MF adopts the inner-product space while CML adopts the Euclidean space. In this paper, we merely consider the DCRS for Euclidean space. The corresponding strategy for the inner-product space is left as future work. 2) In most metrics, GFJA+DCRS could outperform GFJA significantly, which supports the advantages of our proposed DCRS. 3) Compared with M2F, the performance gain of GFJA is sharp on both datasets. This suggests the superiority of our proposed method against the current multi-vector-based competitors.

\begin{table}[htbp]
  \centering
  \caption{Performance comparison of joint accessibility model equipped with DCRS on the Steam-200k and MovieLens-1m datasets.}
    \begin{tabular}{c|cccccccc}
    \toprule
    \multicolumn{9}{c}{Steam-200k} \\
    \midrule
    Method & P@3 & R@3 & NDCG@3 & P@5 & R@5 & NDCG@5 & MAP & MRR \\
    \midrule
    M2F & 11.33 & 5.69 & 11.95 & 11.44 & 5.73 & 12.98 & 6.43 & 25.05 \\
    M2F+DCRS & 10.92 & 5.58 & 11.49 & 10.89 & 5.48 & 12.37 & 6.25 & 24.26 \\
    GFJA & 21.53 & \textbf{12.60} & 22.52 & 20.37 & \textbf{10.16} & 21.49 & 19.32 & 40.69 \\
    GFJA+DCRS & \textbf{21.63} & 12.40 & \textbf{22.72} & \textbf{20.38} & 9.98 & \textbf{21.74} & \textbf{19.53} & \textbf{40.92} \\
    \midrule
    \multicolumn{9}{c}{MovieLens-1m} \\
    \midrule
    M2F & 8.61 & 1.84 & 9.36 & 7.60 & 2.30 & 8.67 & 2.95 & 20.40 \\
    M2F+DCRS & 7.59 & 1.49 & 8.16 & 7.10 & 2.02 & 7.92 & 2.53 & 18.51 \\
    GFJA & 15.79 & 3.19 & 16.11 & 16.02 & 4.77 & 16.66 & 11.04 & 32.54 \\
    GFJA+DCRS & \textbf{16.71} & \textbf{3.54} & \textbf{16.94} & \textbf{17.24} & \textbf{5.27} & \textbf{17.71} & \textbf{11.75} & \textbf{33.87} \\
    \bottomrule
    \end{tabular}%
  \label{tab:reg_for_MF}%
\end{table}%

\end{document}